\documentclass[aps,prd,twocolumn,nofootinbib,superscriptaddress,longbibliography]{revtex4}
\usepackage{latexsym}
\usepackage{hyperref}
\usepackage{amssymb}
\usepackage{epsfig,amsmath,graphics}
\usepackage{pstricks}
\usepackage{color}
\usepackage{dcolumn}
\usepackage{ulem}
\usepackage{xcolor}
\usepackage{placeins}
\usepackage{siunitx}
\usepackage{physics}
\usepackage{float}

\definecolor{linkcolor}{rgb}{0.0, 0.28, 0.67}
\hypersetup{colorlinks=true,
linkcolor=black,
citecolor=linkcolor,
urlcolor=linkcolor,
linktocpage=true
}

\newcommand{\Tstep}{T_{\textrm{step}}}

\newcommand{\Neff}{N_{\rm eff}}
\newcommand{\dNeff}{\triangle N_{\rm eff}}

\newcommand{\neff}{N_{\rm eff}}
\newcommand{\Nint}{N_{\rm int}}

\newcommand{\nudark}{\ensuremath{\nu_{d}}}
\newcommand{\mdark}{m_{\nu d}}

\newcommand{\alphad}{\alpha_{d}}
\newcommand{\eq}{{\rm eq}}
\newcommand{\Tequil}{T_{\eq}}
\newcommand{\gnuint}{g_*^{\nu{\rm int}}}

\newcommand{\Tn}{T_{\nu}}
\newcommand{\Tnz}{T_{\nu0}}
\newcommand{\rhonu}{\rho_{\nu}}
\newcommand{\rhonuint}{\rho_{\nu}^{\rm int}}
\newcommand{\rhonuz}{\rho_{\nu0}}
\newcommand{\rhonuintz}{\rho_{\nu0}^{\rm int}}
\newcommand{\rhod}{\rho_{d}}
\newcommand{\fr}{q}

\usepackage{comment}
\newcommand{\be}{\begin{equation}}
\newcommand{\ee}{\end{equation}}

\newcommand{\kev}{{\rm keV}}
\newcommand{\Mev}{{\rm MeV}}

\def\bea{\begin{eqnarray}}
\def\eea{\end{eqnarray}}
\def\ltap{\ \raise.3ex\hbox{$<$\kern-.75em\lower1ex\hbox{$\sim$}}\ }
\def\gtap{\ \raise.3ex\hbox{$>$\kern-.75em\lower1ex\hbox{$\sim$}}\ }
\def\lsim{\ \raise.3ex\hbox{$<$\kern-.75em\lower1ex\hbox{$\sim$}}\ }
\def\gsim{\ \raise.3ex\hbox{$>$\kern-.75em\lower1ex\hbox{$\sim$}}\ }

\usepackage{slashed}

\usepackage{color}

\newcommand{\ignore}[1]{}
\newcommand{\beq}{\begin{equation}}
\newcommand{\eeq}{\end{equation}}
\newcommand{\bear}{\begin{eqnarray}}
\newcommand{\eear}{\end{eqnarray}}

\def\kev{\,{\rm keV}}
\def\MeV{\,{\rm MeV}}

\def\ev{\,{\rm eV}}
\def\lcdm{$\Lambda{\rm CDM}$}

\begin{document}

\title{Neutrino-Dark Sector Equilibration and Primordial Element Abundances}

\author{Cara Giovanetti}
\affiliation{Center for Cosmology and Particle Physics, Department of Physics, New York University, New York, NY 10003, USA}
\author{Martin Schmaltz}
\affiliation{Physics Department, Boston University, Boston, MA 02215, USA}
\author{Neal Weiner}
\affiliation{Center for Cosmology and Particle Physics, Department of Physics, New York University, New York, NY 10003, USA}
\begin{abstract}
After neutrinos decouple from the photon bath, they can populate a thermal dark sector. If this occurs at a temperature above $\sim 100 \kev$, this can have measurable impacts on light element abundances. We calculate light element abundances in this scenario, studying the impact from rapid cooling of the Standard Model neutrinos, and from an increase in the number of relativistic degrees of freedom $\Neff$, which can occur in the presence of a mass threshold. We incorporate these changes in the publicly available BBN code PRIMAT, using the reaction networks from PRIMAT and from the BBN code PArthENoPE, to calculate Y$_{\rm{P}}$ and D/H. We provide limits from the two different reaction networks as well as with expanded errors to include both results. If electron neutrinos significantly participate in the cooling, we find limits down to temperatures as low as 100 \kev. If electron neutrinos are weakly participating (for instance if only the mass eigenstate $\nu_3$ equilibrates), cooling places no limits. However, if the dark sector undergoes a ``step'' in $\Neff$, there can be additional, $\omega_b$-dependent constraints. These limits can vary from strong (for low values of $\omega_b$) to a mild preference for new physics (for high values of $\omega_b$). Future analyses including upcoming CMB data should improve these limits. 
	
\end{abstract}

\maketitle

\section{Introduction}
One of the great successes of particle cosmology is the consistency of Big Bang cosmology with a current CMB temperature $T=2.7K$, the measured value of $\Omega_b$, the presence of three light neutrinos in the Standard Model (SM), and the measured primordial abundances of helium-3 ($^3$He), helium-4 ($^4$He) and deuterium (D). The formation of these elements is sensitive to physics in temperature ranges of $100 \, \kev$ to $\sim 10\, \Mev$, at times from a few seconds until a few minutes in the life of the Universe.

As measurements of primordial $^4$He and D have achieved percent precision, we are capable of asking questions about the properties of the Universe in that era, with a promise of getting quantitative answers.

One such question concerns the nature of the ``dark radiation'' of the universe. It is now well established both through BBN and the CMB that a sizable fraction of the energy density of the universe at early times is in the form of some dark radiation. The SM provides a natural explanation for this radiation in the form of SM neutrinos, which are in thermal contact with the photon bath until temperatures near a few $\Mev$ at which point the neutrinos decouple.

Beyond simple scientific rigor, there are important reasons to want to test this explanation. For instance, other (near-)massless states in thermal contact with the SM at earlier times will generally add to this dark radiation. Current 95\% constraints limit additional radiation $\dNeff\lsim0.4$ (BBN), $\dNeff\lsim 0.33$ (CMB+BAO for $\Lambda$CDM+$\dNeff$), and $\dNeff\lsim 0.23$ (BBN+CMB) (see, e.g., \cite{Pitrou_2018,Planck:2018vyg,Yeh:2022heq}), allowing ample room for new particles.

Dark states can also remain depopulated until late times when they can come into equilibrium \cite{Berlin_2019,Aloni:2023tff}. For instance, a simple scenario was recently studied in \cite{Aloni:2023tff}, wherein it was demonstrated that a neutral fermion, mixing with SM neutrinos, would naturally thermalize, even for mixing angles as small as $10^{-14}$. The temperature of thermalization is set by the mass of the mixing dark particle, implying that a late $O(1)$ change to $\neff$ is natural as long as some interacting particle with mass below an MeV exists, which mixes with SM neutrinos.\footnote{A late change to $\neff$ can also arise if a massive particle decays at late times, adding energy to the dark sector. Typically this requires a fine tuning for the energy density to be an $O(1)$ change to $\neff$, rather than much larger or smaller.}

If this thermalization happens sufficiently late, it can evade all constraints from BBN, while yielding a signal in the CMB. However, this precise boundary can have important implications for models. 

In this Letter, we shall study the effects of dark sectors equilibrating with neutrinos at temperatures near 1 \Mev. The layout of this paper is as follows: in Sec.~\ref{sec:lnebbn} we describe the process of neutrino equilibration with a dark sector and the physics phenomena that can affect BBN. In Sec.~\ref{sec:limits} we calculate effects and give limits on these new states. Finally, in Sec.~\ref{sec:disc} we conclude, and discuss what future measurements could strengthen these constraints. 

\section{Late Neutrino Equilibration and BBN}
\label{sec:lnebbn}
Neutrinos can equilibrate with a neutral sector via mass mixing; this is well understood \cite{Dolgov:1980cq,Barbieri:1990vx,Enqvist:1990ad,Sigl:1993ctk,McKellar:1992ja,Dodelson_1994,Dasgupta:2021ies}. If neutrinos, mixed with a neutral fermion $\nudark$, oscillate and then scatter at a rate higher than Hubble, they will equilibrate $\nudark$ and any other particles with significant couplings to $\nudark$. Specifically, one requires
\begin{align}
    \Gamma_{\nu \rightarrow d} = \frac{1}{4}\sin^2 2\theta_m \, \Gamma_{scatter} \gsim H,
\end{align}
where $\theta_m$ is the effective mixing angle at finite density and temperature, and $\Gamma_{scatter}$ is the scattering rate off of any background particles. 

As shown by \cite{Aloni:2023tff} (and see Appendix~\ref{app:DS_eq}), if $\nudark$ has a self interaction with a light mediator, this is a highly IR-dominated process, with equilibration temperature
\begin{align}
    \Tequil \simeq \mdark\left(\frac{\sin^2 2\theta_0 M_{pl}}{\mdark}\right)^{1/5}.
\end{align}
Here $\theta_0$ is the mixing angle in vacuum. As a consequence of the $1/5$ power, equilibration typically happens not far above the mass of the dark fermion, $\mdark$. If $\mdark < \Mev$, equilibration can naturally occur after neutrino decoupling. Importantly, neglecting second-order effects, this thermalization does not increase $\Neff$, as the out-of-equilibrium process only redistributes the energy. 

However, there are three primary effects that impact light element abundances. They are
\begin{itemize}
    \item{{\bf Incomplete neutrino decoupling} - Although neutrinos are no longer in equilibrium, there are still relevant processes exchanging energy between the photon/electron bath and neutrino bath (the second-order effects referenced above). Because neutrinos are depleted, the transfer back from the neutrino bath is suppressed compared to SM values, leading to an increase in $\Neff$.}
    \item{{\bf A suppression of $n \nu_e \rightarrow p\, e$} - If electron neutrinos are depleted and cooled, there is reduction in the level of neutron to proton conversion, leading to more neutrons at late times. This is true even if neutrino cooling occurs after neutrons are no longer tracking their equilibrium distributions.}
    \item{{\bf Changing $\dNeff$ through a mass threshold ``step''} - Passing through a mass threshold in the dark sector while in equilibrium naturally leads to an increase in $\dNeff$ (see \cite{Berlin:2019pbq,Aloni_2022}). This will increase Hubble, accelerating freezeout.}
\end{itemize}

The first two of these are a consequence of the neutrino cooling, while the third is a consequence of the overall dark radiation heating up. Thus, we separate these two cases and first describe the physics relevant for each before providing limits in the next section.

In both of these cases, we wish to place constraints on (or someday find evidence for) signs of new physics near or below 1 $\MeV$. However, the limits we find are dependent on the range of $\omega_b$ that is assumed. Because models that modify physics after BBN often change cosmology during recombination, there is no universal value of $\omega_b$ to be used. On the other hand, if we assume a large range of $\omega_b$ (much larger than the error in a typical model), then limits can be washed out. Thus, in order to understand what limits can arise here, we choose three cases for $\omega_b$. First, we consider a ``medium'' value $\omega_b = 0.0221$, which is the value extracted from Planck 2018 \cite{Planck:2018vyg} in the model where $\neff$ and $m_\nu$ can float. Next, we consider a ``low'' value $\omega_b = 0.02154$, which was found in a specific analysis of Early Dark Energy \cite{Hill:2021yec}.\footnote{This should not be taken as representative for EDE fits, as many have higher values of $\omega_b$.} Finally, we consider a ``high'' value $\omega_b = 0.02287$, which was the value found in an analysis of WZDR+\cite{Joseph:2022jsf}, a model of dark matter interacting with dark radiation. We take a uniform error of $0.00022$ in all three cases, so that the different exclusions represent the change in the central values, not varying error bars. 

In what follows, we shall refer to a range of different energy densities, temperatures and other quantities. For compactness, we summarize these in Table \ref{tb:notation}.
\begin{table}[]
\begin{tabular}{|c|p{0.35\textwidth}|}\hline
  \textbf{Quantity}   &  \textbf{Definition}\\ \hline\hline
   $\rhonu$  & Energy density of all SM neutrinos\\
   \hline
   $\rhonuz$ & Energy density of a single, unperturbed neutrino in the SM, tracking standard cosmic evolution, $\rhonuz\propto \Tnz^4 \propto a^{-4}$\\
   \hline
   $\Tnz$ & Temperature of an unperturbed neutrino in the SM, tracking standard cosmic evolution\\
   \hline
   $\rhonuint$ & Energy density of the SM neutrino states which are oscillating and scattering into the dark sector\\
   \hline
   $\Nint$& Number of SM neutrino states (1,2 or 3) thermalizing with the dark sector\\
   \hline
   $\rhod$& Energy density in the dark sector\\
   \hline
\end{tabular}
    \caption{Definitions of various symbols used.}
    \label{tb:notation}
\end{table}

\subsection{Neutrino cooling}
The consequences of neutrino cooling are flavor-dependent, and vary depending on the fraction of the electron neutrinos which are removed. We will consider a few different cases.

As discussed in \cite{Aloni:2023tff}, if the neutrinos equilibrating are not mass eigenstates, {\it flavor} (not SM-dark) oscillations will equilibrate the flavors of the SM. I.e., all $\nu_e$, $\nu_\mu$, $\nu_\tau$ will have their temperature reduced as the dark sector equilibrates. However, if dominantly a single mass eigenstate mixes with the dark sector then only the flavor content of that eigenstate will equilibrate. In other words, if a single mass eigenstate has an electron fraction $f_e$ (i.e., $|\bra{\nu_e}\ket{\psi}|^2 = f_e$), and that mass eigenstate equilibrates with the dark sector while the other mass eigenstates do not, then $1-f_e$ of the electron neutrinos will retain their old distribution, and $f_e$ will equilibrate. In the case that two mass eigenstates mix, we care about the sum, i.e., $\sum |\bra{\nu_e}\ket{\psi_i}|^2 = f_e$

Consequently, we need to distinguish between possible scenarios where $f_e \ne 1$ is thermalized. Recent measurements of the PMNS matrix imply the electron component of the different mass eigenstates are approximately $0.02\lsim|U_{e3}|^2 \lsim 0.024$, $0.26\lsim |U_{e2}|^2 \lsim 0.34$ and $0.64\lsim |U_{e1}|^2 \sim 0.72$ \cite{Esteban:2020cvm,nufit52}.
We  consider the following four scenarios for equilibration: ({\it i}) only $v_3$ (mass eigenstate 3, $f_e=.02$) equilibrates, ({\it ii}) both $v_2$ and $v_3$ ($f_e=0.35$), ({\it iii})  only $\nu_1$ ($f_e=0.65$) and ({\it iv}) all eigenstates ($f_e=1$). Our results do not depend significantly on the precise values chosen for $f_e$. The value of $f_e$ will factor into the calculation as we will describe.
\subsubsection{Neutrino Cooling and BBN}

The conversion rate for SM neutrinos to convert to dark sector particles is \cite{Aloni:2023tff}
\begin{align}
\langle \Gamma_{\nu\rightarrow d}\rangle = \frac{ \frac14\sin^2{2\theta_0}(3c_\Gamma \Tnz^5 G_F^2 + \alphad^2 \frac{T_d^2}{\Tnz})}{\left(\cos{2\theta_0} + \alphad \frac{T_d^2}{\mdark^2}+18c_V\frac{G_F^2 \Tnz^6}{\mdark^2} \right)^2+\sin^2{2\theta_0}} \,.
\label{eq:gammarate}
\end{align}
This appears in the Boltzmann equation for energy in the dark sector
\begin{align}
\label{eq:Boltzmann}
\frac{d}{d\log a} \left(a^4 \rhod\right) = \frac{\langle \Gamma_{\nu\rightarrow d}\rangle }{H} a^4
\left(\rhonuint -\left.\frac{\rhonuint}{\rhod}\right|_{\eq} \rhod \right)\ .
\end{align}

In a broad region of parameter space, the dark sector will come into equilibrium with the neutrinos. Assuming the dark sector remains self-thermalized, one can show over much of the parameter space that the fraction of the neutrino energy density which has been transferred to the dark sector from $\Nint$ interacting (equilibrating) neutrinos, $\fr\equiv \rhod/(\Nint\, \rhonuz)$ grows as
\bea
\fr(\Tnz)=\fr_{\eq}\,(1-e^{-(\Tequil/\Tnz)^{10/3}})\ .\label{eq:q_intext}
\eea

\begin{figure}[t]
   \hspace*{0.35cm}
    \includegraphics[width=0.41\textwidth]{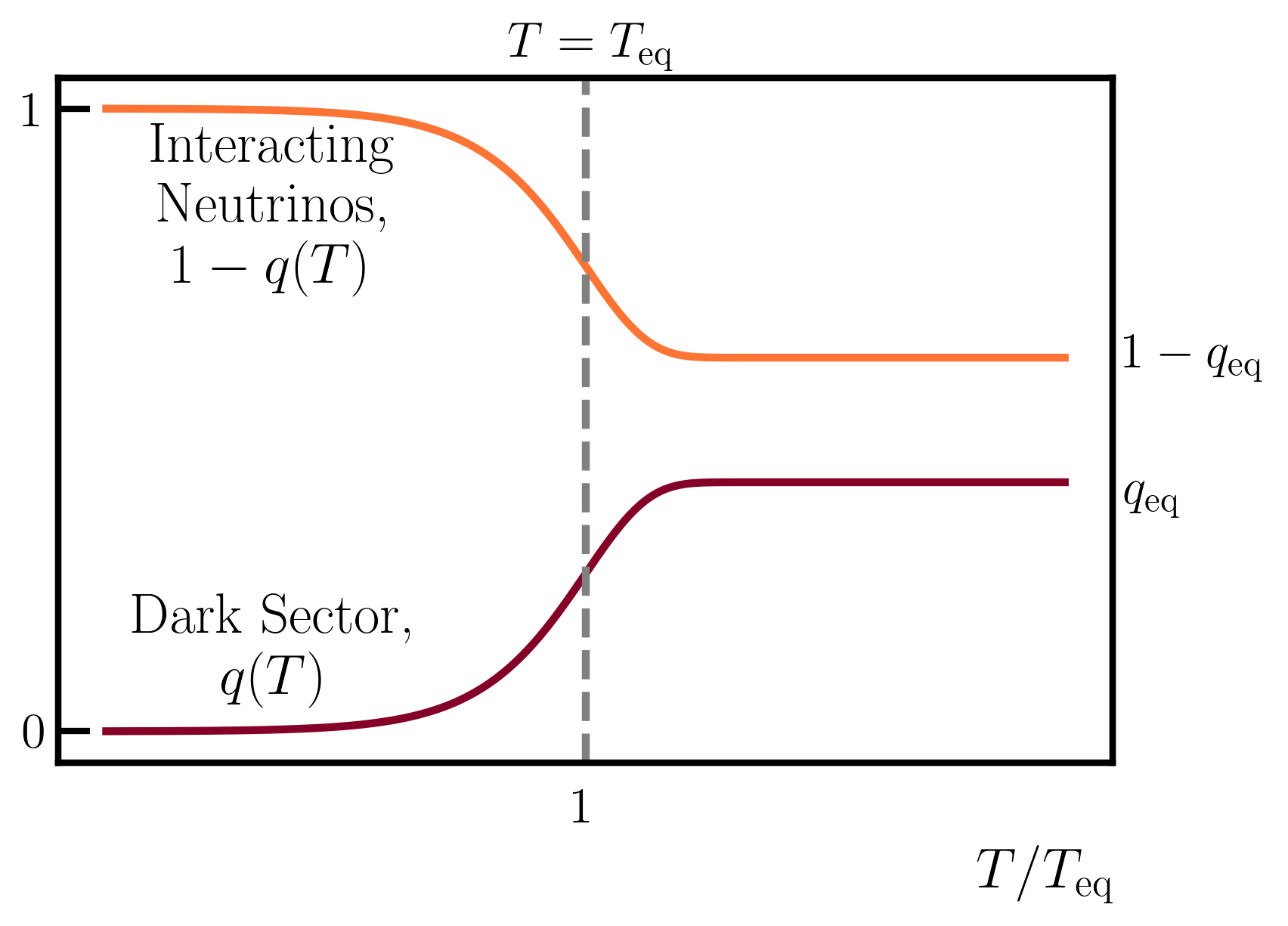}\hspace*{1.5cm}
    \caption{Evolution of the dark radiation and interacting (thermalizing) neutrinos versus background neutrino temperature.  $q(T)$ and $1-q(T)$ describe the energy densities of the equilibrating fluids, with $q(T) = \rho_d/(N_{\rm{int}}\rho_{\nu 0})$ and $1-q(T)=\rho_{\nu}^{\rm{int}}/(N_{\rm{int}}\rho_{\nu 0})$.}
    \label{fig:qevolve}
\end{figure}

The energy densities of the separate fluids then evolve according to Eq.~\ref{eq:q_intext} and
\bea
    \rho_d(\Tnz)&=&\Nint\, \rhonuz(\Tnz)\, \fr(\Tnz) \\
    \rho^{\rm{int}}_{\nu}(\Tnz)&=&\Nint\, \rhonuz(\Tnz) (1-\fr(\Tnz))
\eea
We show this evolution in Fig. \ref{fig:qevolve}.

As the dark sector comes into equilibrium with the SM, the first relevant effect is the cooling of neutrinos, and that is what we shall first explore. As discussed above, we assume that a fraction of the electron neutrinos $f_e$ are cooled to a thermal distribution at lower temperature via equilibration with the dark sector, while the remaining portion $1-f_e$ retains its previous thermal distribution.
In reality, the conversion probability is dependent on the energy of the neutrino, so this is only an approximation, although generally a very good one.  When this equilibration happens at the temperature $\Tequil$ the energy density in dark radiation becomes large, and neutrinos are correspondingly cooled.  The net effect to first order is no change in $\Neff$; the energy density depletion of the neutrino species is exactly compensated by the newly-equilibrated dark radiation.  

\begin{figure}[t]
    \centering
   \includegraphics[width=0.35\textwidth]{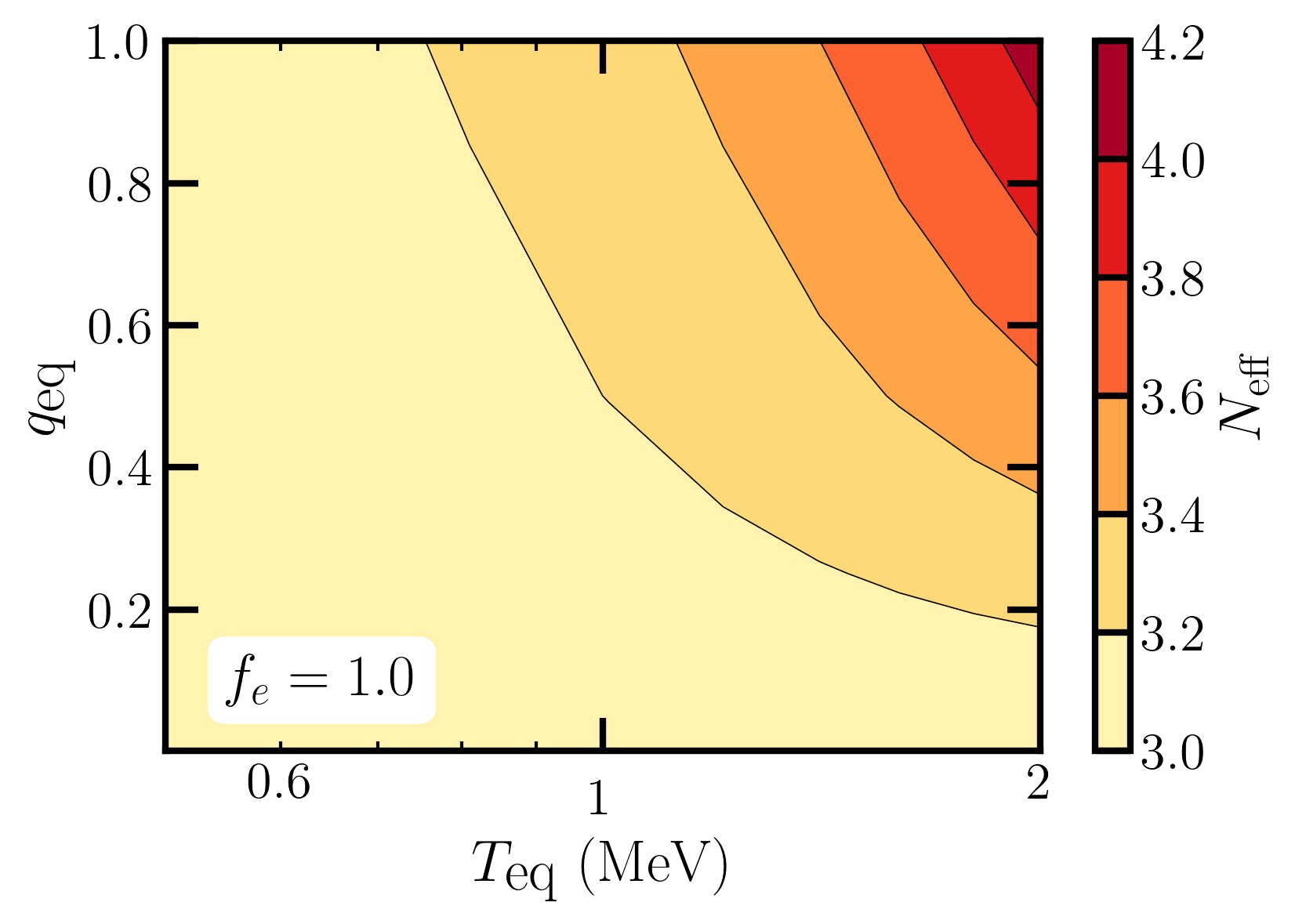}
    \caption{Change in $\Neff$ after neutrino cooling at $T=\Tequil$ as a function of $q_\eq$. Recall $q_\eq=1$ corresponds to complete cooling of the equilibrating neutrinos and $q_\eq=0$ is the SM.}
    \label{fig:nuheat}
\end{figure}

However, if $\Tequil$ is larger than an MeV, there may be noticeable increases in $\Neff$ due to incomplete neutrino decoupling.  The electron neutrino can still exchange energy with electrons even after decoupling, and therefore early cooling of electron neutrinos can subsequently cool the photon plasma, leading to a sometimes noticeable increase in $\Neff$. Although this effect is contained in our full limits, it is instructive to separate out this effect. We show in Fig. \ref{fig:nuheat} the change to $\Neff$ from neutrino cooling. This effect is usually subdominant, but not irrelevant, in the final changes to light element abundances. 

\begin{figure}[t]
    \centering
    \includegraphics[width=0.235\textwidth]{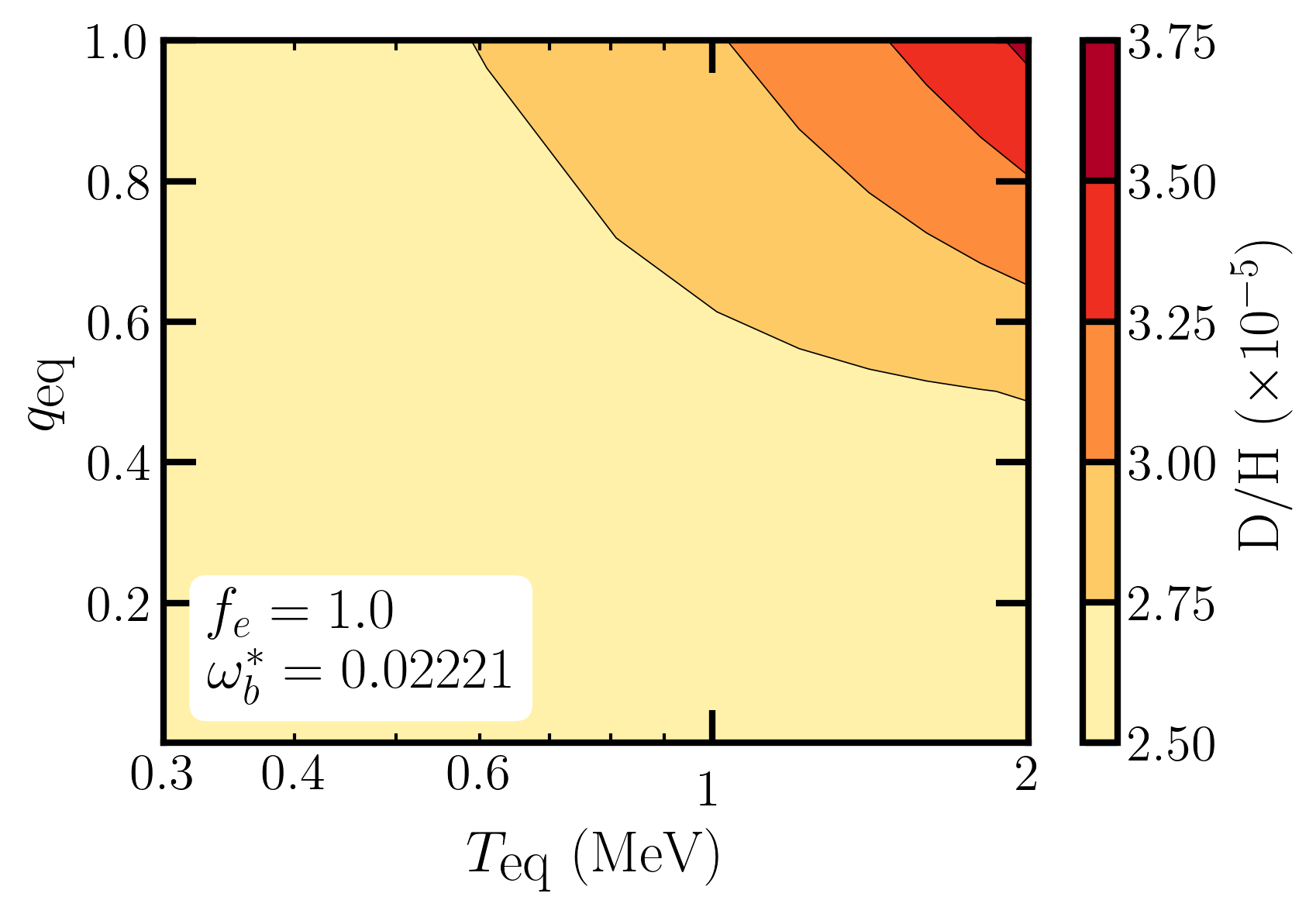}
    \includegraphics[width=0.235\textwidth]{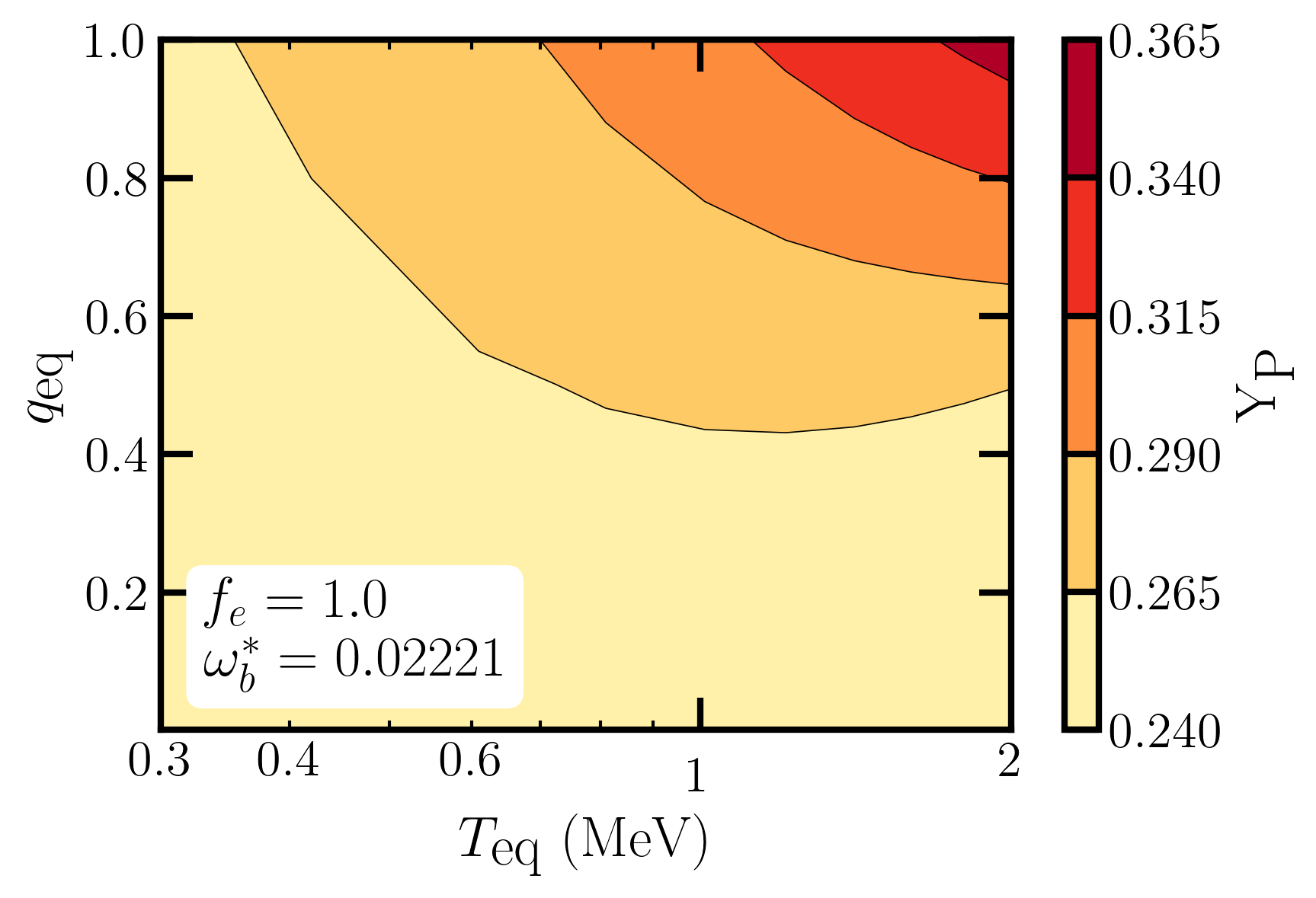}
    \caption{Changes in abundances due to neutrino cooling.  The reaction network used to calculate these abundances encompasses the predictions of PRIMAT~\cite{Pitrou_2018} and PArthENoPE~\cite{Consiglio_2018}; see Section~\ref{sec:limits}.  To be compared with the experimental values $\textrm{Y}_{\textrm{P}}=0.245 \pm 0.003$ \cite{Workman_2022} and $10^5\textrm{D/H}=2.527\pm 0.030$ \cite{Cooke_2018}.}
    \label{fig:abundances}
\end{figure}

A more important consequence of cooling neutrinos during BBN is the effect on proton/neutron interconversion.  Even if $\Neff$ remains largely unchanged, $p-n$ processes are still affected directly by changes in the neutrino temperature.  A reduction in $T_{\nu}$ suppresses the non-equilibrium (post $p-n$ freezeout) process of $\nu_e\, n \rightarrow p\, e$, favoring a larger abundance of neutrons.  This change in the relative abundances of protons and neutrons directly translates to a change in abundances of light nuclei, shown in Fig.~\ref{fig:abundances} for $^4$He and D (depicted as the $^4$He mass fraction, Y$_{\textrm{P}}$, and the ratio of deuterium to hydrogen, D/H). 

\subsection{Steps in $\neff$}
If the dark sector equilibrates with only $\mu$ or $\tau$ neutrinos, the effects we have just described do not arise, and there is no immediate effect on the light element abundances. However, if the equilibrated dark sector contains a massive particle, then as the temperature drops below its mass, $\neff$ increases, increasing the Hubble rate compared to the case without a step. This accelerates freezeout processes, impacting BBN\@. 

There is no \textit{a priori} reason to exclude additional electron neutrino cooling in this scenario,  but then the effects from the previous section and from the step would combine and be difficult to separate. In order to decouple the consequences of this step in $\Neff$ from the effects of electron neutrino cooling, we study signals under the assumption that the electron neutrino cooling is negligible. As we shall see, this is likely to be a reasonable limit up to temperatures of $2 \MeV$. In principle, changes in Hubble are possible above this temperature, and we show steps at higher temperatures as well, although achieving this without influencing electron neutrinos may be difficult to realize in the specific scenario we have described. 

The change in $\Neff$ can have different implications depending on the temperature at which the step occurs, but primarily it accelerates freezeout of ongoing processes. At early times, this can accelerate $n-p$ freezeout, leading to a higher abundance of neutrons and therefore also Helium. (This is similar to a simple time-independent $\triangle\Neff$.)    If the step occurs at lower temperatures, well after weak freezeout but before the end of BBN, then Y$_{\rm{P}}$ is still fixed by the number of neutrons leftover from weak freezeout, and does not feel an effect of the step. In contrast, 
D/H is enhanced because of an earlier freezeout of deuterium annihilation. 

\section{Limits}
\label{sec:limits}
With the physical effects understood, we can proceed to study the regions of parameter space where either of the two scenarios described above are permissible or preferred by the data.  
To obtain these limits, we use a modified version of \texttt{nudec\_BSM} \cite{Escudero_2019}, which we use to calculate the relevant thermodynamic quantities ($T_{\gamma}$, $T_{\nu}$, Hubble, and scale factor) in a given scenario.  
Then, we feed the precalculated thermodynamic history into the BBN code PRIMAT \cite{Pitrou_2018}, which has been modified to read in the thermodynamic output of \texttt{nudec\_BSM}.  
The modifications to these codes are similar to those that appear in \cite{Giovanetti_2022}.  
We further modified PRIMAT to account for equilibration with a fraction $f_e$ of neutrino species, as it is essential to specifically track the temperature of electron neutrinos in PRIMAT's calculation of proton/neutron interconversion.

PRIMAT uses results from \cite{Gomez_2017} for D(D,n)$^3$He and D(D,p)$^3$H in its reaction network.\footnote{Our default PRIMAT reaction network is \texttt{BBNRatesAC2021.dat}, which uses the results from \cite{Mossa_2020} for D(p,$\gamma$)$^3$He.} Another public BBN code, PArthENoPE \cite{Consiglio_2018} uses rates calculated in \cite{Pisanti_2021}, leading to a different prediction for the abundance of deuterium. Depending on $\omega_b\equiv\Omega_b h^2$ this can yield qualitatively different results, with PArthENoPE being largely consistent with observed D/H, while PRIMAT had a $\sim 2 \sigma$ tension.  This discrepancy was explored in \cite{Pitrou_2021,Pitrou_2021b,Pisanti_2021}.  In our analyses, since we are interested in scenarios where the physics at recombination can differ from that of \lcdm, $\omega_b$ can naturally be different and this tension is less pronounced, as we shall see.

Nonetheless, with the awareness of potential discrepancies between the two reaction networks, we include both in our study. To remain agnostic about which choice of reaction rates is more ``correct", we created an additional rate file that includes the PArthENoPE rates for  D(D,n)$^3$He and D(D,p)$^3$H, and have included results for both the default PRIMAT network and this PArthENoPE network.  To estimate the uncertainties in our results, we employ linear covariance estimation, as detailed in \cite{Fiorentini_1998} and implemented for PRIMAT in \cite{Giovanetti_2022}.

We also include results for a ``combined" network, in which we perform an unweighted average to select a central value for our BBN predictions somewhere in between the predictions of these two independent networks.  We then define the $1\sigma$ region around these central values to encompass the predictions of both networks.  This procedure allows exploration of conservative constraints on new physics during BBN, without needing to assert a preference for a particular reaction network.

Throughout, we compare to the experimental average for Y$_{\textrm{P}}$ presented in \cite{Workman_2022} and the value for D/H obtained in \cite{Cooke_2018},
\begin{align}
    \textrm{Y}_{\textrm{P}}&=0.245 \pm 0.003\label{eq:Yp_exp}\\
    10^5\textrm{D/H}&=2.527\pm 0.030\label{eq:DH_exp}.
\end{align}

\subsection{Cooling limits}
We first consider the case where the only relevant effect is the cooling of neutrinos as they equilibrate. The relevant quantities are $\Tequil$, where the dark sector equilibrates, $f_e$, the cumulative fraction of electron neutrino in the equilibrating neutrino states, and $q_\eq$, the asymptotic fraction of the equilibrating energy density that ends up in the dark sector. 

\begin{figure}[t]
    \centering
\includegraphics[width=0.23\textwidth]{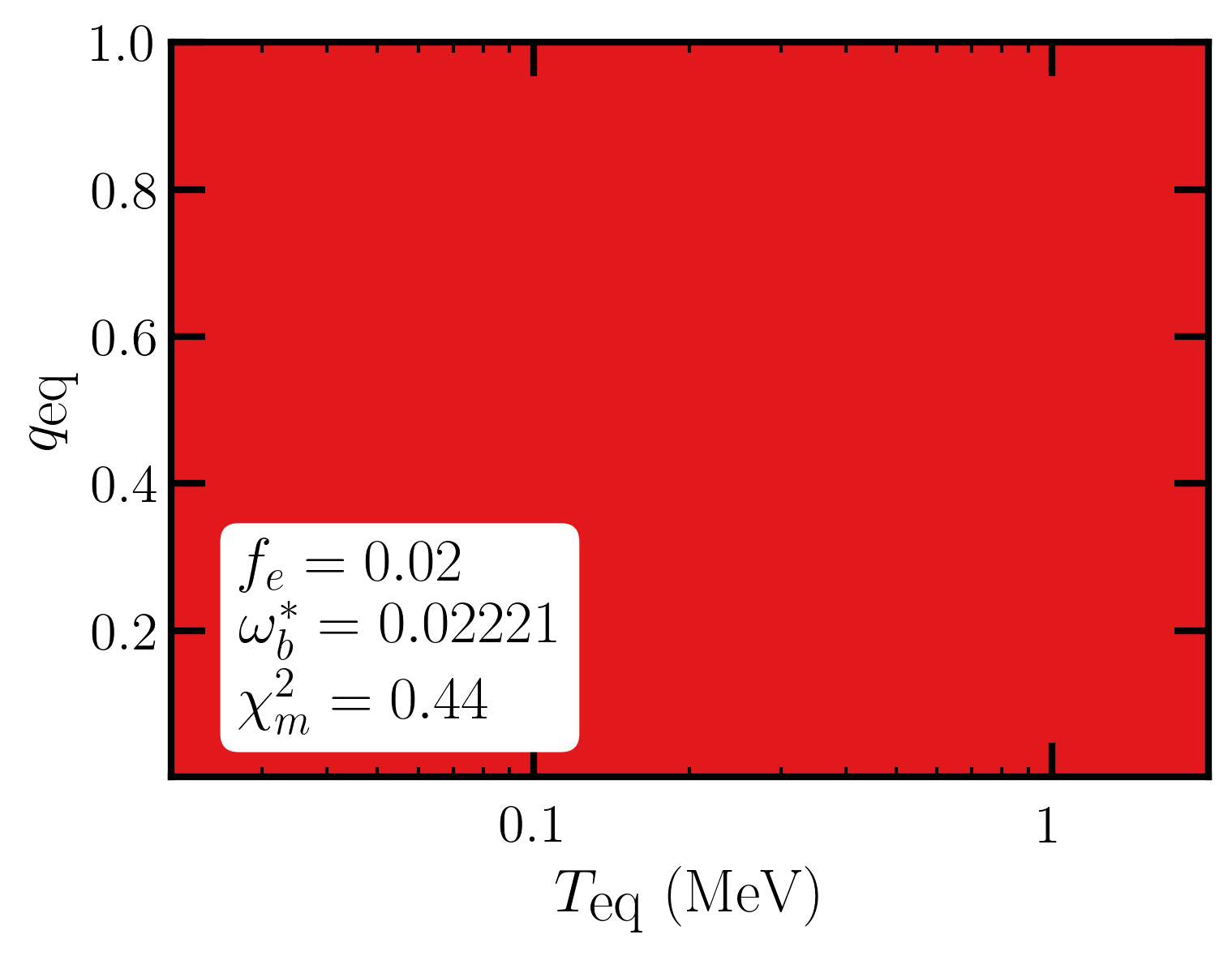}
    \includegraphics[width=0.23\textwidth]{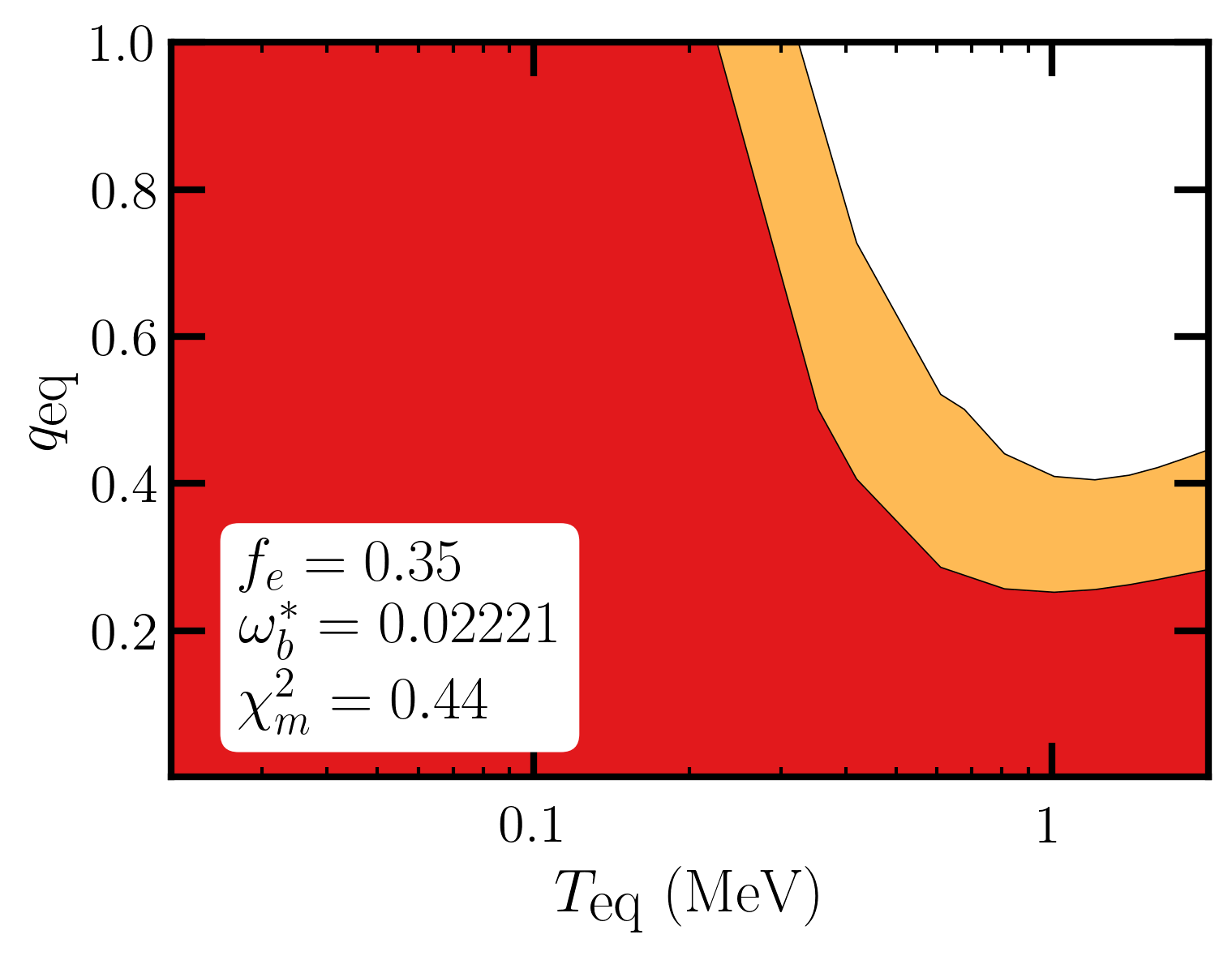}
    \includegraphics[width=0.23\textwidth]{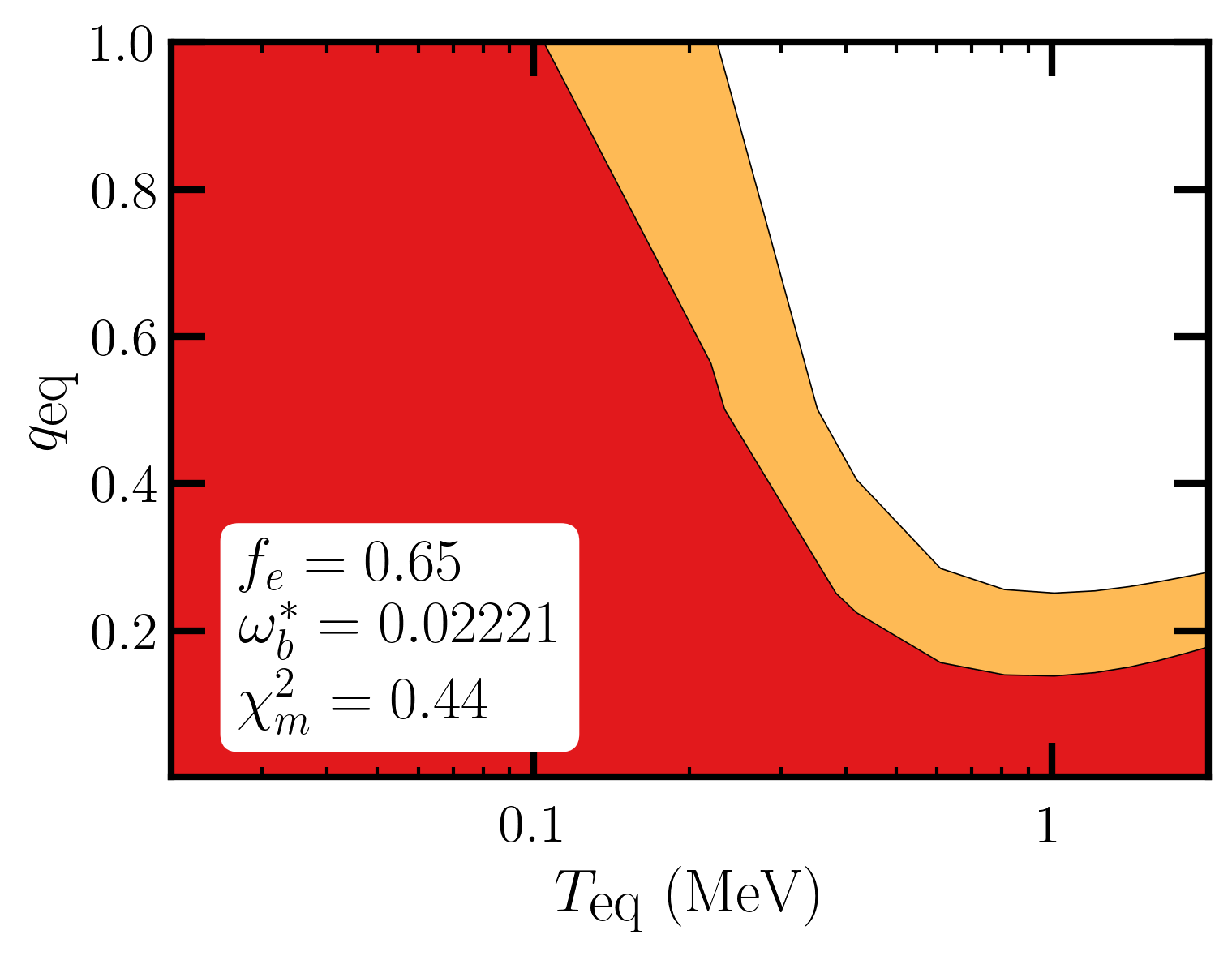}
    \includegraphics[width=0.23\textwidth]{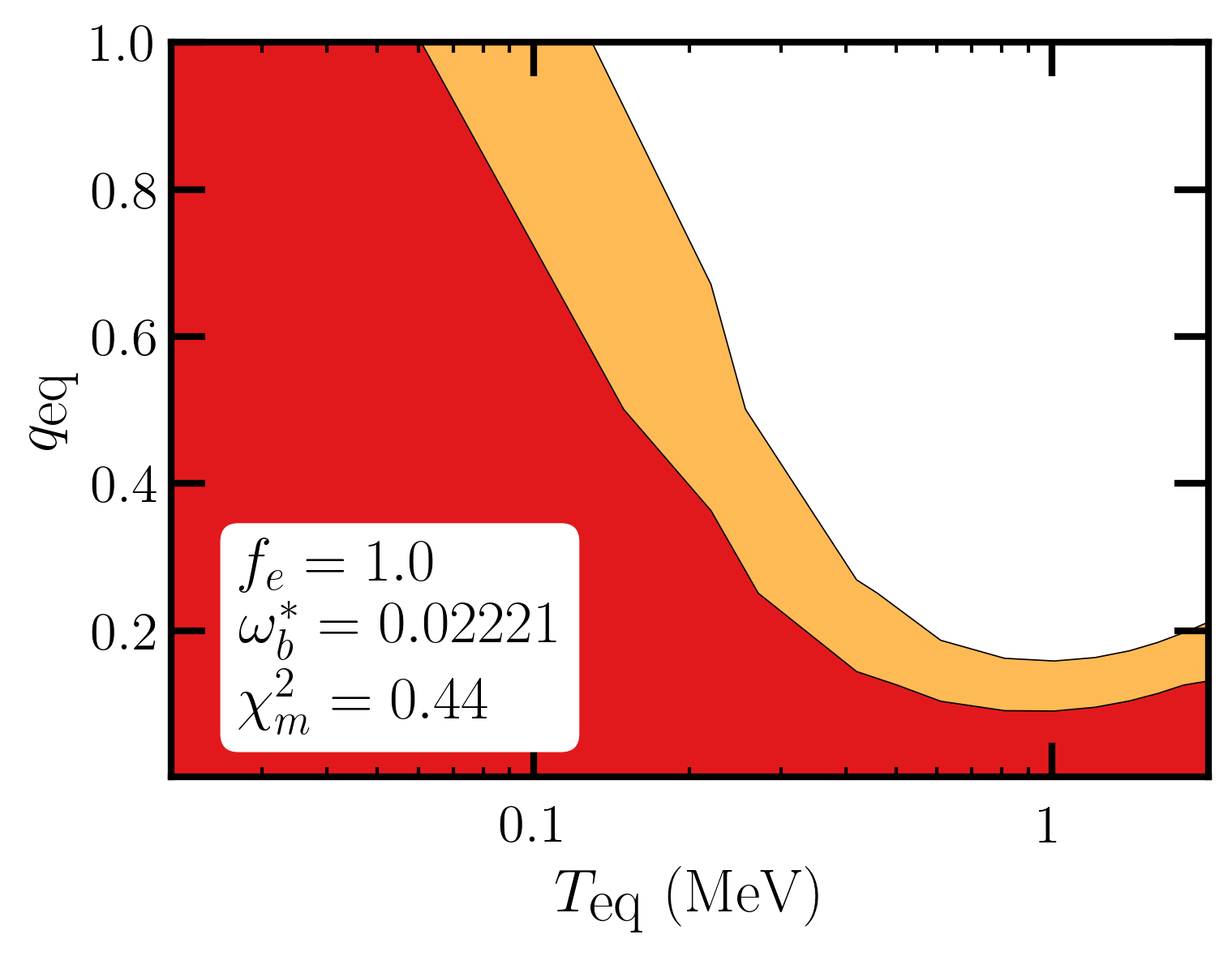}
    \caption{The regions allowed at $1\sigma$ (red) and $2\sigma$ (yellow) in the cooling scenario, calculated in the ``combined" network, for different values of $f_e$.  $\omega_b^*$ is the central value of $\omega_b$ assumed to calculate $\chi^2$, and $\chi^2_m$ is the minimum $\chi^2$ attained by the model in this region of parameter space.  $f_e$ is fixed, while $\omega_b$ is profiled over (see Appendix~\ref{app:full_results}), with $\sigma_{\omega_b^*}=0.00022$.}
    \label{fig:cooling_both}
\end{figure}

Modeling this physics as described in the previous section, we find limits that do not qualitatively differ between PRIMAT and PArthENoPE. We thus show limits from the ``combined'' network in Figure~\ref{fig:cooling_both}.

As is clear from the figures, the limits can go from irrelevant (for $f_e=0.02$) to quite strong (for $f_e=1$). Over much of the parameter space, the limits stretch below $1\MeV$, providing the first direct constraints on such neutrino conversion at late times.

\subsection{Limits on a step in $\neff$}
If $\neff$ changes after neutron freezeout, the corresponding increase in $H$ can influence deuterium freezeout. In practice, the dominant parameters are $\dNeff$ and $\Tstep$. There are small additional dependencies on what fraction of the energy density is participating in the step, but in practice this dependence is negligible. For concreteness, we use the step model explored in \cite{Aloni_2022}. We assume the dark sector has a single scalar and Majorana fermion, and that the step arises as the scalar becomes non-relativistic and deposits its entropy into the fermion.

We assume that $\neff=N_{\rm{dark}}+N_{\nu}=3.044$ (the canonical \lcdm\ value) before the step, and $N_{\rm{dark}}=\Nint\, \fr_\eq$. 
After the scalar depopulates, the dark sector will have increased its contribution to $(1+8/7)^{1/3} \Nint\, \fr_\eq$, or, equivalently, $\dNeff =( (1+8/7)^{1/3}-1)\Nint\, \fr_\eq \approx 0.29\, \Nint\, \fr_\eq$. Thus, there is a $1-1$ map between the more physically observable parameter $\dNeff$ and the model parameters $\Nint\, \fr_\eq$.  The specific shape of the step is detailed in Appendix~\ref{app:step_shape}.

\begin{figure}[t]
    \centering
    \includegraphics[width=0.23\textwidth]{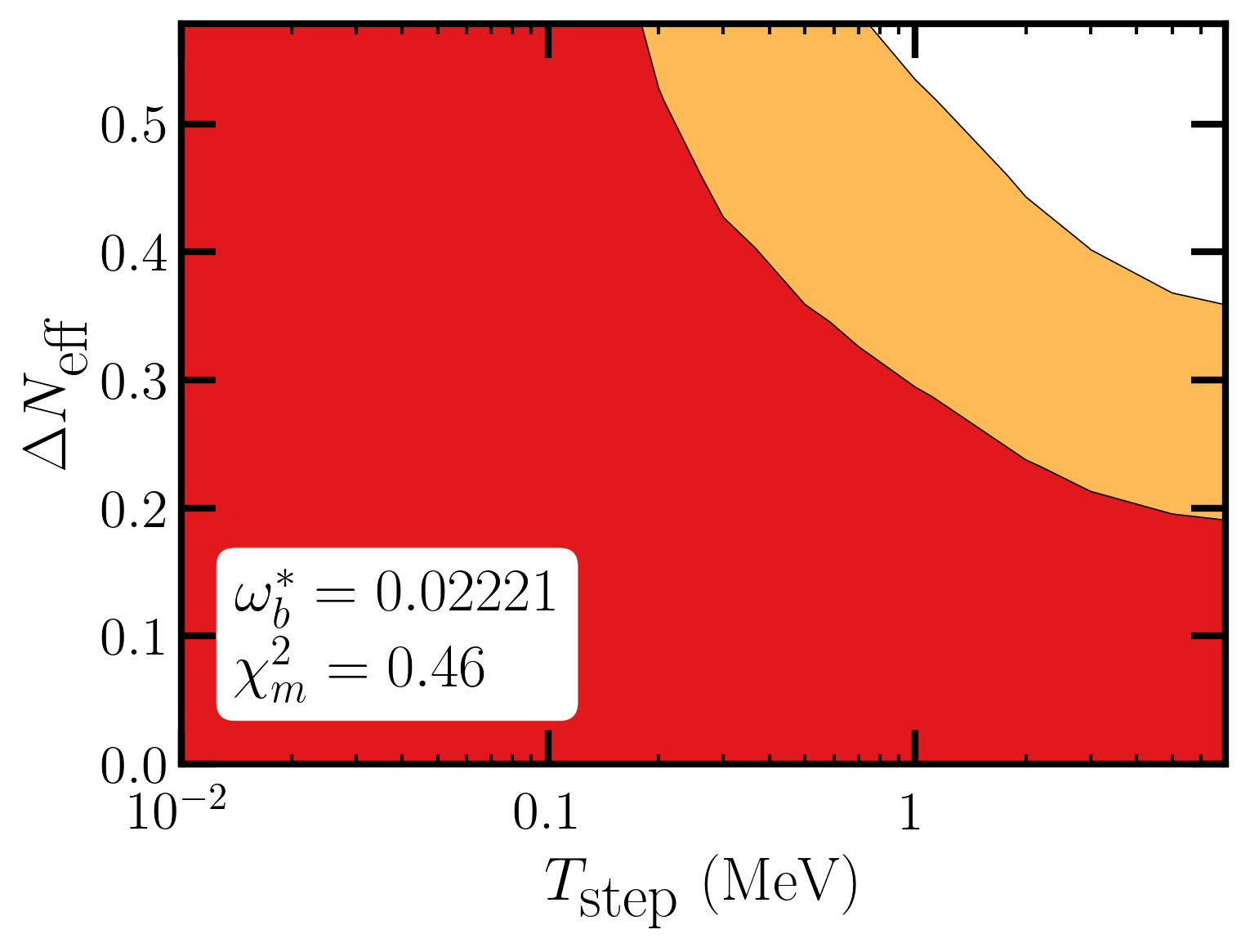}
    \includegraphics[width=0.23\textwidth]{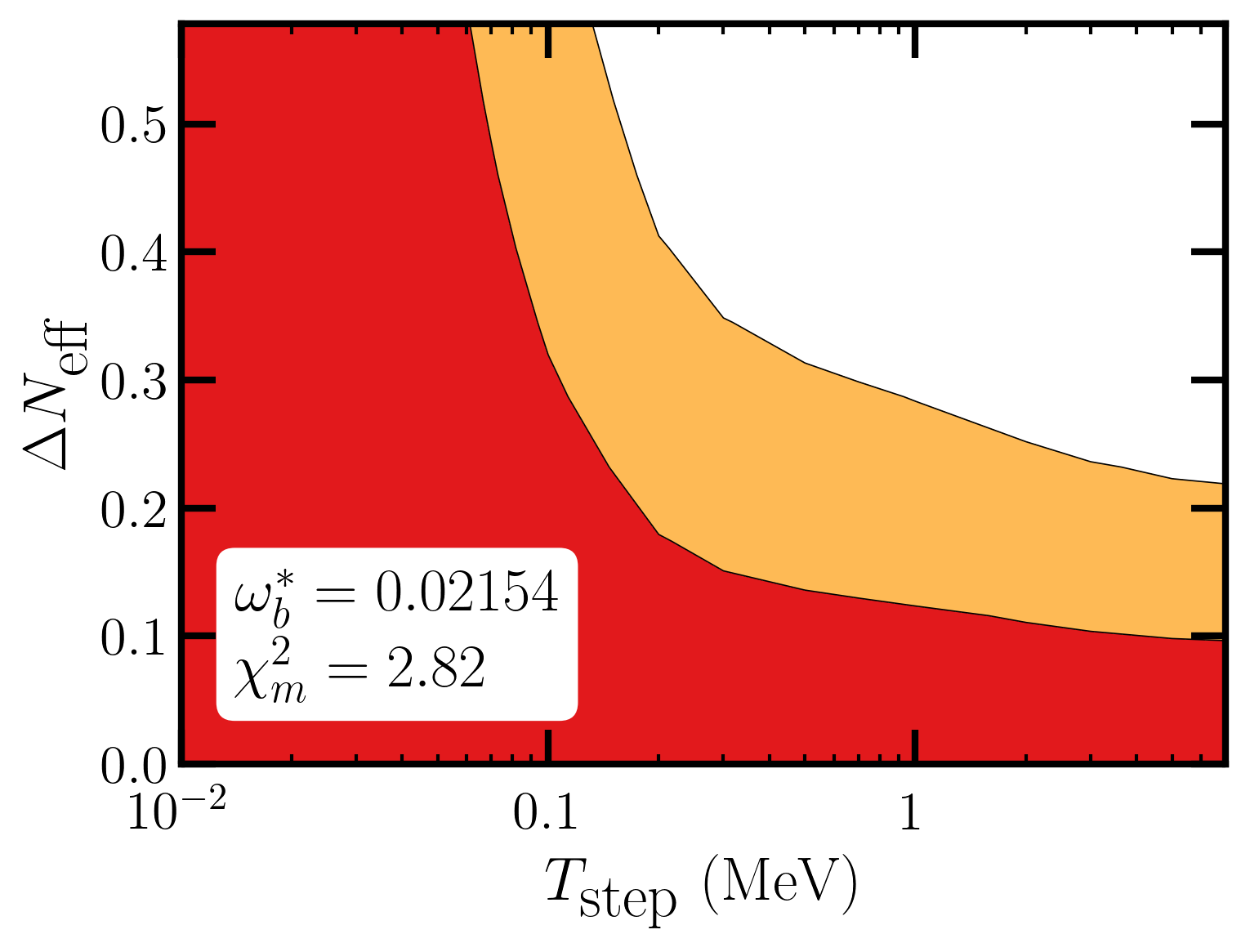}
    \includegraphics[width=0.23\textwidth]{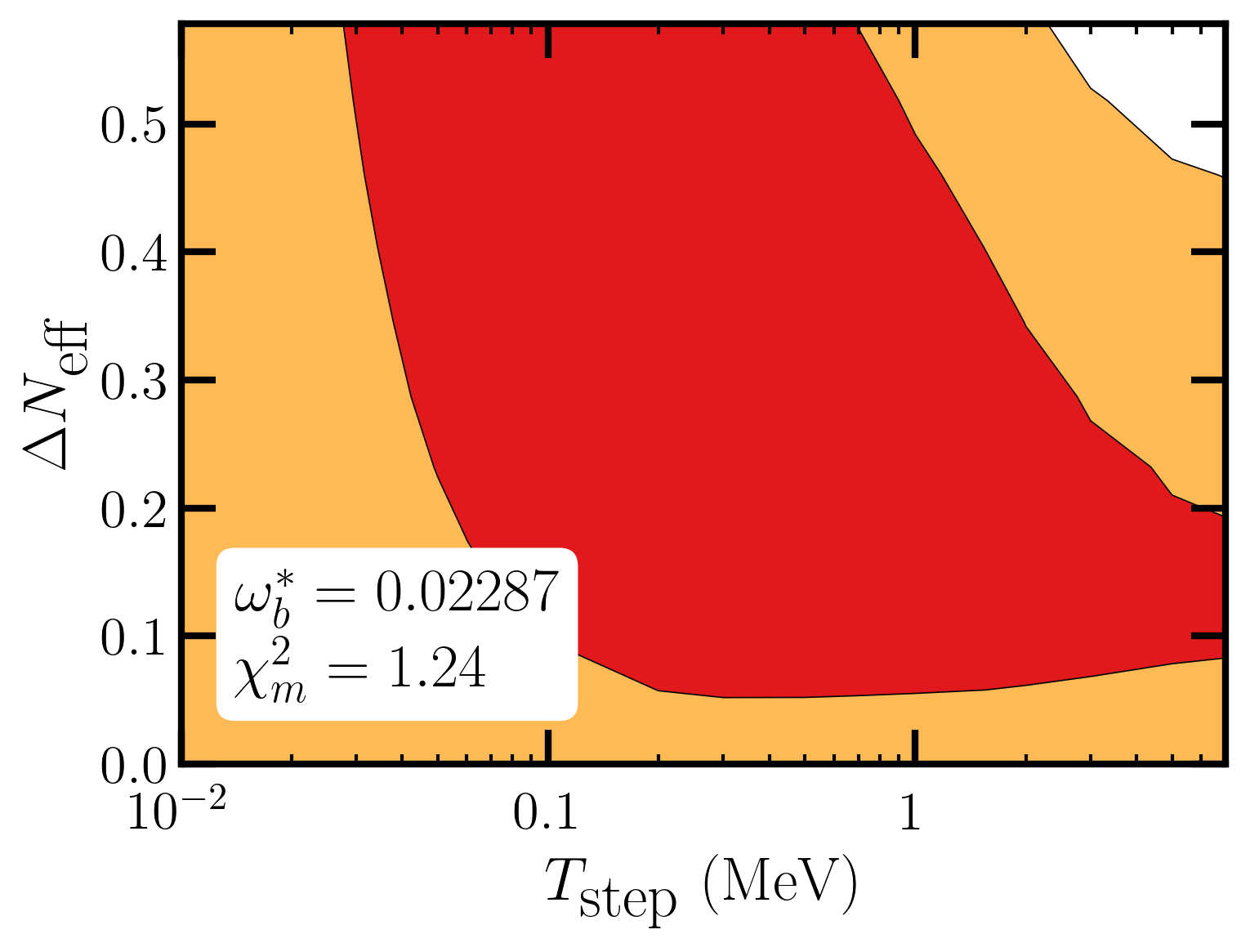}
    \caption{The regions allowed at $1\sigma$ (red) and $2\sigma$ (yellow) in the step scenario, calculated in the ``combined" network, for different values of  $\omega_b^*$. See Figure~\ref{fig:cooling_both} caption for definitions of $\omega_b^*$ and $\chi^2_m$.}
    \label{fig:step}
\end{figure}

We show the results of our analysis of a step in Figure~\ref{fig:step}.

As we see, the constraints are very dependent on $\omega_b$. For the middle value of $\omega_b$, there are only significant limits on the step if it occurs above an MeV. On the other hand, for the low value of $\omega_b$, there are very strong limits, even for $\Tstep$ below 200 \kev. For a high value of $\omega_b$, there is a mild {\it preference} for a step below an $\Mev$ (which grows beyond $2 \sigma$ in the PRIMAT network, see Figure~\ref{fig:PRIMAT_step} in Appendix~\ref{app:full_results}). 

\section{Discussion}
\label{sec:disc}
As new data from numerous cosmological experiments loom, we can anticipate unprecedented precision in our knowledge of the dark radiation that pervades the universe. Before any deviation from the \lcdm\ prediction of SM neutrinos appears, it behooves us to consider what constraints we can garner from other sources.

BBN has long been pointed to as tightly constraining the number of light species near an MeV. However, the precise knowledge on how late neutrinos must act like SM neutrinos has not been quantified. We have here explored, for the first time, what constraints can arise from a dark sector equilibrating with neutrinos at sub-MeV temperatures.

We have shown that, if the neutrinos equilibrating with a dark sector contain a significant electron content, there are strong constraints that arise from cooling of the SM neutrinos. Limits exist generically well below an $\Mev$, including in some cases as low as $\sim 200 \kev$. In contrast, if there is negligible electron neutrino content (such as if only $\nu_3$ the third neutrino mass eigenstate equilibrates), there are essentially no constraints on this process below 2 $\MeV$. The limits are only weakly dependent on $\omega_b$.

If a dark sector is equilibrated, and it subsequently passes through a mass threshold where $\Neff$ changes, there can be signals for BBN by accelerating deuterium freezeout. These limits, however, are very $\omega_b$-dependent. As an example, some models attempting to modify cosmology to solve the Hubble tension predict lower values of $\omega_b$ than \lcdm, while others predict a higher value. For the canonical \lcdm\ value of $\omega_b$, the limits are quite weak, allowing large changes to $\dNeff$ at 2 $\MeV$. For low values of $\omega_b$, such a step must be late ($\sim 200 \kev$). For high values of $\omega_b$, there is a mild preference for a change in $\dNeff$. Consequently, as future CMB data appear, and as these alternative cosmologies become more constrained, we may be able to make more definitive statements about the expansion rate of the universe in this $T\sim 100\kev - 1 \Mev$ period. 

We note that these results could have implications for the $\sim2\sigma$ deuterium tension reported in \cite{Pitrou_2021b}, since we provide a mechanism to alter the prediction for D/H without significant impacts to Y$_{\rm{P}}$.  However, whether a tension appears at all depends on assumptions about the underlying cosmology and the resulting measured value of $\omega_b$.  Due to the sensitivity of our results to the value of $\omega_b$, it is difficult to offer a concrete resolution to this tension.  Nevertheless, improvements in measurements of deuterium annihilation rates can improve these limits as well.

In summary, there is significant -- but not unlimited -- freedom to modify the dynamics of the SM neutrinos below $1\ \Mev$. As our understanding of the history of the universe in the $\ev - \Mev$ era improves, these constraints will help us understand what modifications to the dark radiation are possible.

\section*{Acknowledgements}
We thank Julien Froustey and Zilu Zhou for helpful conversations.  N.W. is supported by NSF under award PHY-2210498, by the BSF under grant 2018140, and by the Simons Foundation.  This material is based upon work supported by the NSF Graduate Research Fellowship under Grant No. DGE1839302.  This research was supported in part by Perimeter Institute for Theoretical Physics. Research at Perimeter Institute is supported by the Government of Canada through the Department of Innovation, Science and Economic Development and by the Province of Ontario through the Ministry of Research, Innovation and Science. M.S. thanks the CCPP at NYU and the Institut for Theoretische Physik at University of G\"ottingen for their hospitality. His research for this project was supported by the U.S. Department of Energy (DOE) under Award No. DE-SC0015845 and by a Research Award from the Alexander von Humboldt Foundation.

\bibliographystyle{apsrev}
\bibliography{references}
\appendix
\section{Dark Sector Equilibration}\label{app:DS_eq}
Although the precise form of equilibration is not expected to influence our results significantly, it is still important to have a prescription of the equilibration process. This both makes concrete what we mean by $\Tequil$ and can help connect it to specific model scenarios. 
To do this, we compute the evolution of the dark sector energy density as it is equilibrating with some or all of the thermal neutrinos. 

We begin with the Boltzmann equation for the dark sector energy density, Eq.~(\ref{eq:Boltzmann}),
\bea
a \frac{d}{da}(a^4 \rhod)=\frac{\langle \Gamma_{\nu\rightarrow d}\rangle }{H} a^4
\left(\rhonuint -\left.\frac{\rhonuint}{\rhod}\right|_{\eq} \rhod \right)\, ,
\eea
 where $\rhod$ is the energy density in the dark sector and $\rhonuint$ is the energy density of the neutrinos which participate in the equilibration process. After equilibration, the ratio of the energy density remaining in the neutrinos over the energy density in the dark sector is given by the ratio $(\rhonuint/\rhod)_{\eq} =\gnuint/g_*^d \equiv r_{dof} $. Here $\gnuint=\Nint 7/4$ reflects how many SM neutrino degrees of freedom participate in the equilibration. 
 
We rewrite the energy densities above in terms of the energy density of the interacting neutrinos in absence of mixing with a dark sector (i.e. in \lcdm) $\rhonuintz\equiv \Nint\, \rhonuz$. This is convenient because the evolution of the neutrinos in \lcdm\ is simple, their temperature redshifts as $\Tnz \propto 1/a$ and $\rhonuz=7\pi^2/120\, \Tnz^4 $.
We change variables from $a$ to $\Tnz$ and from $a^4 \rhod$ to $\fr\equiv \rhod/\rhonuintz$, the quotient $\fr$ corresponds to the fraction of the interacting neutrino energy density which has been transferred to the dark sector.

Energy conservation in the equilibration process corresponds to $\rhonuint + \rhod =\rhonuintz$ which implies that the fraction of the energy density which remains in the neutrinos is $\rhonuint/\rhonuintz =1-\fr$.  Using the new variables, the Boltzmann equation becomes
 \bea
- \Tnz \frac{d}{d \Tnz} \fr&=&
\frac{\langle \Gamma_{\nu\rightarrow d}\rangle }{H}\left(1 - \frac{\fr}{\fr_{\eq}}\right) \, ,
\eea
where we also defined the equilibrium ratio $\fr_{\eq}\equiv 1/(1+r_{dof})$.
This is the equation which we need to solve. Here $\fr$ is a function of $\Tnz$ only but $\langle \Gamma_{\nu\rightarrow d}\rangle /H$ is a model-dependent function of both $\Tnz$ and $T_d$, see Eq.~(\ref{eq:gammarate}). 
%

Since we are interested in equilibration well after the Dodelson-Widrow era $\Tnz\sim$ 100 MeV the weak interactions are suppressed, and in most of parameter space the terms proportional to $\alpha_d$ dominate \cite{Aloni:2023tff} in both the numerator and denominator of Eq.~(\ref{eq:gammarate}). Then  
\bea
\langle \Gamma_{\nu\rightarrow d}\rangle  \simeq  \frac{\theta_0^2 m_d^4}{\Tnz T_d^2} \ .
\eea
For Hubble we have $H= \sqrt{\rho_{\rm{rad}} /3}/ M_{pl}=c_H\, \Tnz^2/M_{pl} $ where $c_H$ is a slowly varying function of $\Tnz$ which is equal to the constant $\sqrt{\pi^2/90\ 43/4}\sim 1.09$ above the electron annihilation threshold and smoothly increases to $\sim 1.19$ below the threshold due to the electron entropy dump into photons. We ignore this small scale dependence and use $c_H=1.1$.

Finally, we can eliminate $T_d$ by using that
$\fr=(g_*^d T_d^4)/(\gnuint \Tnz^4)$ and thus $(T_d/\Tnz)^4=r_{dof}\, \fr$.
Putting everything together, we obtain
\bea
\frac{\langle \Gamma_{\nu\rightarrow d}\rangle }{H} \simeq  
\frac{\theta_0^2 m_d^4 M_{pl}}{ c_H \sqrt{r_{dof}}} \frac{1}{\sqrt{\fr}\Tnz^5} \equiv \kappa \frac{1}{\sqrt{\fr}\Tnz^5}\ ,
\eea
and the Boltzmann equation for $\fr(\Tnz)$ becomes 
 \bea
 \label{eq:boltzmann}
- \Tnz \frac{d}{d \Tnz} \fr=\frac{\kappa}{\sqrt{\fr}\Tnz^5}  \left(1 - \frac{\fr}{\fr_{\eq}}\right) \, ,
\eea
with the initial condition that $\fr/\fr_{\eq}\ll 1$ at early times. We could solve this numerically, though we prefer to obtain a useful analytical approximation iteratively, as follows. In a first pass we approximate the factor 
$1 - \fr/\fr_{\eq} \rightarrow 1$ to obtain a solution for $\fr$ which is accurate at early times. We then substitute this ``first pass" solution back into the $1/\sqrt{\fr}$ in Eq.~(\ref{eq:boltzmann}) and solve again to obtain a ``second pass" solution which is accurate both well before and well after equilibration.  This solution differs from the numeric solution by a only few percent in the transition region. 
%

For the ``first pass" solution, we find
\bea
\fr^{1/2}=\left(\frac{3\kappa}{10}\right)^{1/3} \Tnz^{-5/3} \ .
\eea
Plugging this into the $1/\sqrt{\fr}$-factor and solving again, we obtain
\bea
\fr&=&\fr_{\eq}\,(1-e^{-(\frac{3\kappa}{10})^{2/3} \fr_{\eq}^{-1} \Tnz^{-10/3}})\\ \nonumber
&\equiv& \fr_{\eq}\,(1-e^{-(\Tequil/\Tnz)^{10/3}})\ \label{eq:tenthirds},
\eea
with
\bea
\Tequil&=&\fr_{\eq}^{-3/10}( \frac{3\kappa}{10})^{1/5} =
\fr_{\eq}^{-3/10}\left( 
\frac{3\,\theta_0^2\, m_d^4 M_{pl}}{10\, c_H \sqrt{r_{dof}}}
\right)^{1/5} \\ \nonumber
& = & 
\frac{0.77 }{\fr_{\eq}^{1/5}\, (1-\fr_{\eq})^{1/10} }
\, m_d \left( \frac{\theta_0^2\, M_{pl}}{m_d} \right)^{1/5}. 
\eea
Note that this is very close to the simple expression derived in \cite{Aloni:2023tff}.

For completeness, we also give a second approximate solution for $\fr$ which applies for smaller $\alpha_d$ when equilibration occurs while the weak interaction term in the numerator of Eq.~(\ref{eq:gammarate}) dominates.
For that case we find
\bea
\langle \Gamma_{\nu\rightarrow d}\rangle  \equiv \kappa' \frac{1}{\fr \Tnz}\ ,
\eea
with 
\bea
\kappa'=\frac{3 c_\Gamma \theta_0^2 m_d^4 M_{pl} G_F^2}{ c_H r_{dof} \alpha_d^2} \ .
\eea
This produces the Boltzmann equation 
\bea
 \label{eq:boltzmann2}
- \Tnz \frac{d}{d \Tnz} \fr= \frac{\kappa'}{\fr \Tnz} \left(1 - \frac{\fr}{\fr_{\eq}}\right) \ ,
\eea
which has a similar iterative solution
\bea
\fr&=&\fr_{\eq}\,(1-e^{-(2\kappa')^{1/2} \fr_{\eq}^{-1} \Tnz^{-1/2}})\\
&\equiv& \fr_{\eq}\,(1-e^{-(\Tequil'/\Tnz)^{1/2}})\ , \nonumber
\eea
with
\bea
\Tequil'=\frac{2\kappa'}{\fr_{\eq}^2}
= \frac{6 c_\Gamma}{c_H \fr_{\eq}(1-\fr_{\eq})}\ \frac{\theta_0^2 m_d^4 \, M_{pl} G_F^2}{\alpha_d^2}\ .
\eea

A reasonable approximation which captures both limits is obtained by adding the two exponents in a joint solution
\bea
\fr=\fr_{\eq}\,(1-e^{-(\Tequil'/\Tnz)^{1/2}-(\Tequil/\Tnz)^{10/3}})\ ,
\eea
which equilibrates at the larger of the two would-be equilibration temperatures with the correct power law. The ultimate results are not highly sensitive to this, and this is primarily useful if one is plotting in a parameter space of model parameters. For our plot in physically relevant parameters, we use only the $10/3$-power scaling.

\subsection{Comments about approximate neutrino phase space distributions}

\vskip.2in
In the equilibration process the neutrinos are initially free and cannot maintain an equilibrium distribution when perturbed through oscillation and scattering. Furthermore, the conversion rate is energy dependent $\Gamma \sim E_\nu^{-1}$ so that the shape of the original thermal distribution of the neutrinos is not maintained. Thus, strictly speaking, during the equilibration process it is neither correct to assume that the neutrinos maintain equilibrium with a dropping temperature nor to assume that they maintain the energy spectrum of the original distribution. In principle, the phase space distribution is calculable while the neutrinos remain free by solving an energy-dependent Boltzmann equation, though this is quite inconvenient and introduces more variables.

As an alternative, we take a reasonable approximate treatment to determine the neutrino distribution function which we use for our numerical code:
we assume that the neutrinos are in self-equilibrium throughout the equilibration. Then their temperature is determined via $T_\nu/\Tnz=(\rhonuint/\rhonuintz)^{1/4}=(1-\fr)^{1/4}$ and their distribution function is always a thermal one
\bea
{\cal{F}}[\Tn]={\cal{F}_{\rm th}}[(1-\fr)^{1/4}\Tnz] \ ,
\eea
where ${\cal{F}_{\rm th}(T)}$ is simply a thermal distribution with temperature $T$. This has the advantage that it smoothly interpolates between the initial distribution and the equilibrium distribution at the end of equilibration. 
%


What does this imply for the distribution function of the electron neutrinos? 
 In the case where all neutrinos equilibrate with the dark sector the electron neutrino distribution function is the above thermal approximation. 

Assuming the equilibrating eigenstate (or states) has (or have) probability $f_e$ to be an electron neutrino, then the electron neutrino distribution function is a sum of two components
\bea
{{\cal{F}}}_{\nu_e} = (1-f_e)\, {\cal{F}}_{th}[\Tnz]+f_e\, {\cal{F}}_{th}[(1-\fr)^{1/4}\Tnz]
\eea

We also note that our formula for the neutrino conversion rate $\Gamma$ assumed that the neutrino distribution is unaffected by the conversions, i.e. we ignored feedback.

\section{Step Shape}\label{app:step_shape}

In the step scenario, the shape of the step in energy density is given by
\begin{equation}
    N(x)=\frac{N_{\textrm{IR}}\left(1+\frac{8}{7}\left(\frac{x^2}{2}K_2(x)+\frac{x^3}{6}K_1(x)\right)\right)}{\left(1+\frac{8}{7}\left(\frac{x^2}{2}K_2(x)+ \frac{x^3}{8}K_1(x)\right)\right)^{4/3}},
\end{equation}
where $K_i(x)$ are modified Bessel functions of the second kind.  $x=T_{\textrm{step}}/T_d$, $T_{\textrm{step}}$ being the characteristic temperature of the step.  $T_d$ is the temperature of the dark sector, it is approximately given by 
\begin{equation}
    T_d = T_{\nu}\left(\frac{1+8/7}{1+\frac{8/7}{1+T_{\textrm{step}}/T_{\nu}}}\right)^{1/3}.
\end{equation}
where $T_\nu$ is the temperature of the interacting neutrinos after they thermalized the dark sector. For definiteness, in these formulas we assume that the dark sector decouples from the neutrinos before the step. 
The step in the dark sector goes from $N_{\textrm{UV}}=\Nint\, \fr_\eq$ to 
\begin{equation}
    N_{\textrm{IR}}=\Nint\, \fr_\eq\left(1+\frac{8}{7}\right)^{1/3}.
\end{equation}

\vskip.4in

\section{Complete Results}\label{app:full_results}

\vskip.2in
\noindent 
Here we present complete results from parameter scans.  These explore both the PRIMAT and PArthENoPE networks.  Throughout, $\omega_b^*$ is the central value of $\omega_b$ assumed to calculate $\chi^2$, and $\chi^2_m$ is the minimum $\chi^2$ attained by the model in this region of parameter space.  The $1\sigma$ allowed contour is shown in red, while the $2\sigma$ allowed contour is shown in yellow.  Additional details are indicated on the plots or in the captions.  

To determine our constraints, we calculate a profile likelihood ratio similar to the one described in~\cite{Giovanetti_2022}.  To summarize this method, for parameters $\boldsymbol{\theta}$, we first construct a Gaussian likelihood of the parameters and $\omega_b$ we call $L(\boldsymbol \theta, \omega_b)$, which is determined by the measured values of Y$_{\rm{P}}$ and D/H.  Then we can calculate $\lambda_p(\boldsymbol{\theta}) = L_p(\boldsymbol{\theta}) / \hat{L}_p$, where $L_p(\boldsymbol \theta) = \max_{\omega_b} L(\boldsymbol \theta, \omega_b)$ and $\hat{L}_p = \max_{\boldsymbol{\theta}, \omega_b} L(\boldsymbol \theta, \omega_b)$.  In both scenarios described in text, we consider two-parameter models, and so in both cases the 68\% or 95\% confidence limits for $\boldsymbol \theta$ are set by $-2 \log \lambda_p(\boldsymbol{\theta}) = 2.30$ or $6.18$, respectively.

\begin{figure*}[h]
    \centering
    \includegraphics[width=0.2\textwidth]{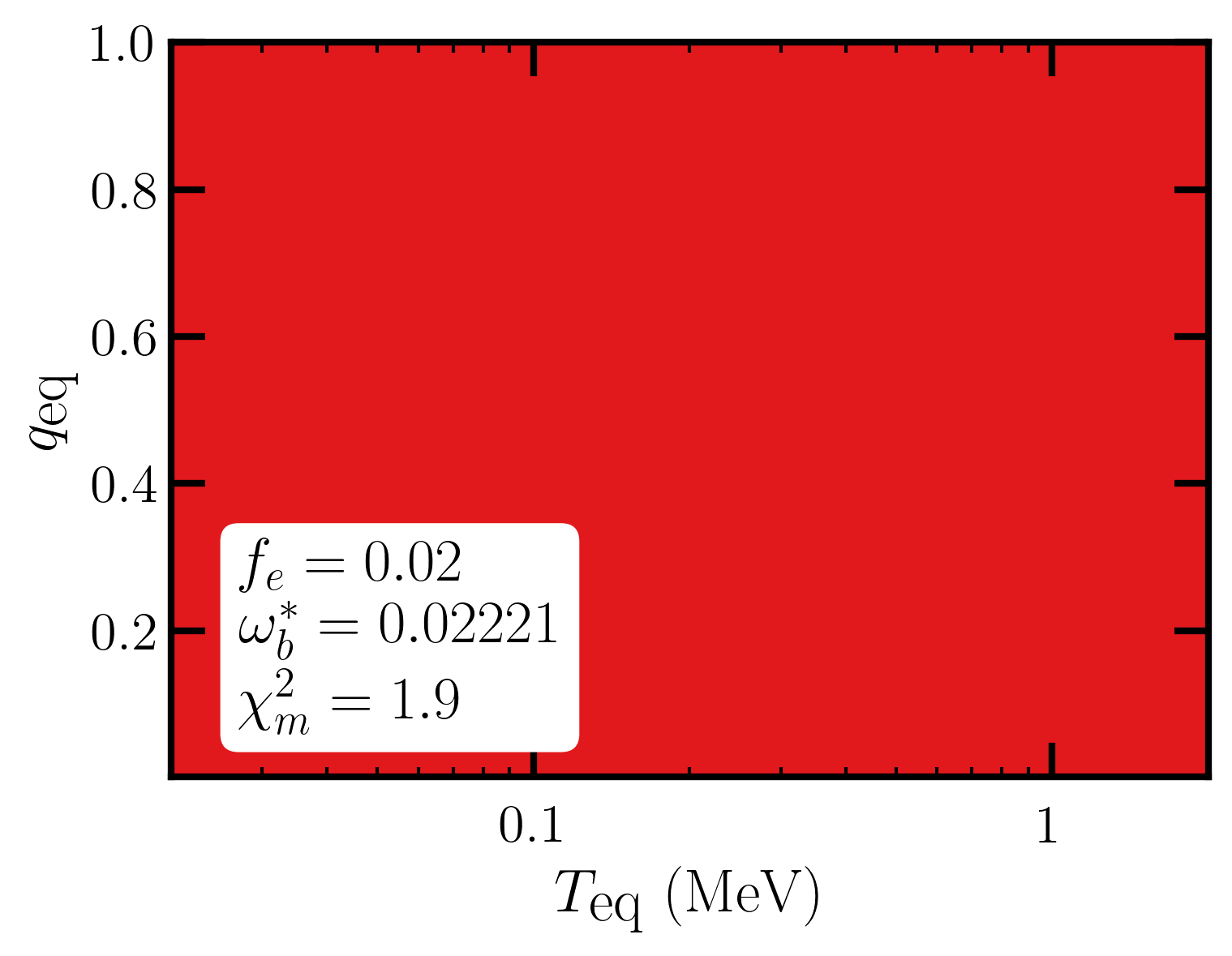}
    \includegraphics[width=0.2\textwidth]{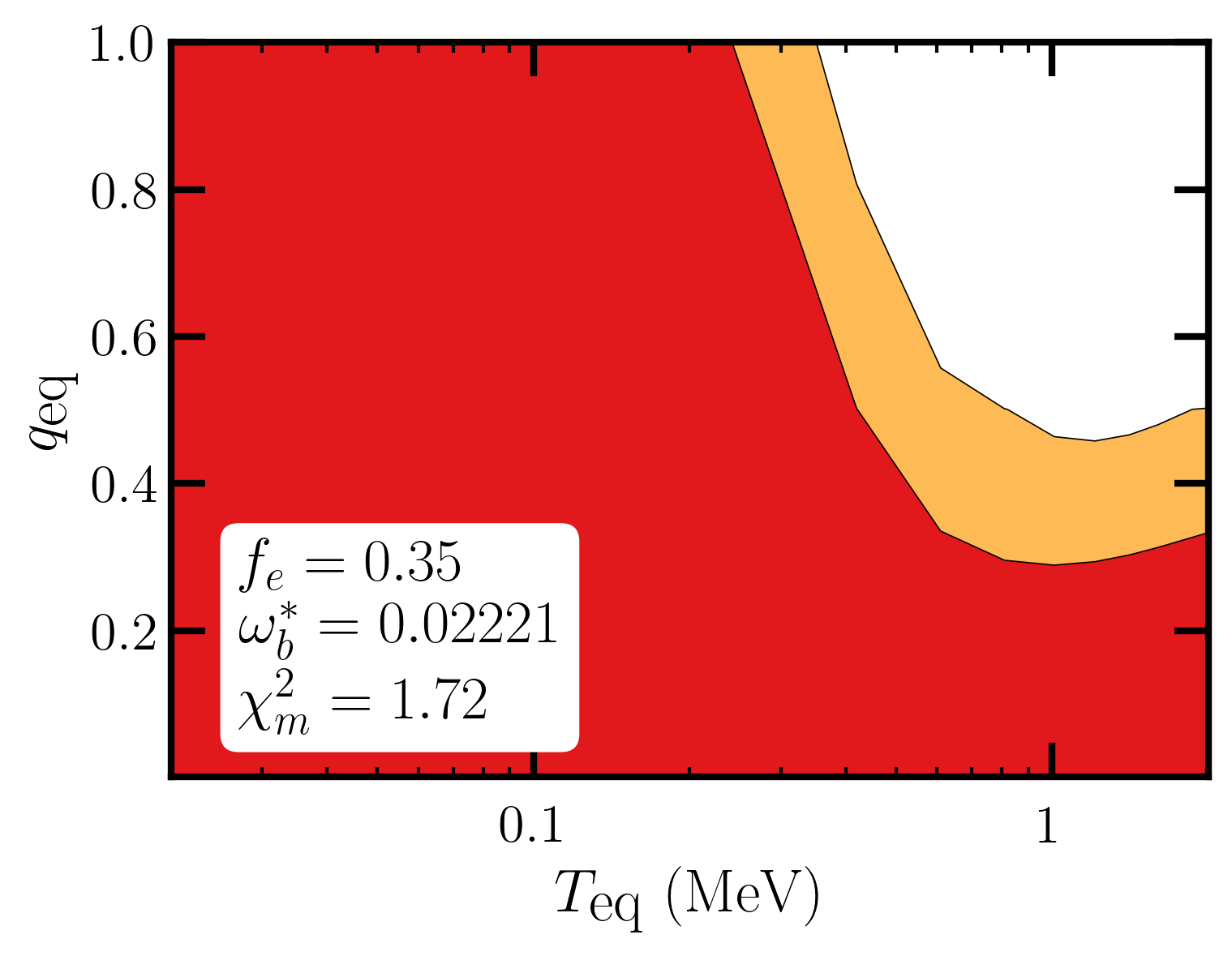}
    \includegraphics[width=0.2\textwidth]{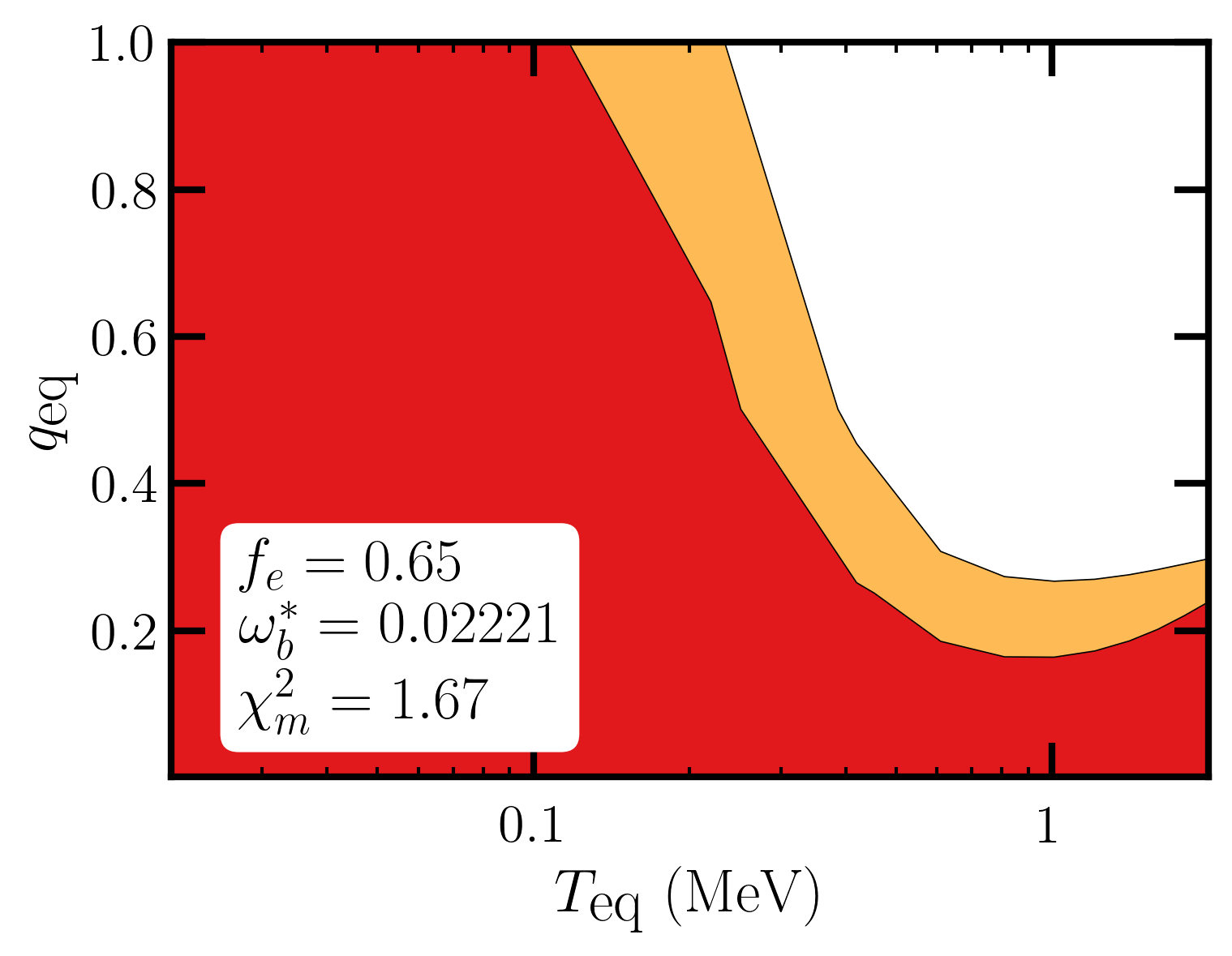}
    \includegraphics[width=0.2\textwidth]{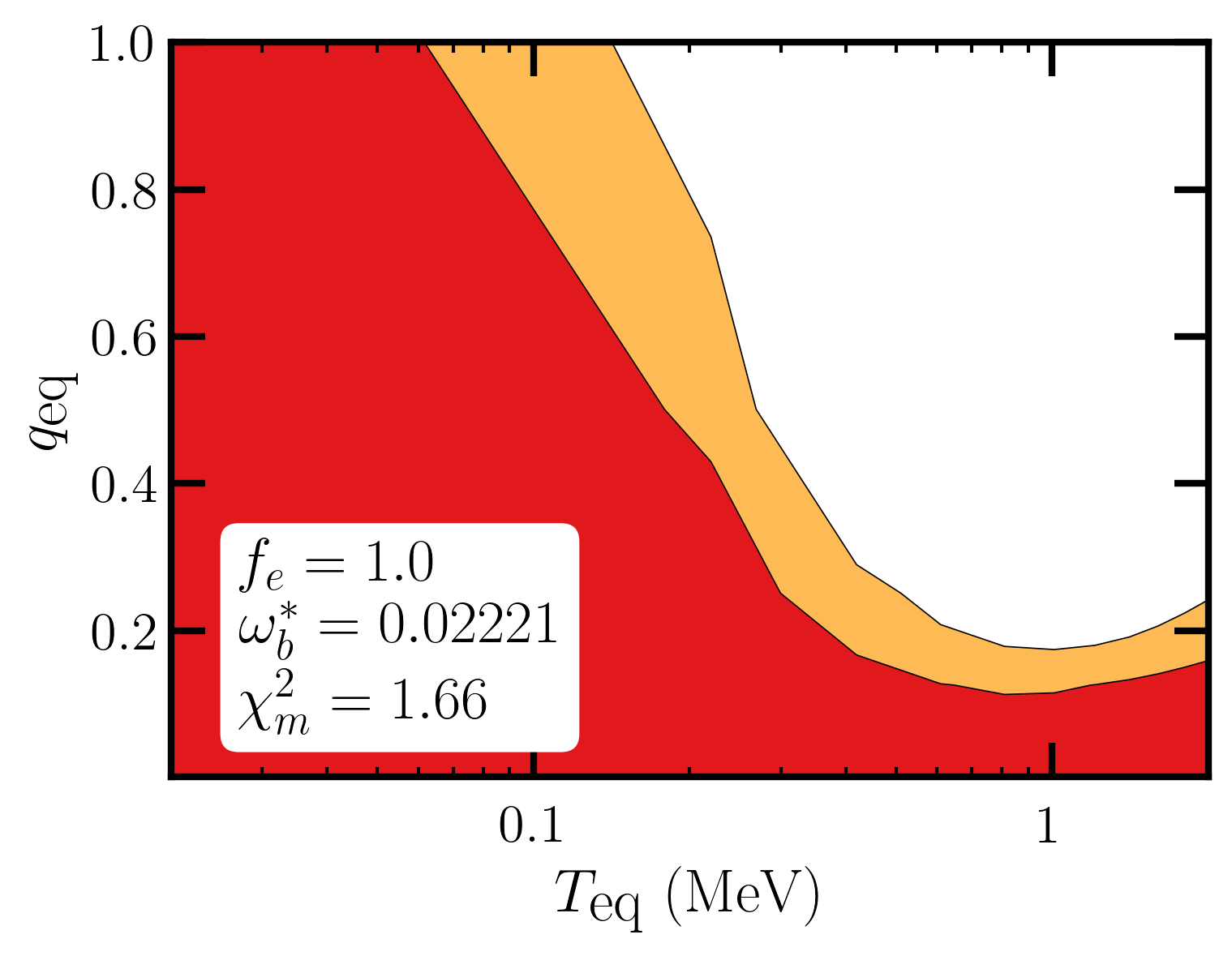}
    \caption{Limits on the cooling scenario, calculated with the PRIMAT network, using $\omega_b^*=0.02221$, for different values of $f_e$.}
\end{figure*}

\begin{figure*}[h]
    \centering
    \includegraphics[width=0.2\textwidth]{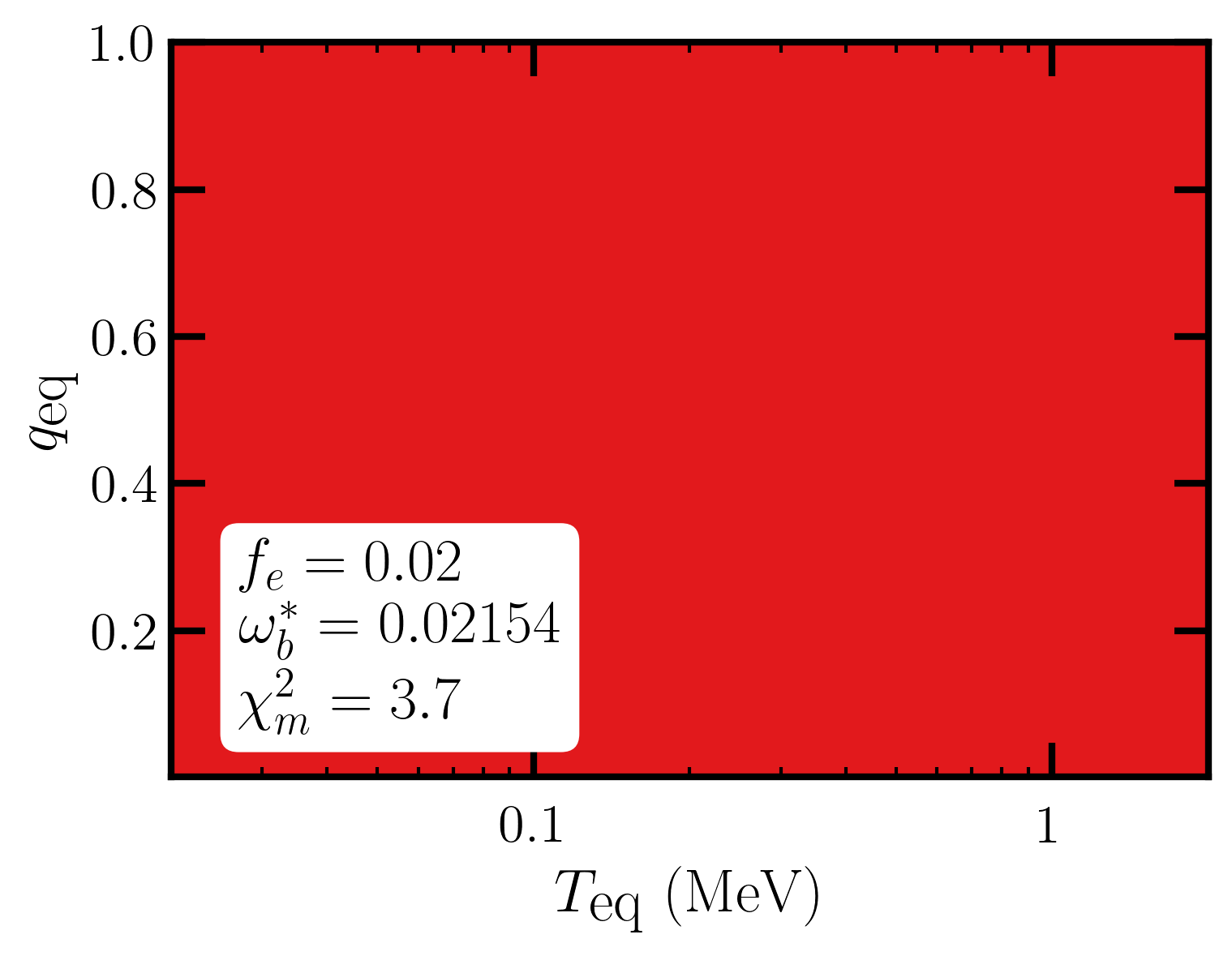}
    \includegraphics[width=0.2\textwidth]{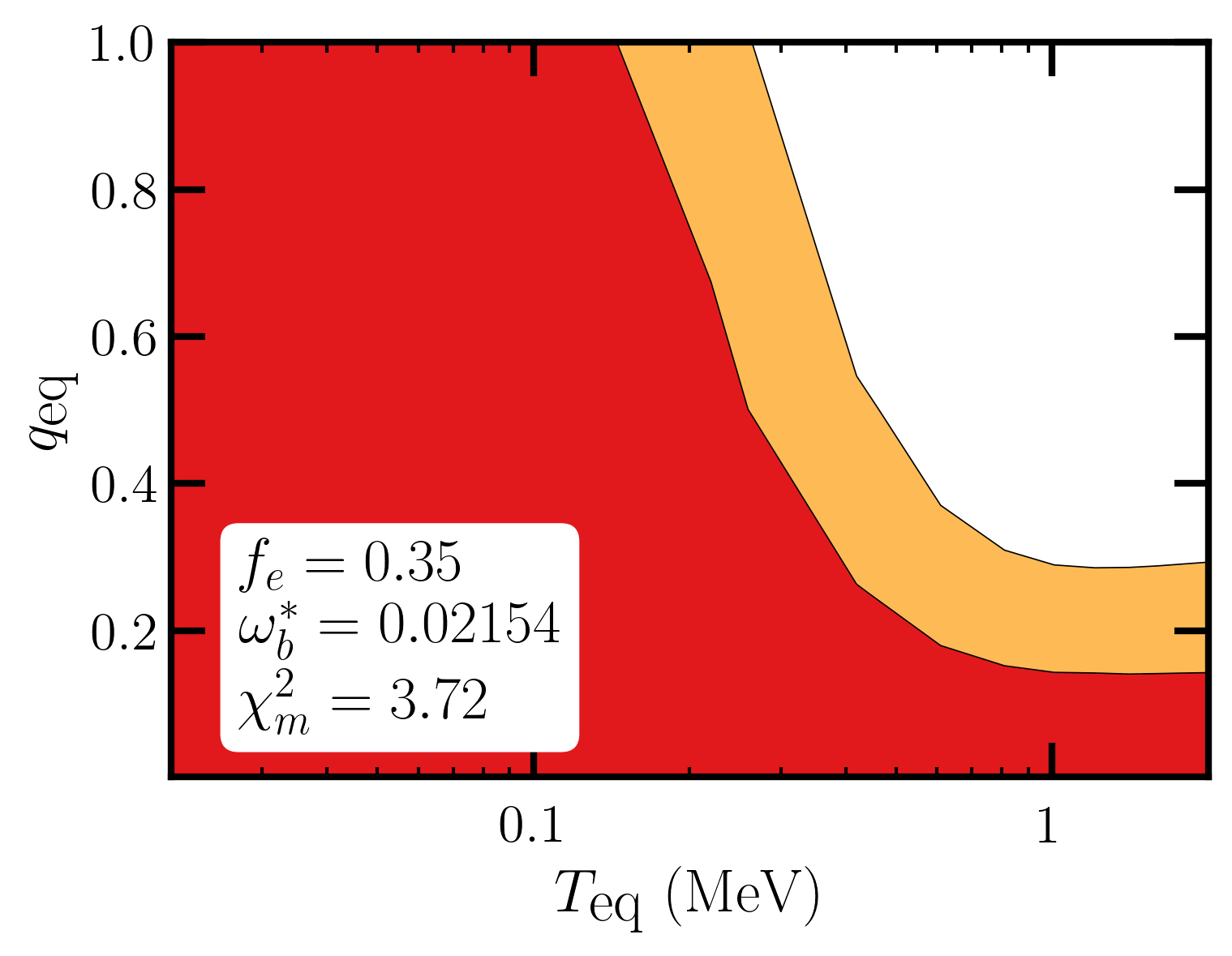}
    \includegraphics[width=0.2\textwidth]{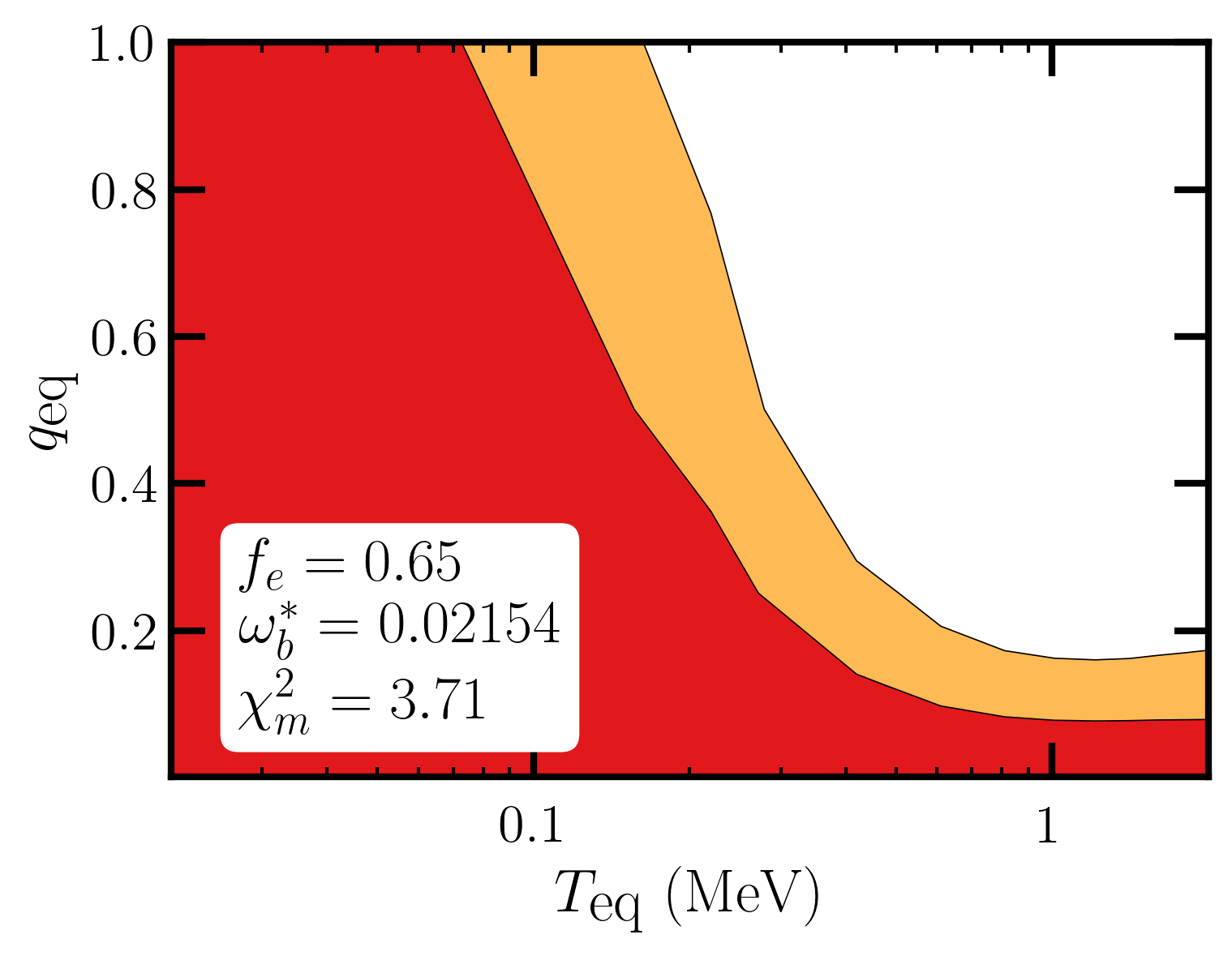}
    \includegraphics[width=0.2\textwidth]{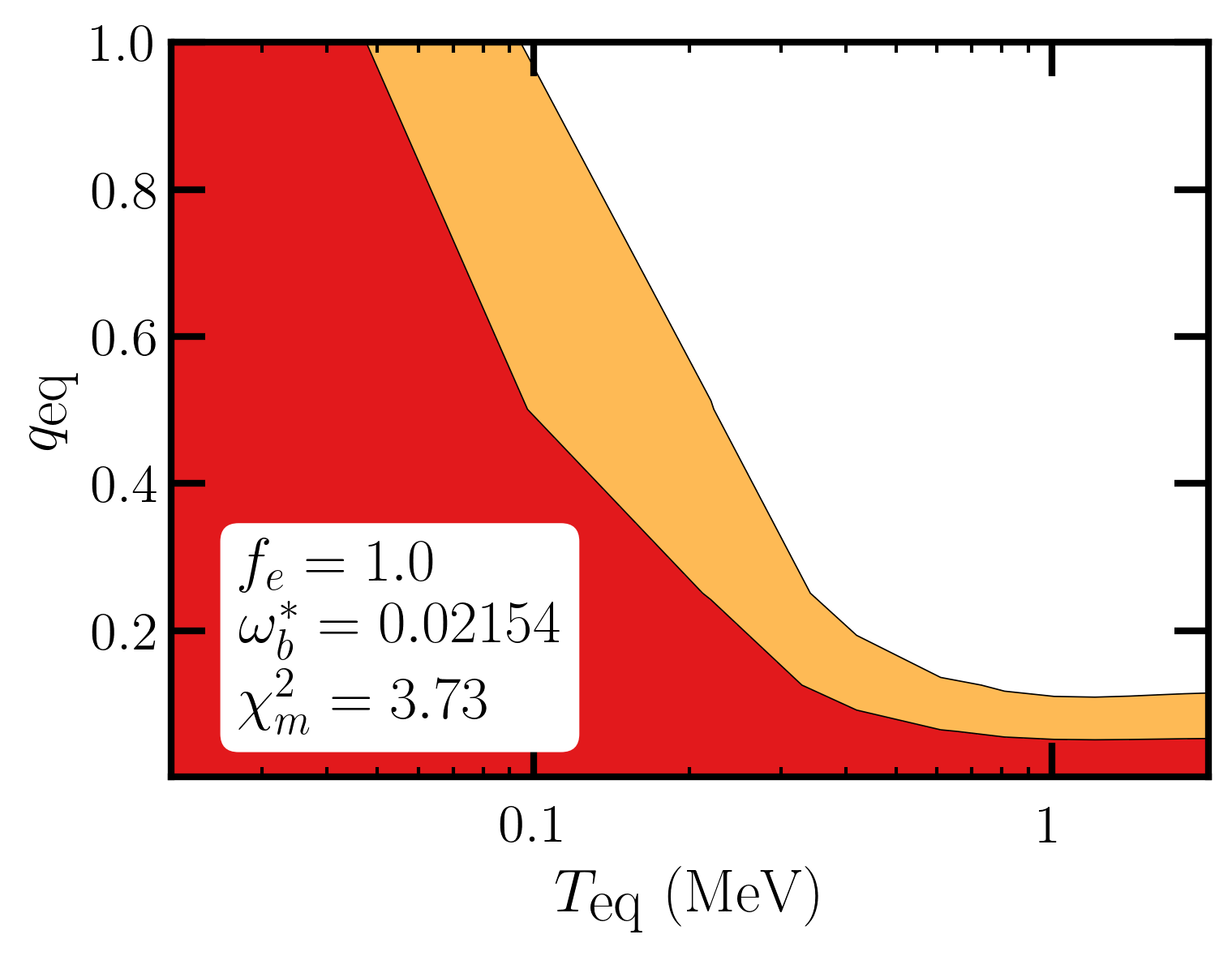}
    \caption{Limits on the cooling scenario, calculated with the PRIMAT network, using $\omega_b^*=0.02154$, for different values of $f_e$.}
\end{figure*}

\begin{figure*}[h]
    \centering
    \includegraphics[width=0.2\textwidth]{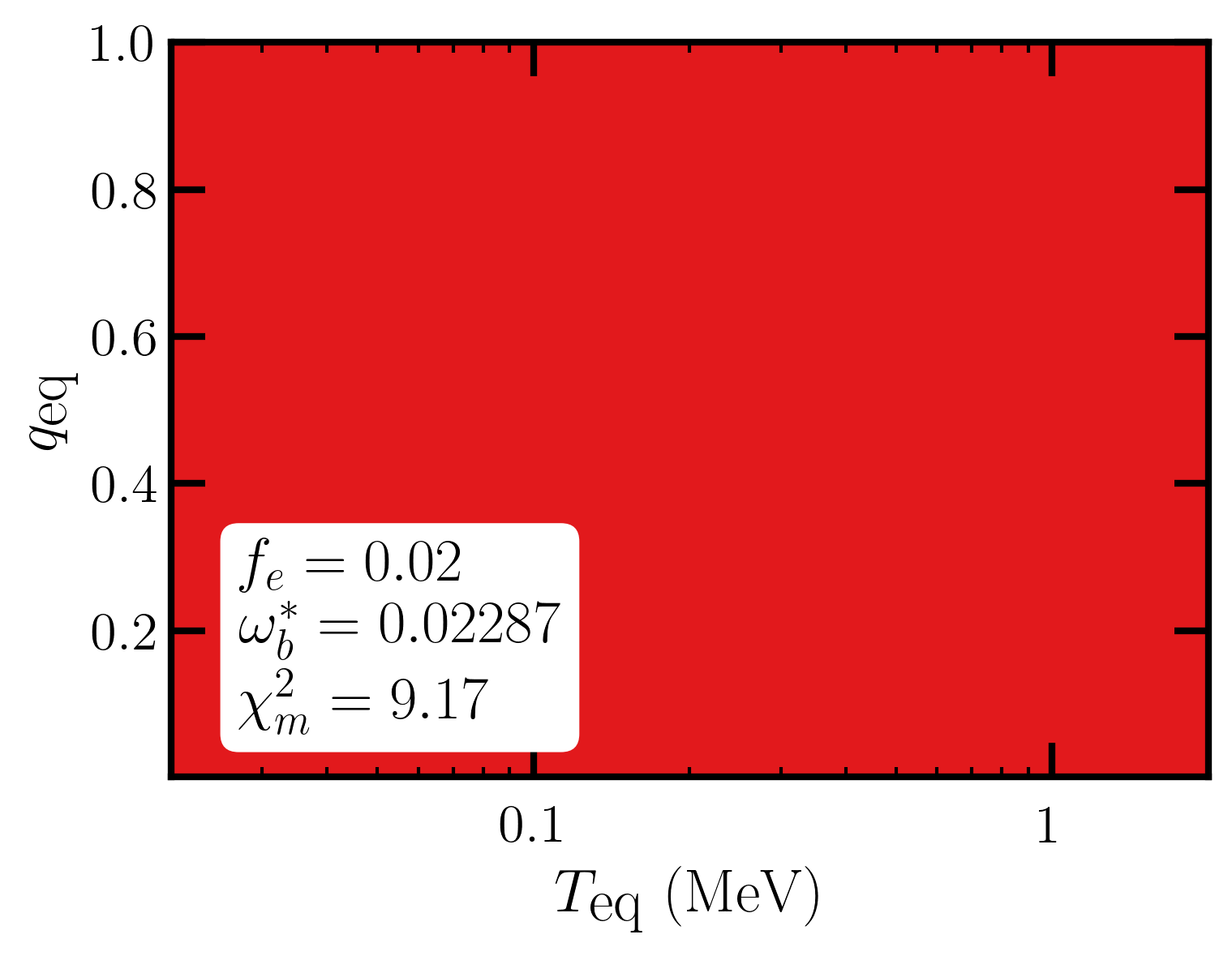}
    \includegraphics[width=0.2\textwidth]{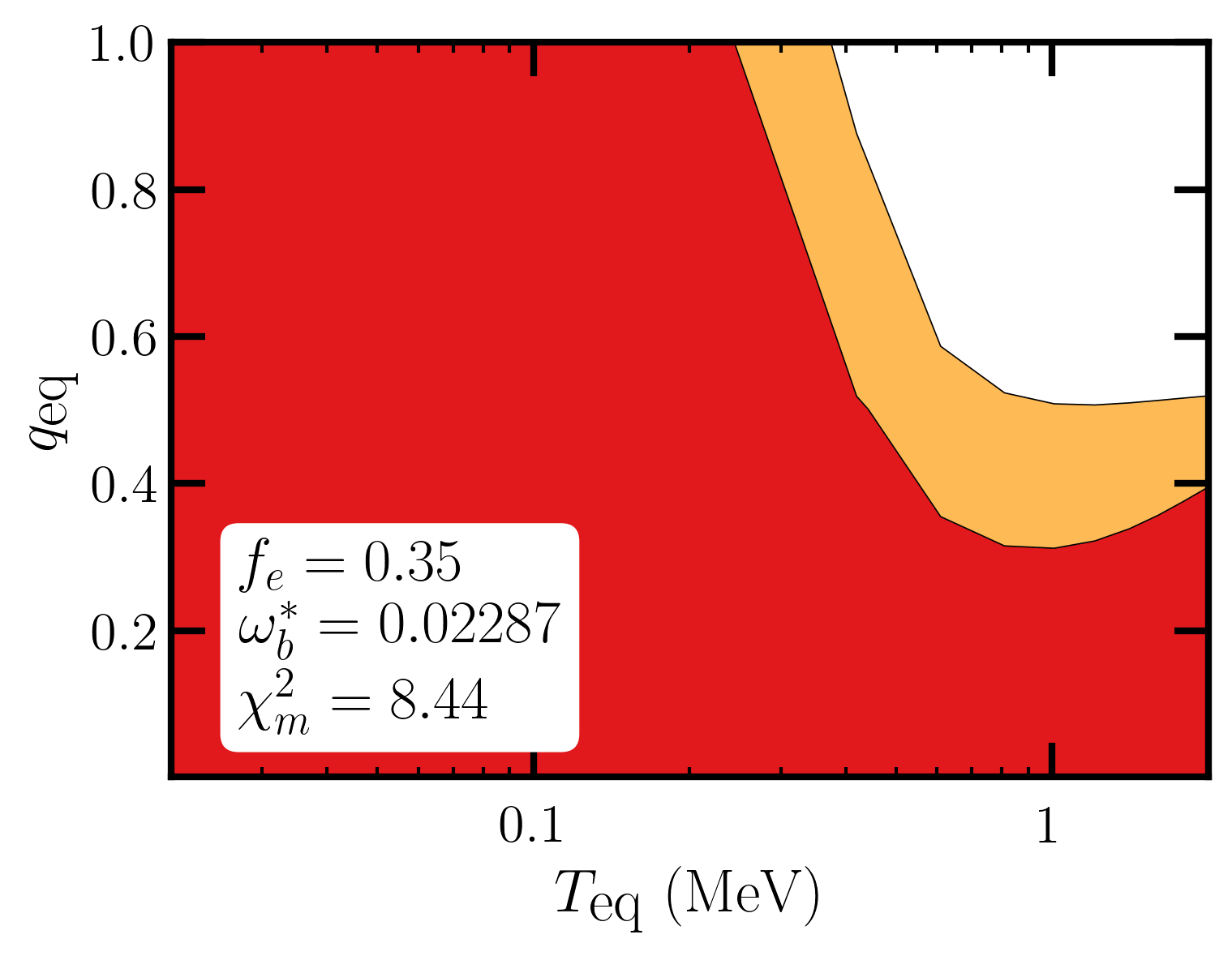}
    \includegraphics[width=0.2\textwidth]{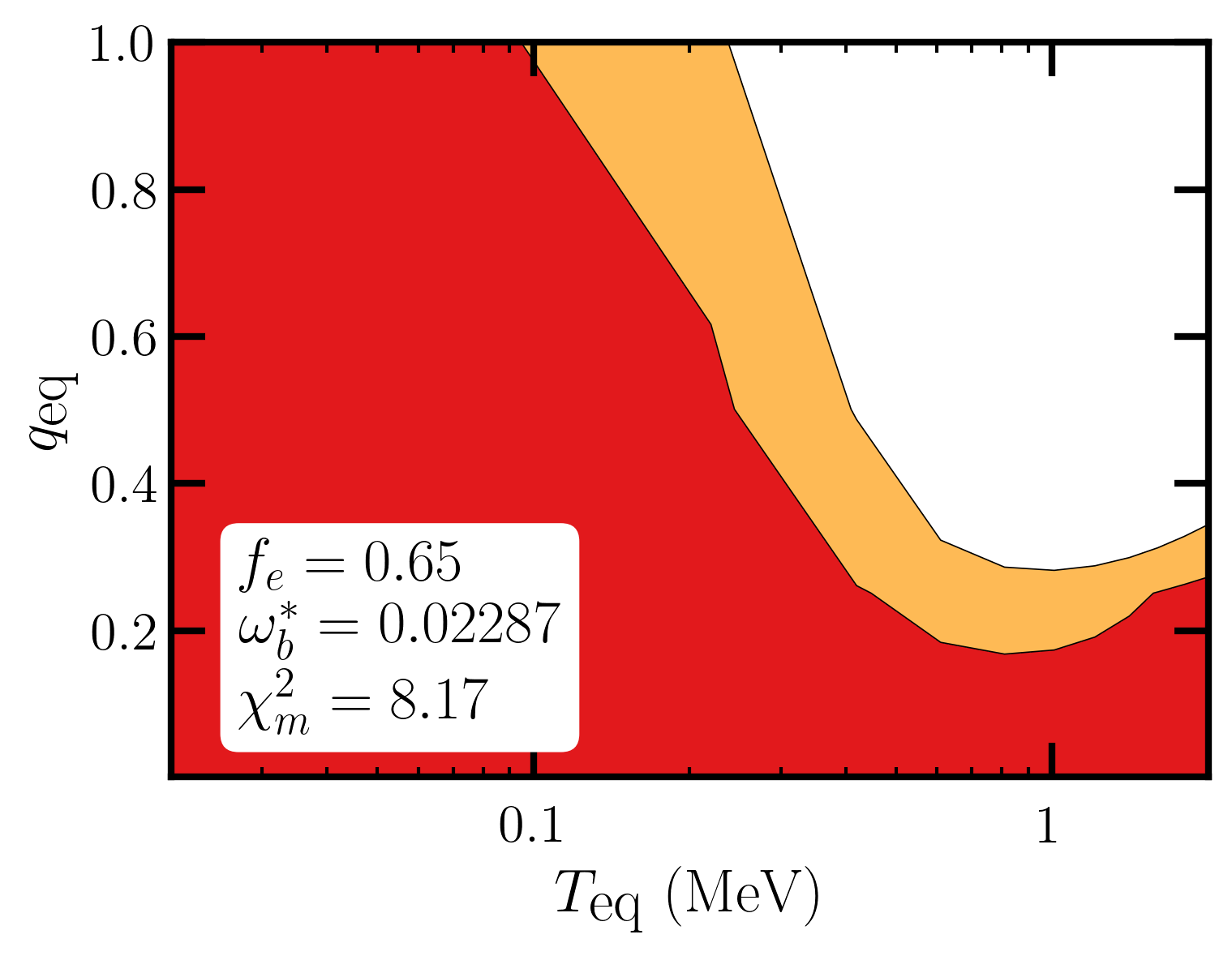}
    \includegraphics[width=0.2\textwidth]{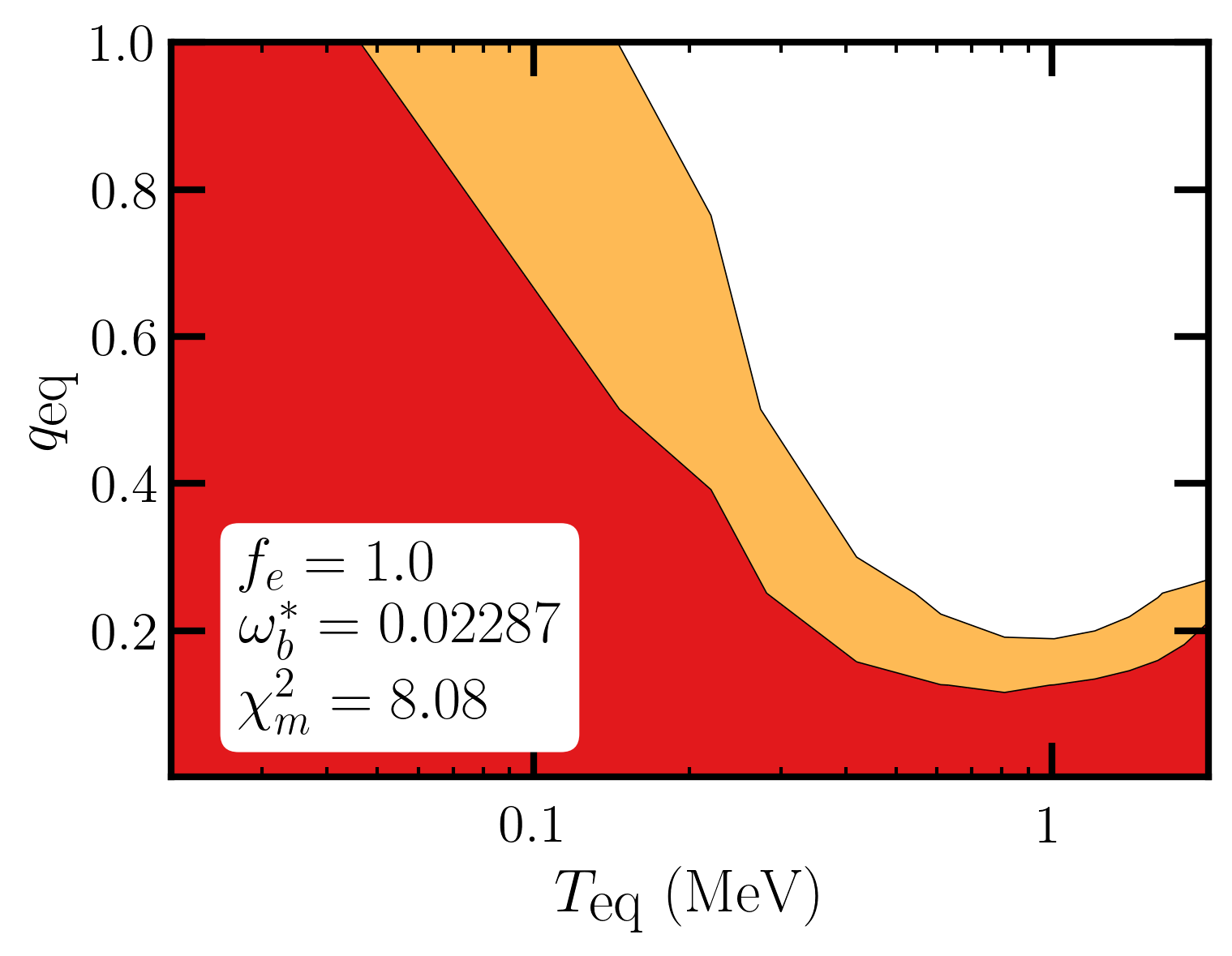}
    \caption{Limits on the cooling scenario, calculated with the PRIMAT network, using $\omega_b^*=0.02287$, for different values of $f_e$.}
\end{figure*}

\begin{figure*}[h]
    \centering
    \includegraphics[width=0.2\textwidth]{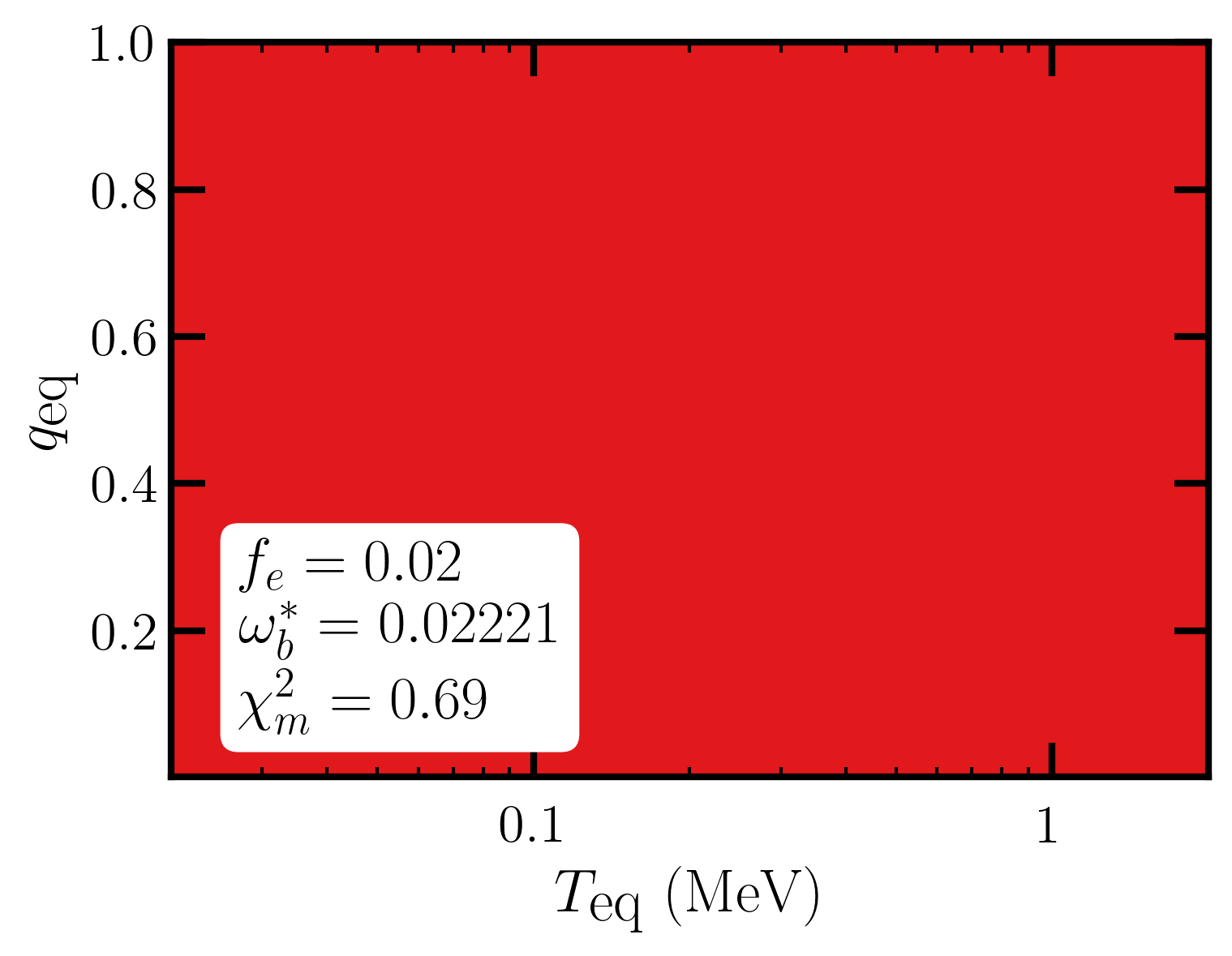}
    \includegraphics[width=0.2\textwidth]{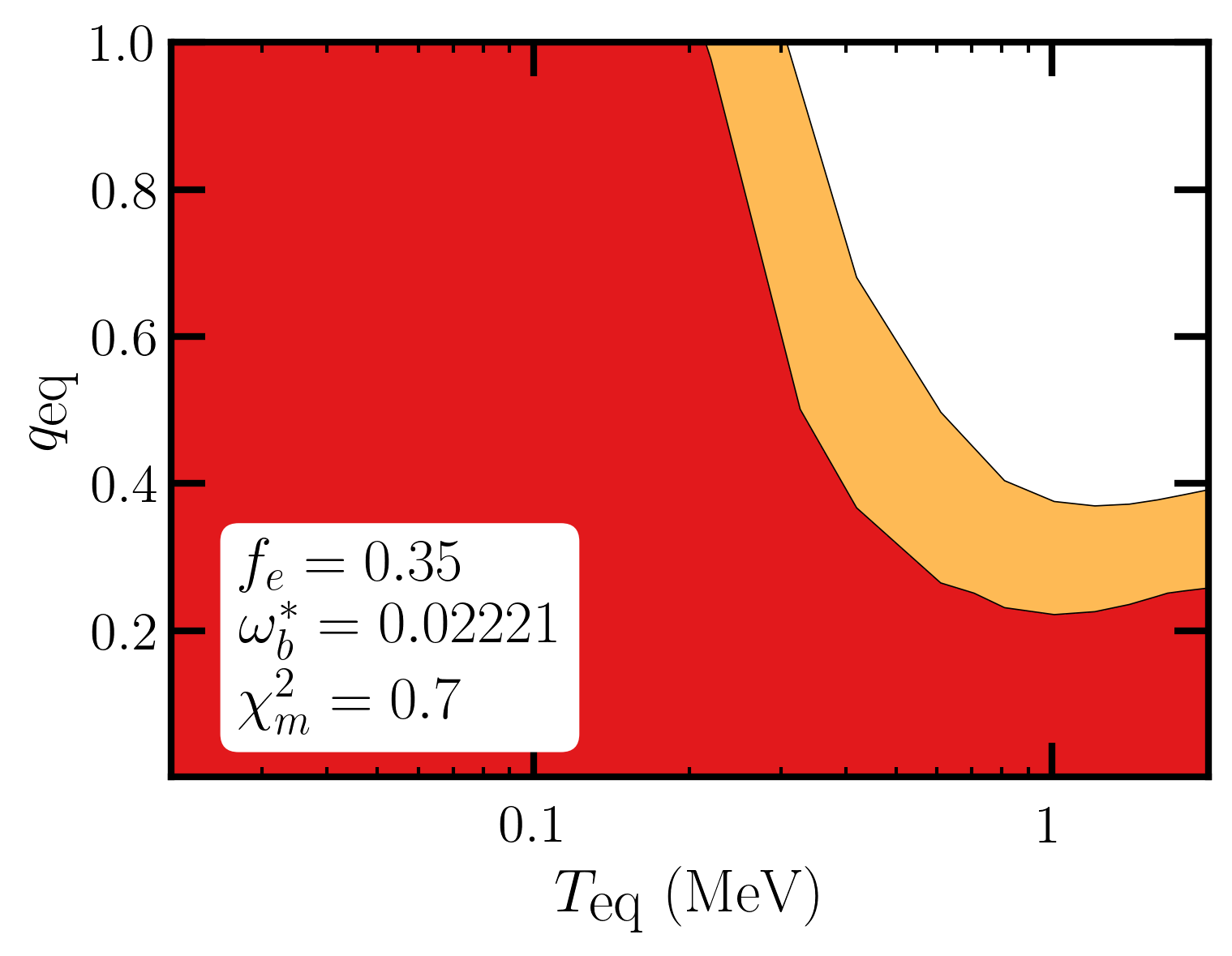}
    \includegraphics[width=0.2\textwidth]{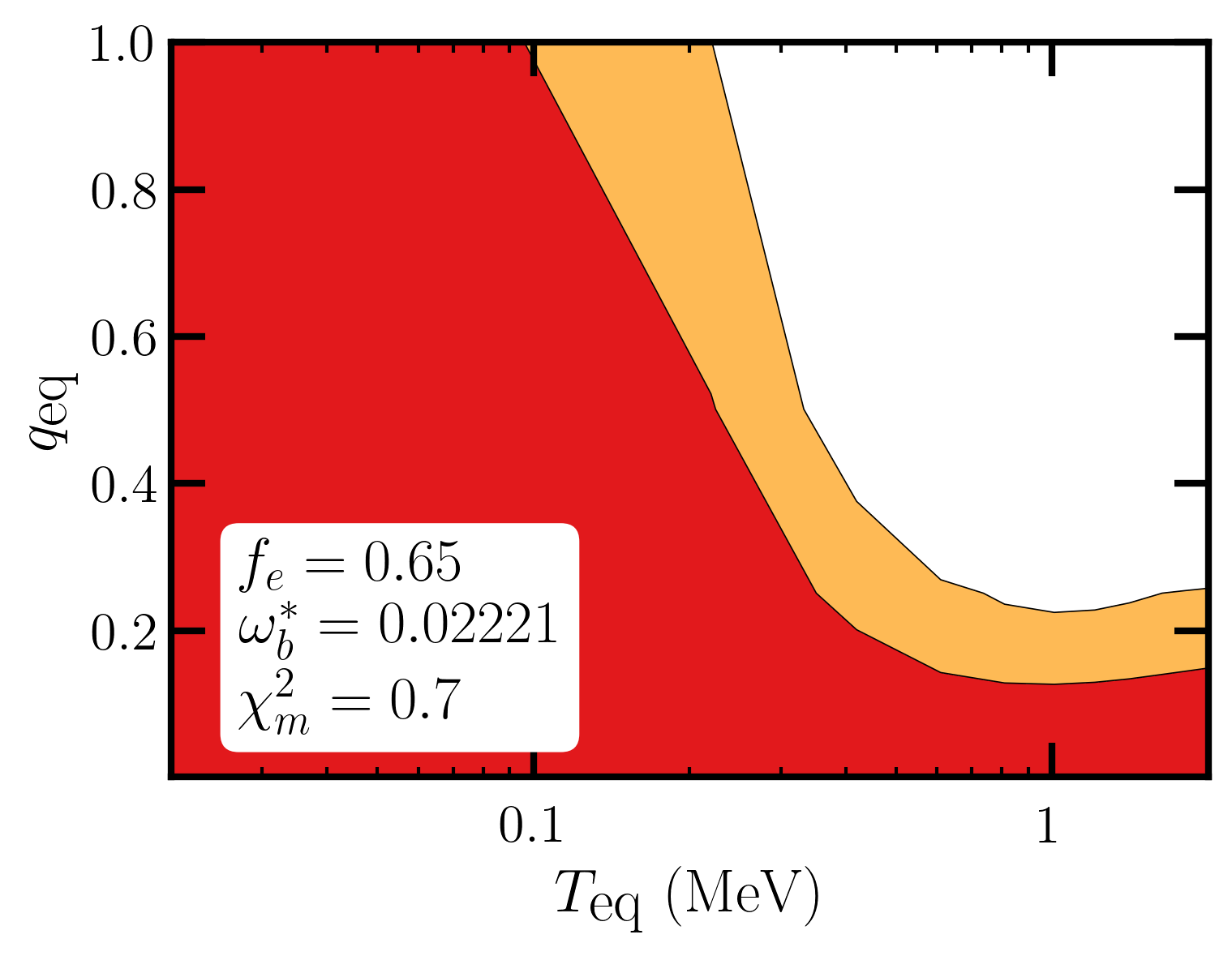}
    \includegraphics[width=0.2\textwidth]{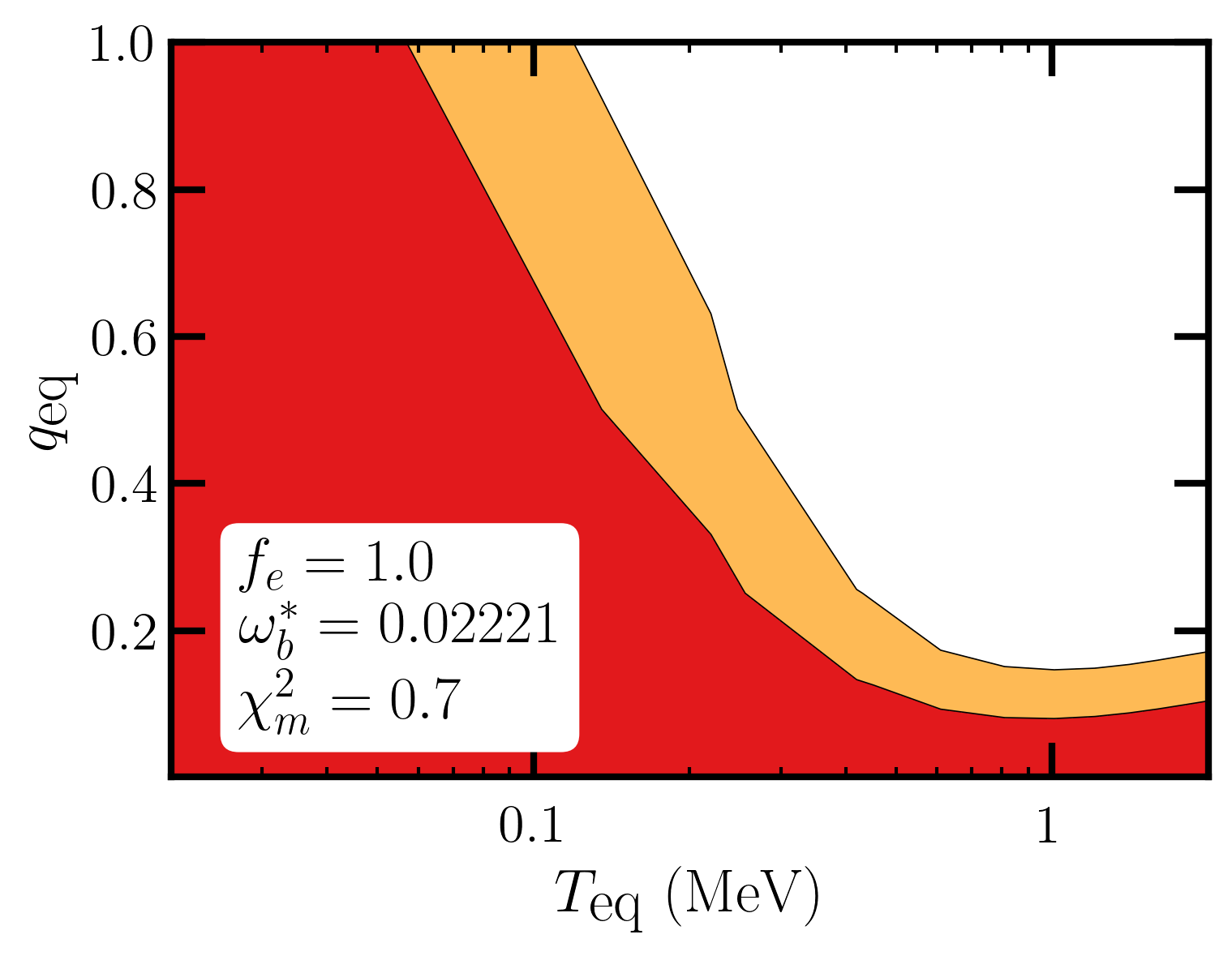}
     \caption{Limits on the cooling scenario, calculated with the PArthENoPE network, using $\omega_b^*=0.02221$, for different values of $f_e$.}
\end{figure*}

\begin{figure*}[h]
    \centering
    \includegraphics[width=0.2\textwidth]{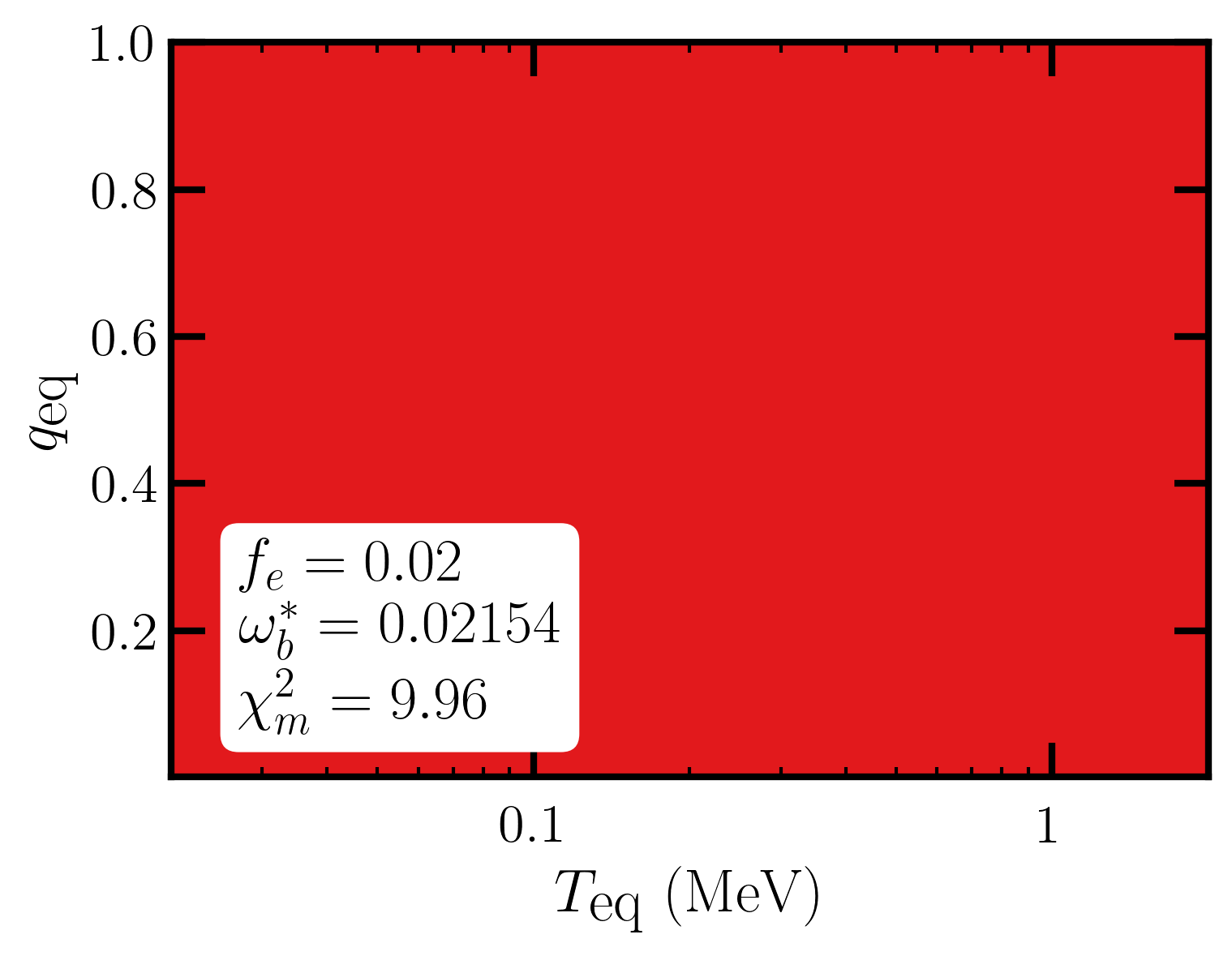}
    \includegraphics[width=0.2\textwidth]{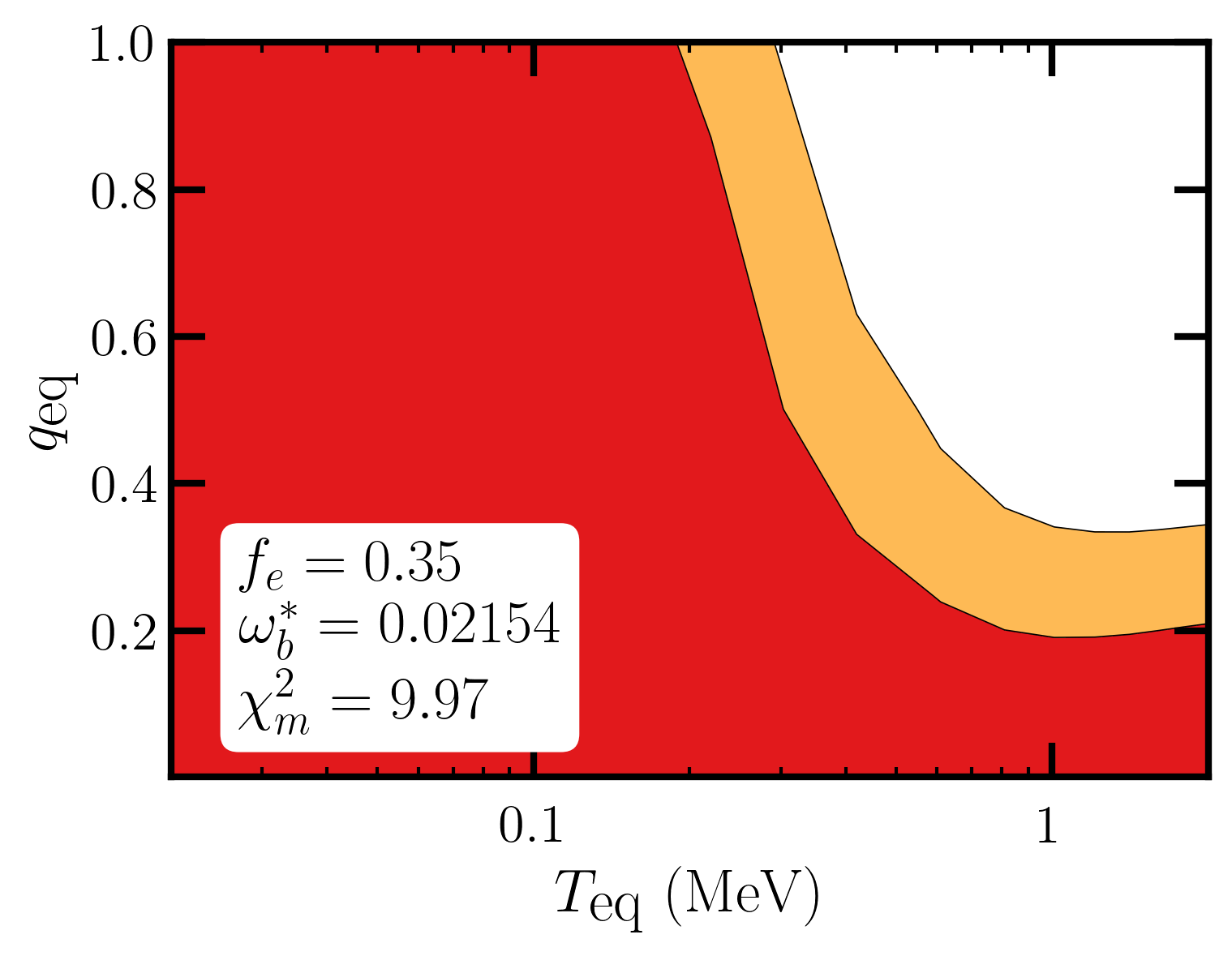}
    \includegraphics[width=0.2\textwidth]{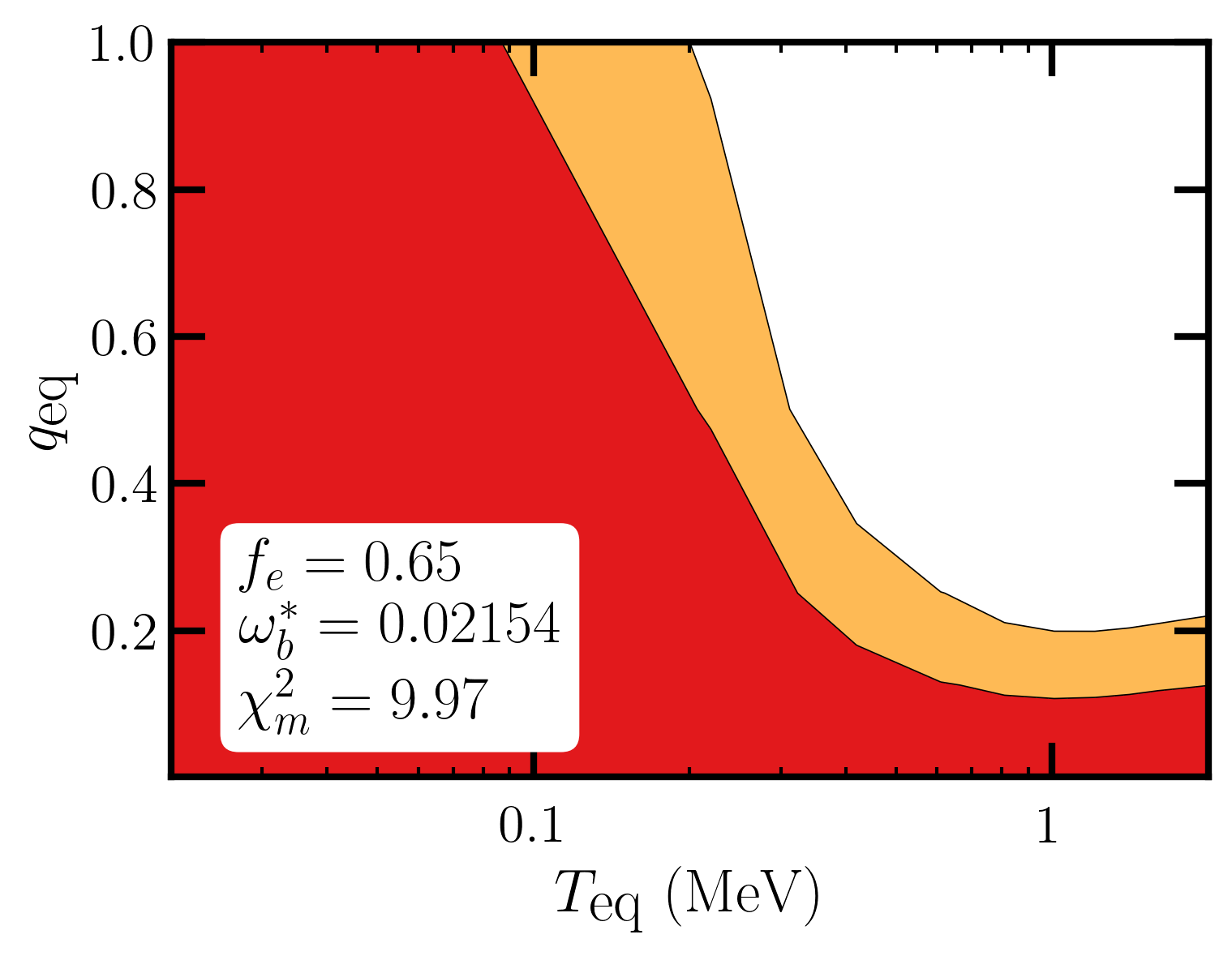}
    \includegraphics[width=0.2\textwidth]{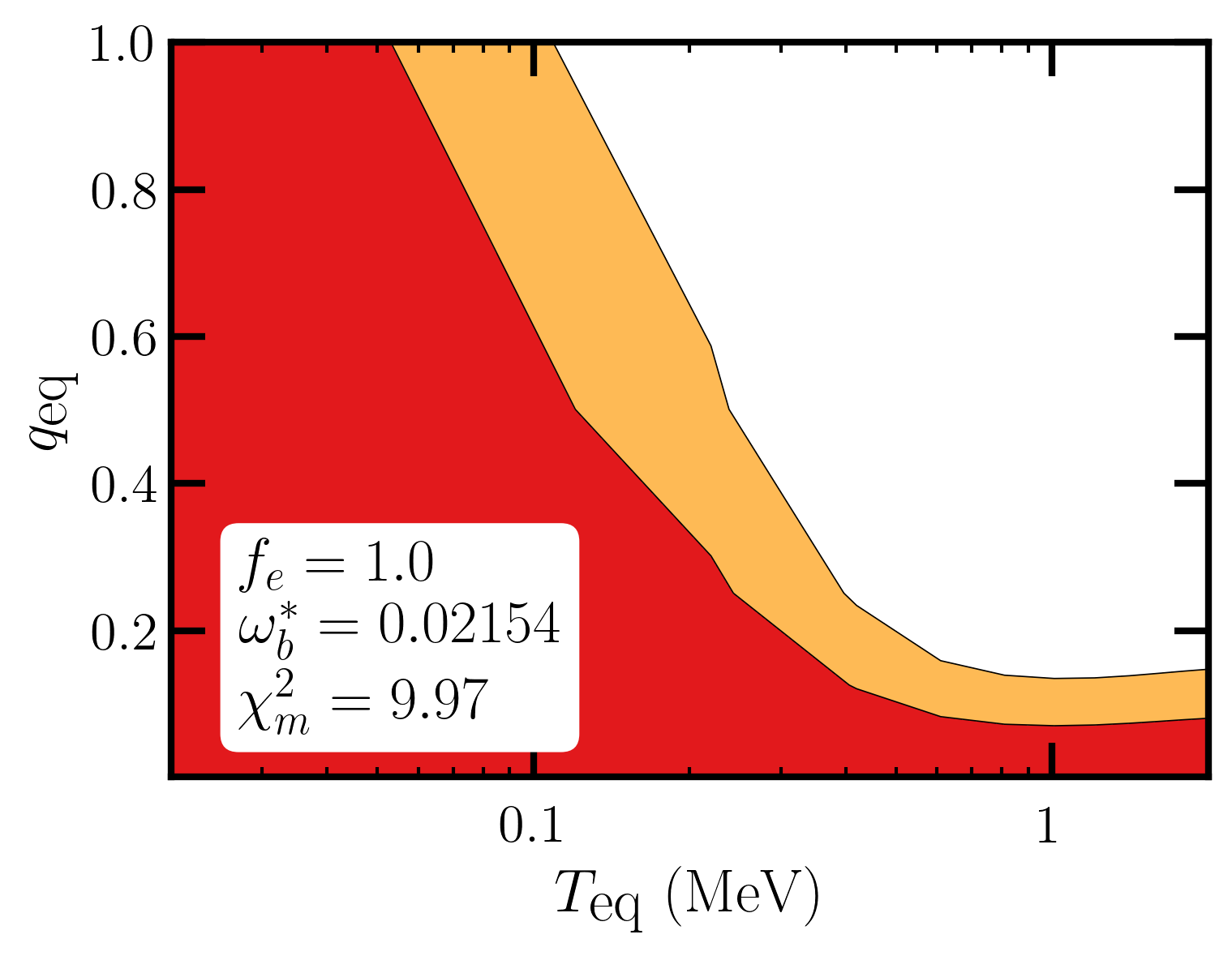}
    \caption{Limits on the cooling scenario, calculated with the PArthENoPE network, using $\omega_b^*=0.02154$, for different values of $f_e$.}
\end{figure*}

\vspace*{\fill}

\begin{figure*}[h]
    \centering
    \includegraphics[width=0.2\textwidth]{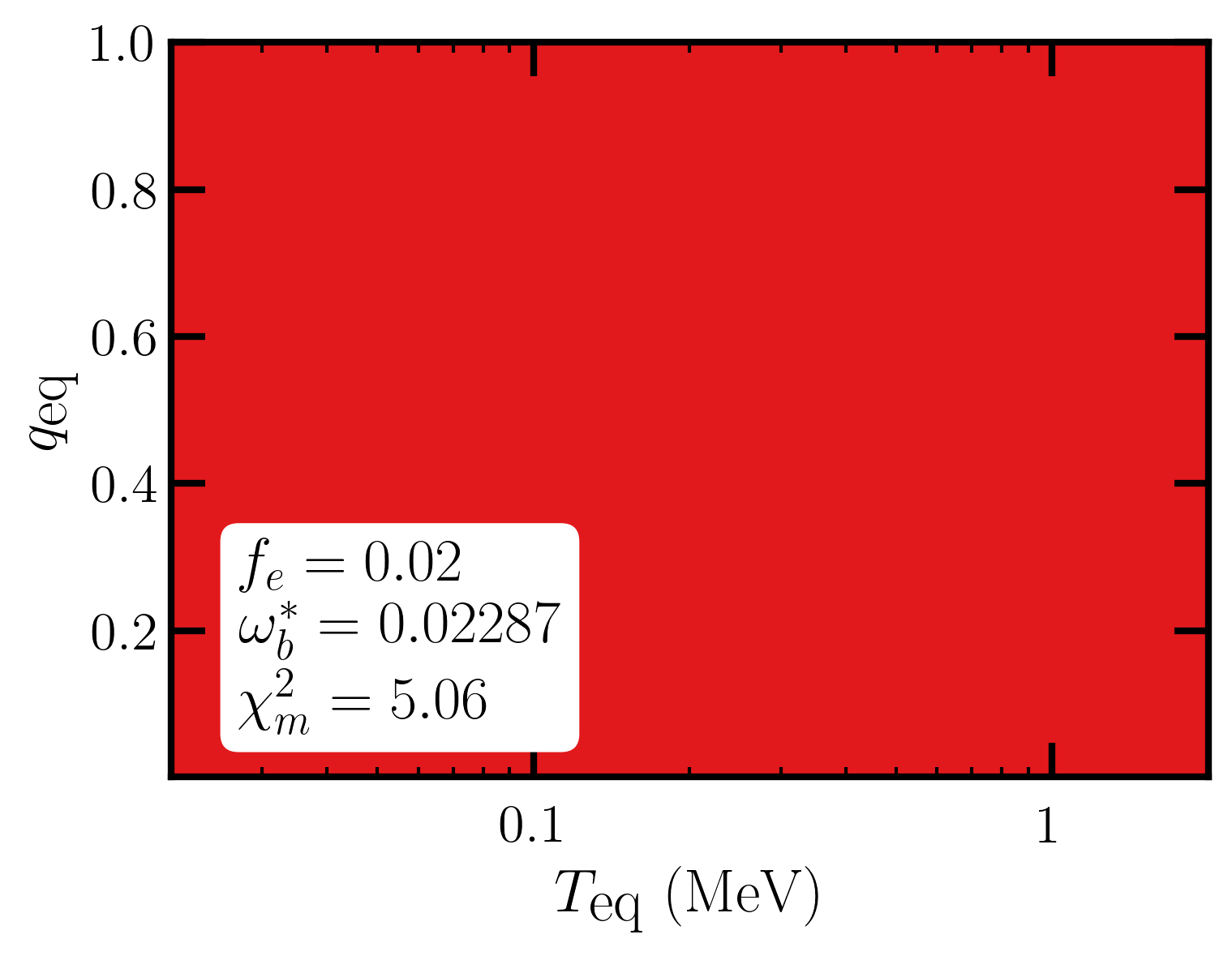}
    \includegraphics[width=0.2\textwidth]{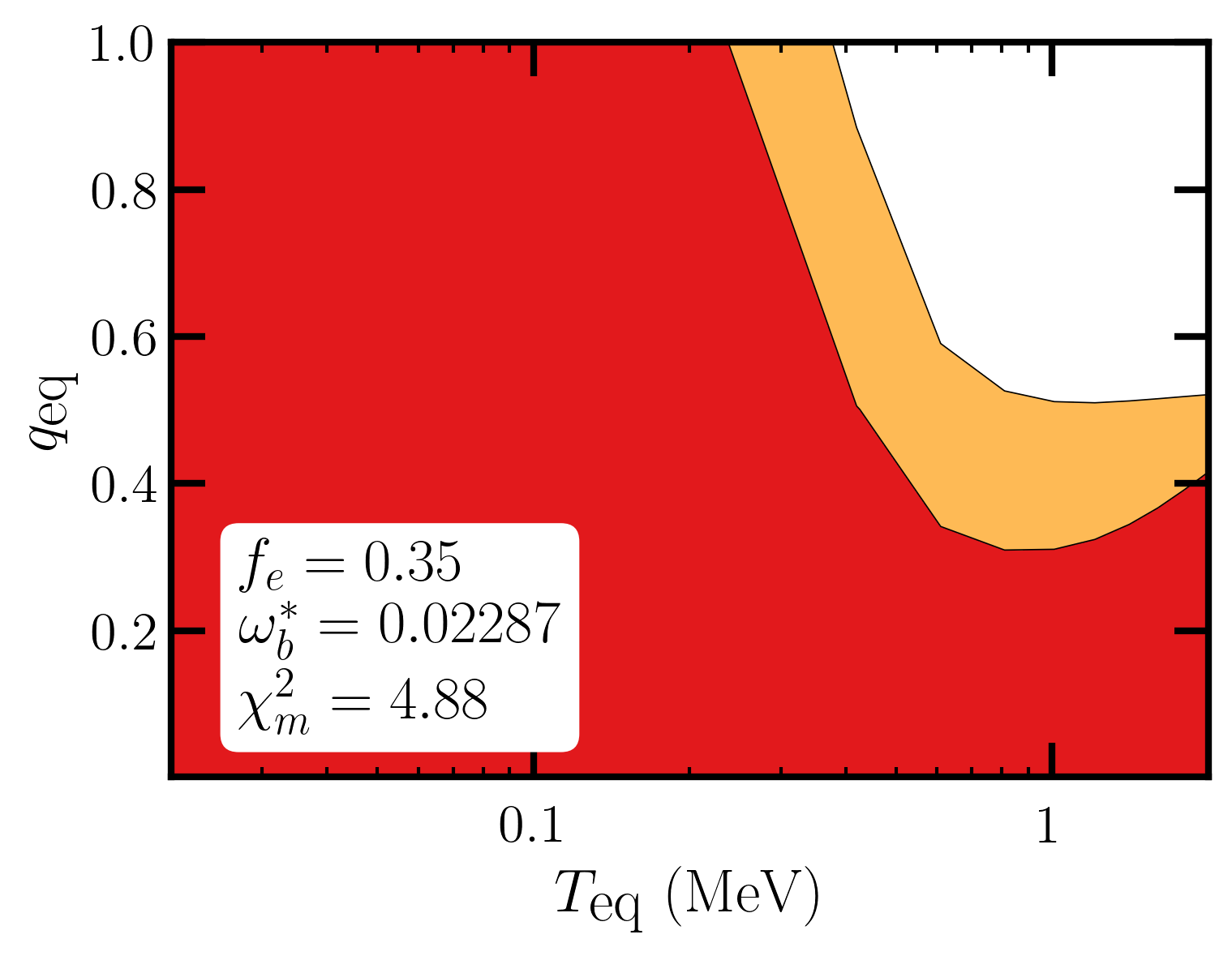}
    \includegraphics[width=0.2\textwidth]{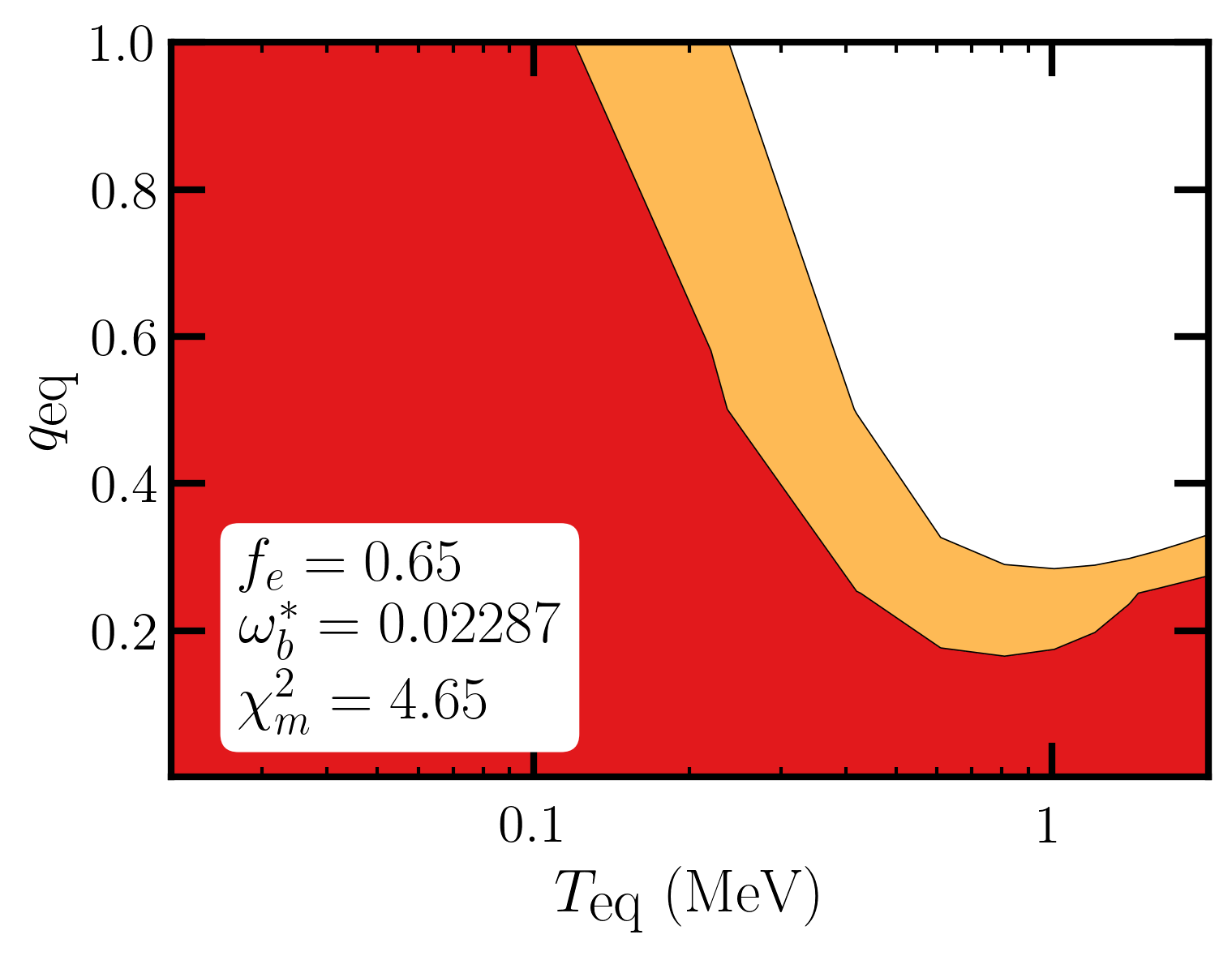}
    \includegraphics[width=0.2\textwidth]{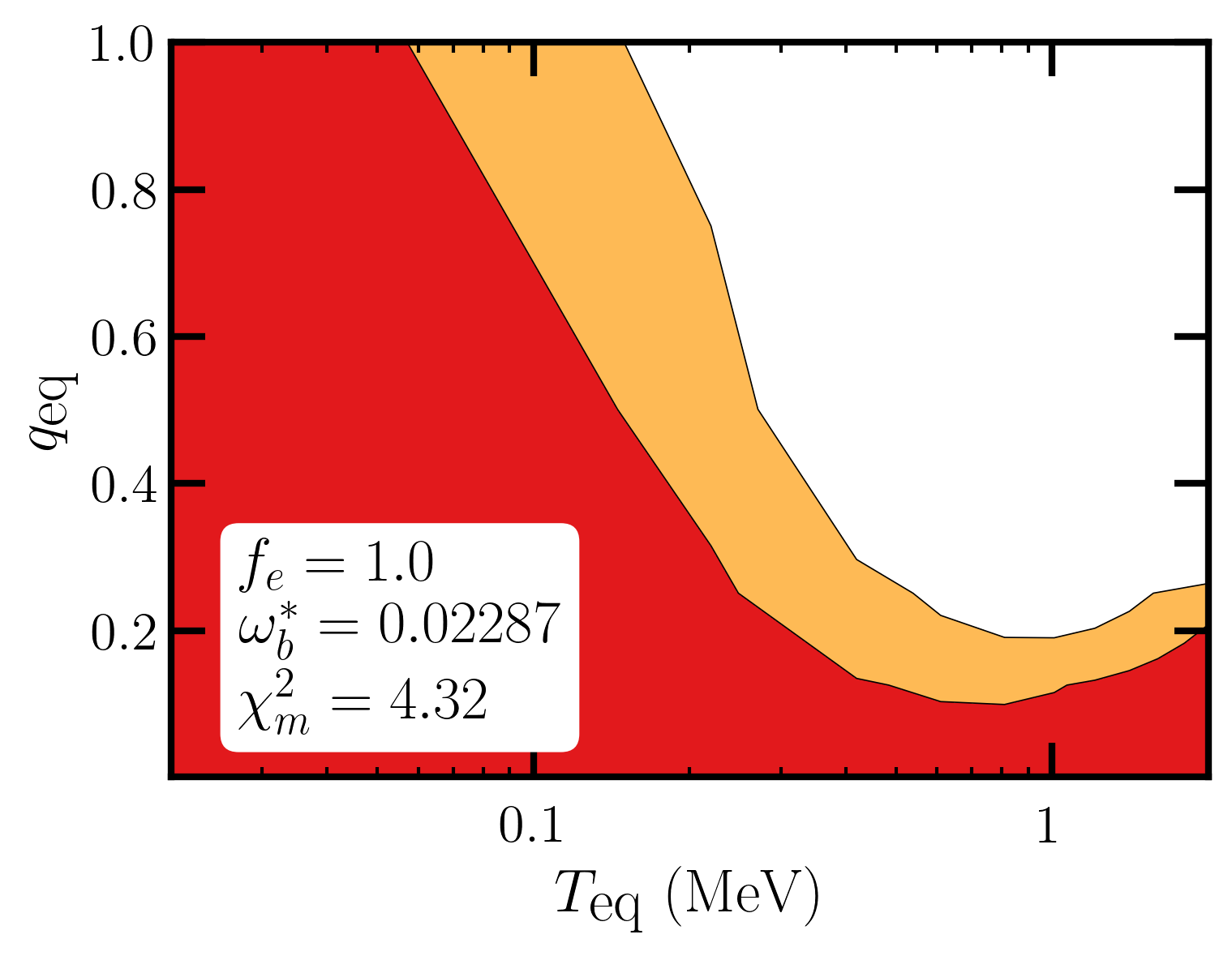}
    \caption{Limits on the cooling scenario, calculated with the PArthENoPE network, using $\omega_b^*=0.02287$, for different values of $f_e$.}
\end{figure*}

\begin{figure*}[h]
    \centering
    \includegraphics[width=0.2\textwidth]{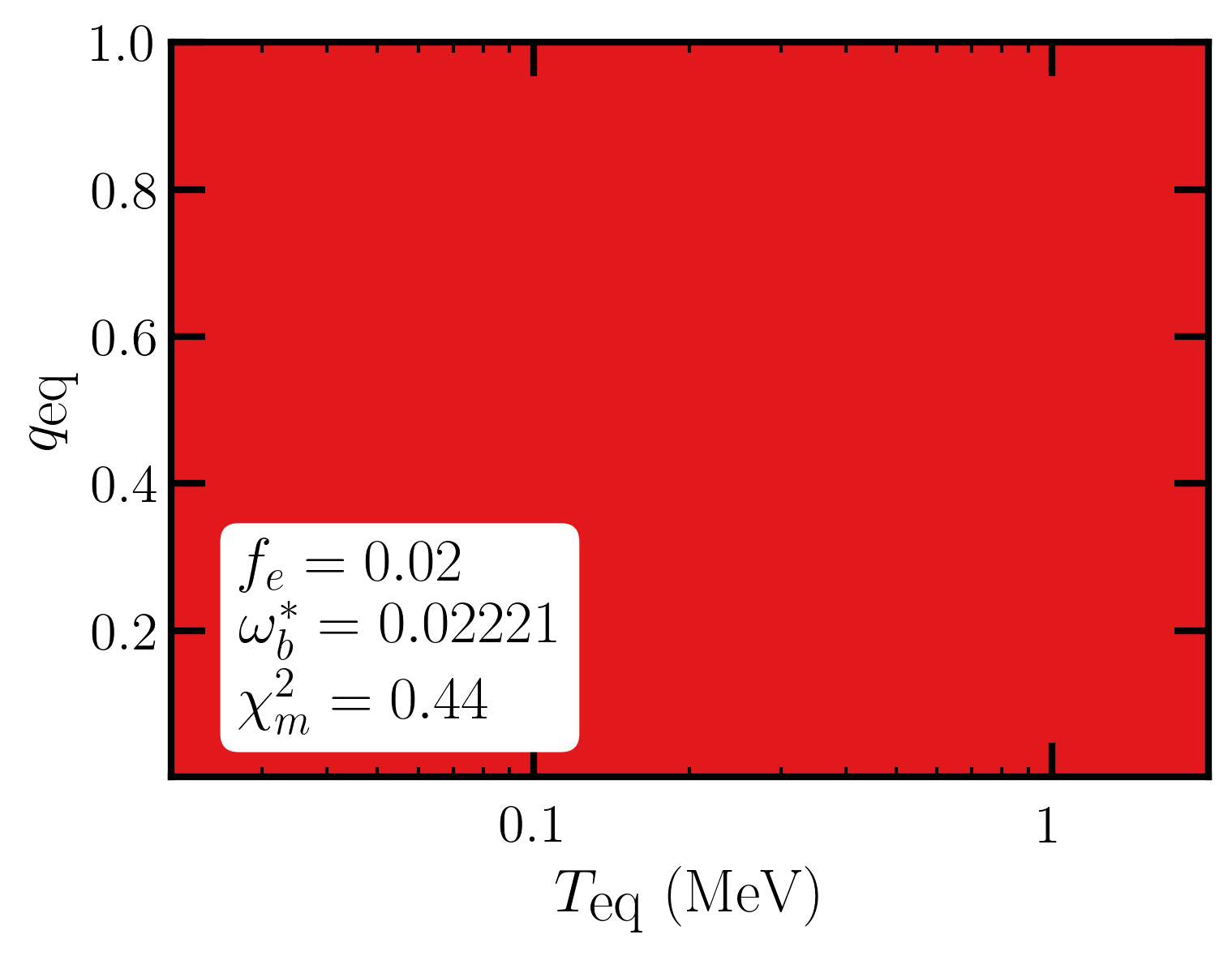}
    \includegraphics[width=0.2\textwidth]{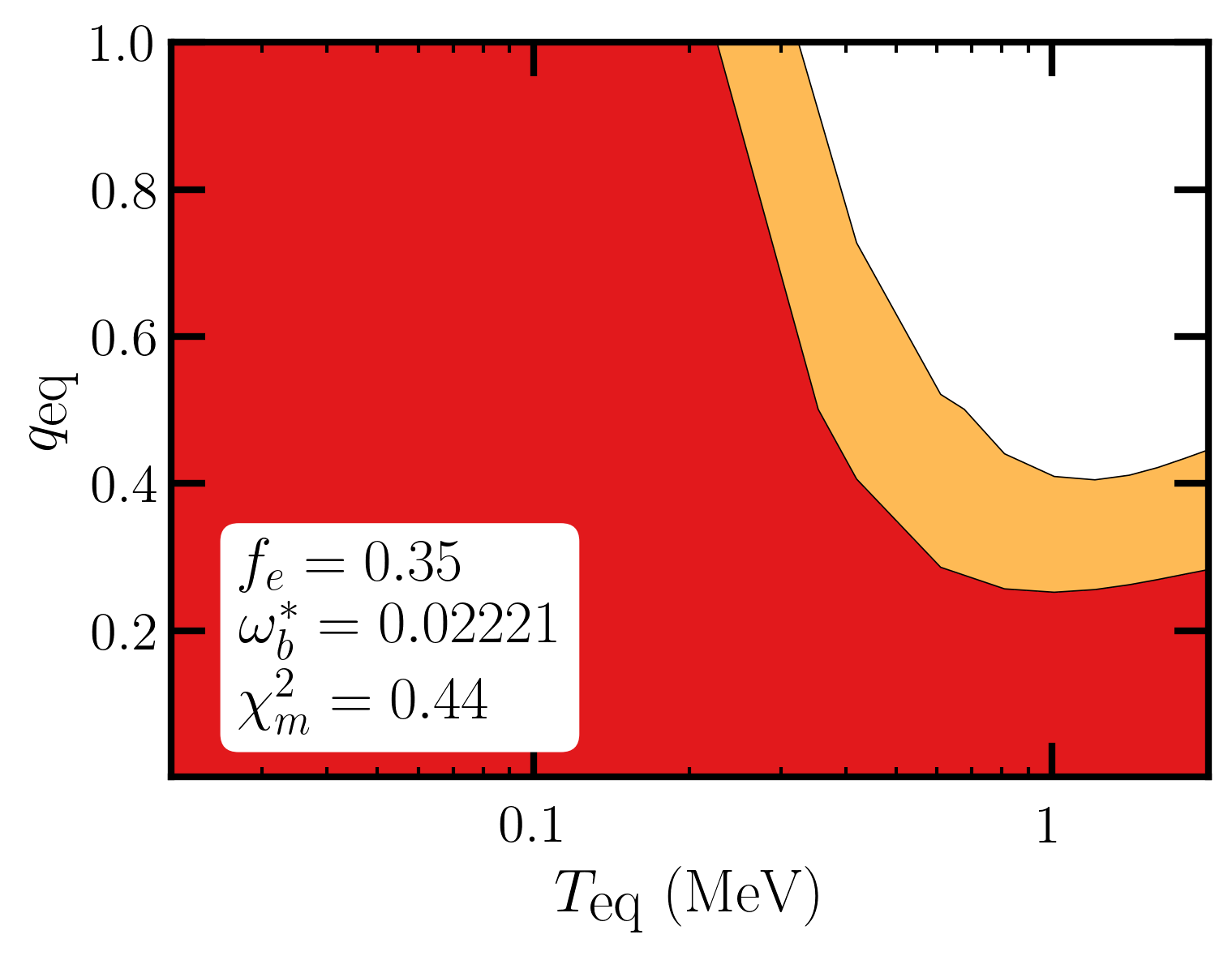}
    \includegraphics[width=0.2\textwidth]{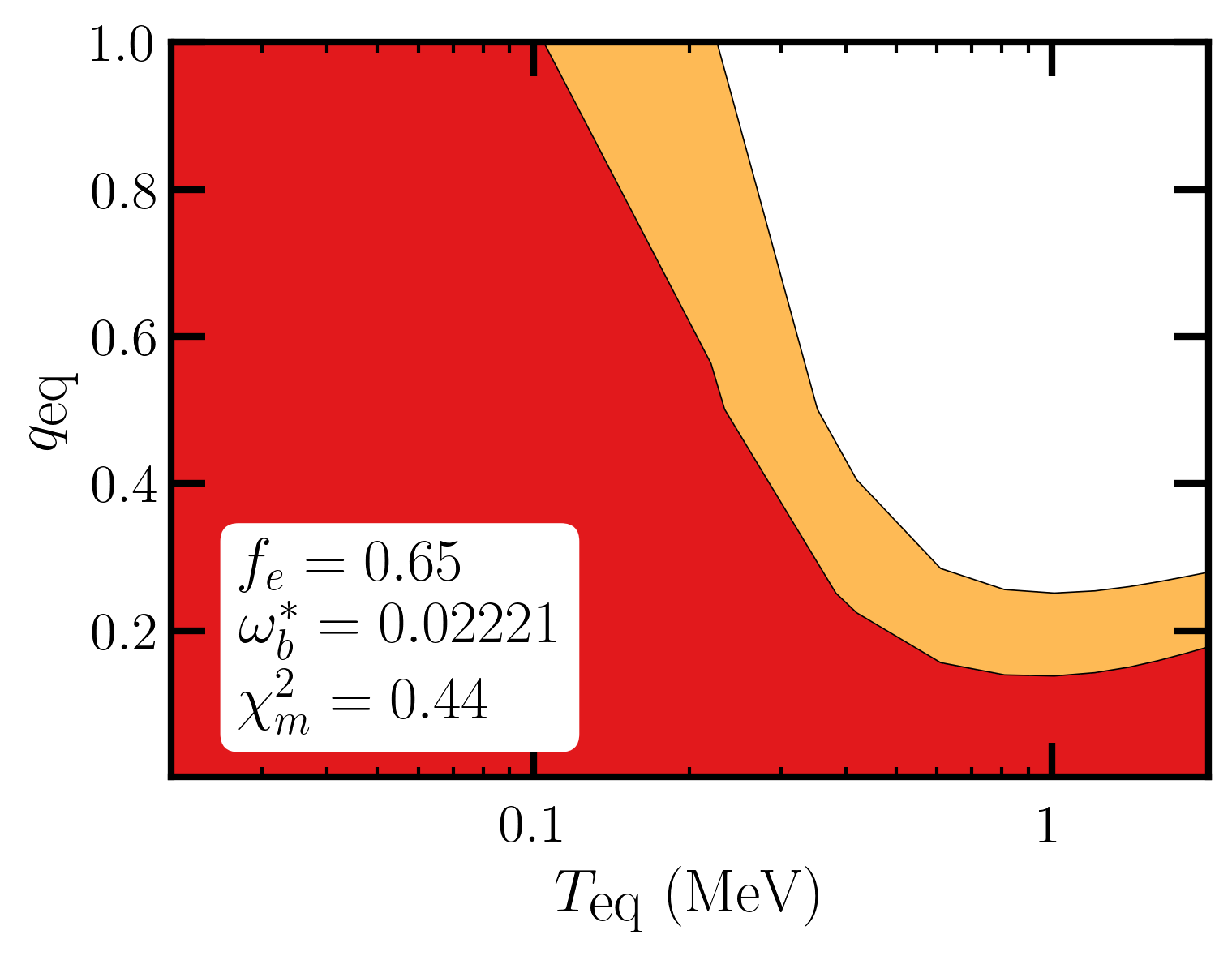}
    \includegraphics[width=0.2\textwidth]{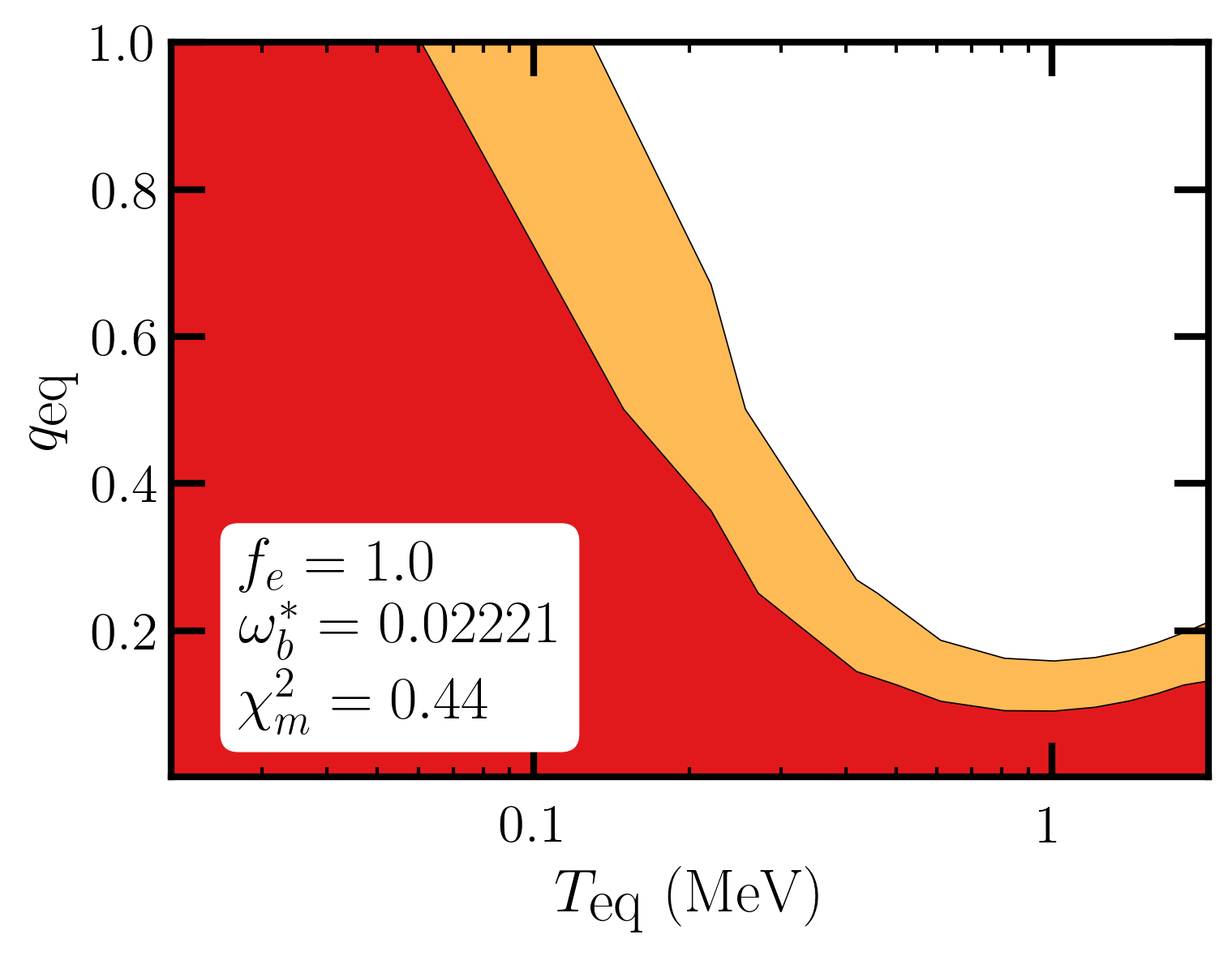}
    \caption{Limits on the cooling scenario, calculated with the ``combined" network, using $\omega_b^*=0.02221$, for different values of $f_e$.}
\end{figure*}

\begin{figure*}[h]
    \centering
    \includegraphics[width=0.2\textwidth]{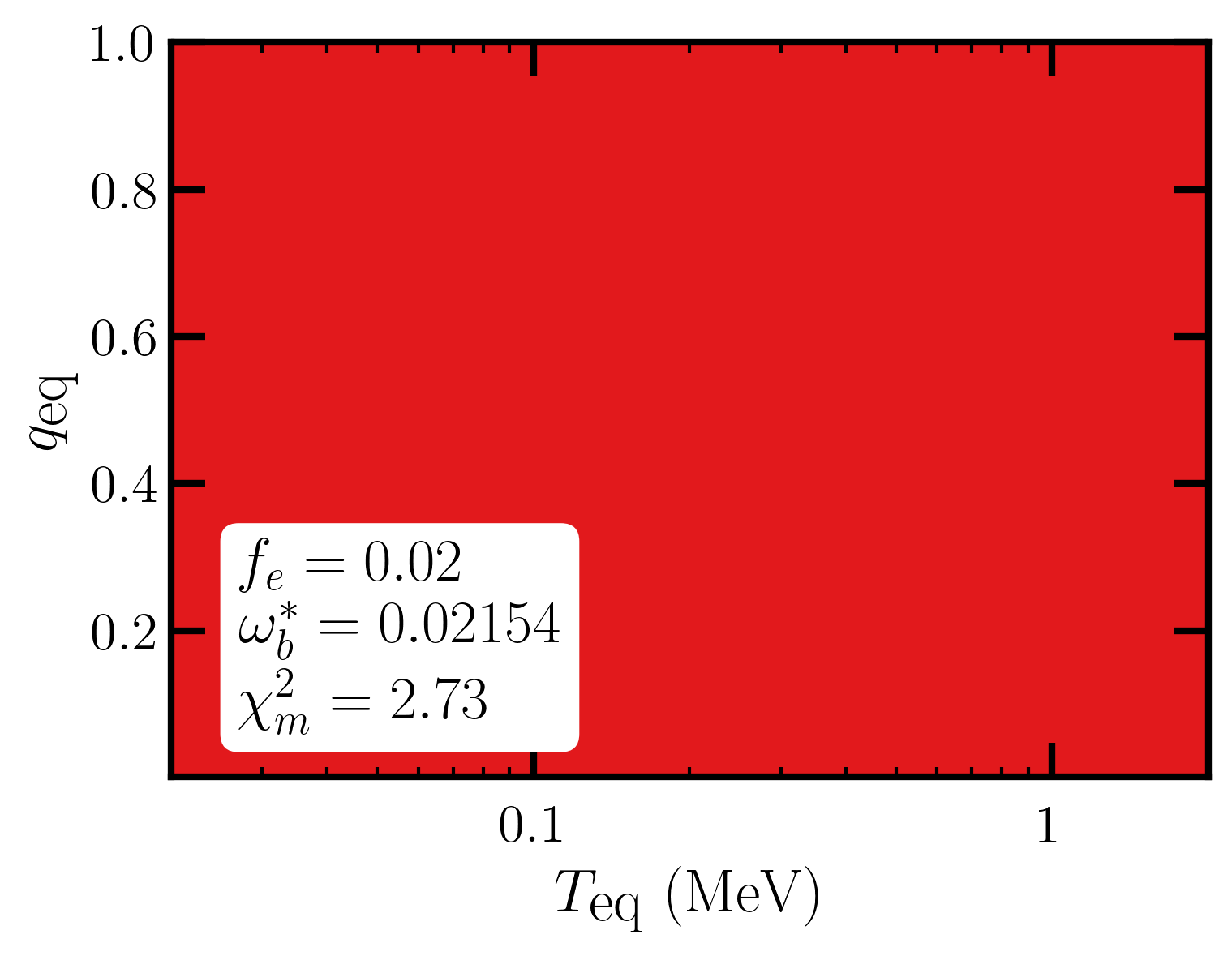}
    \includegraphics[width=0.2\textwidth]{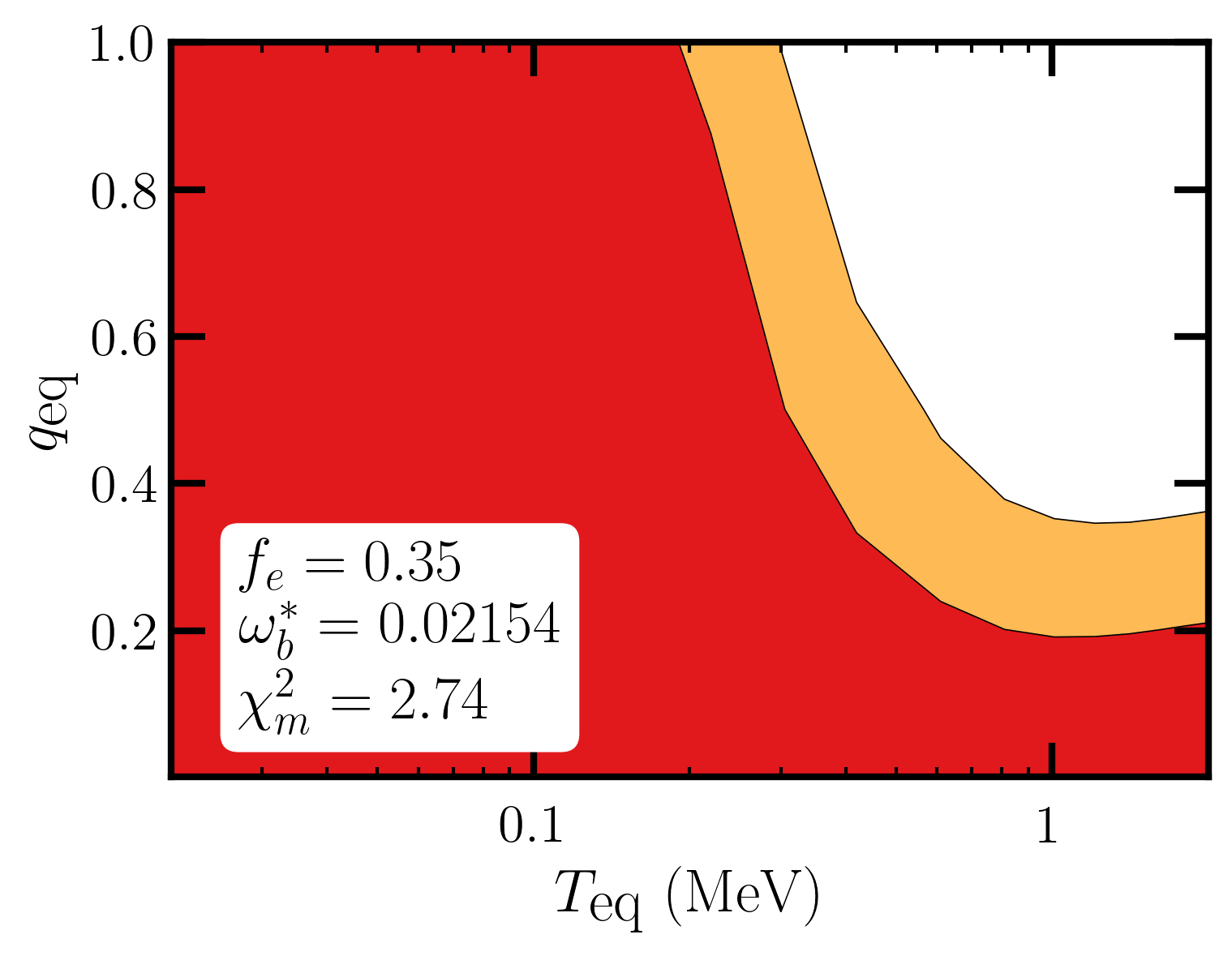}
    \includegraphics[width=0.2\textwidth]{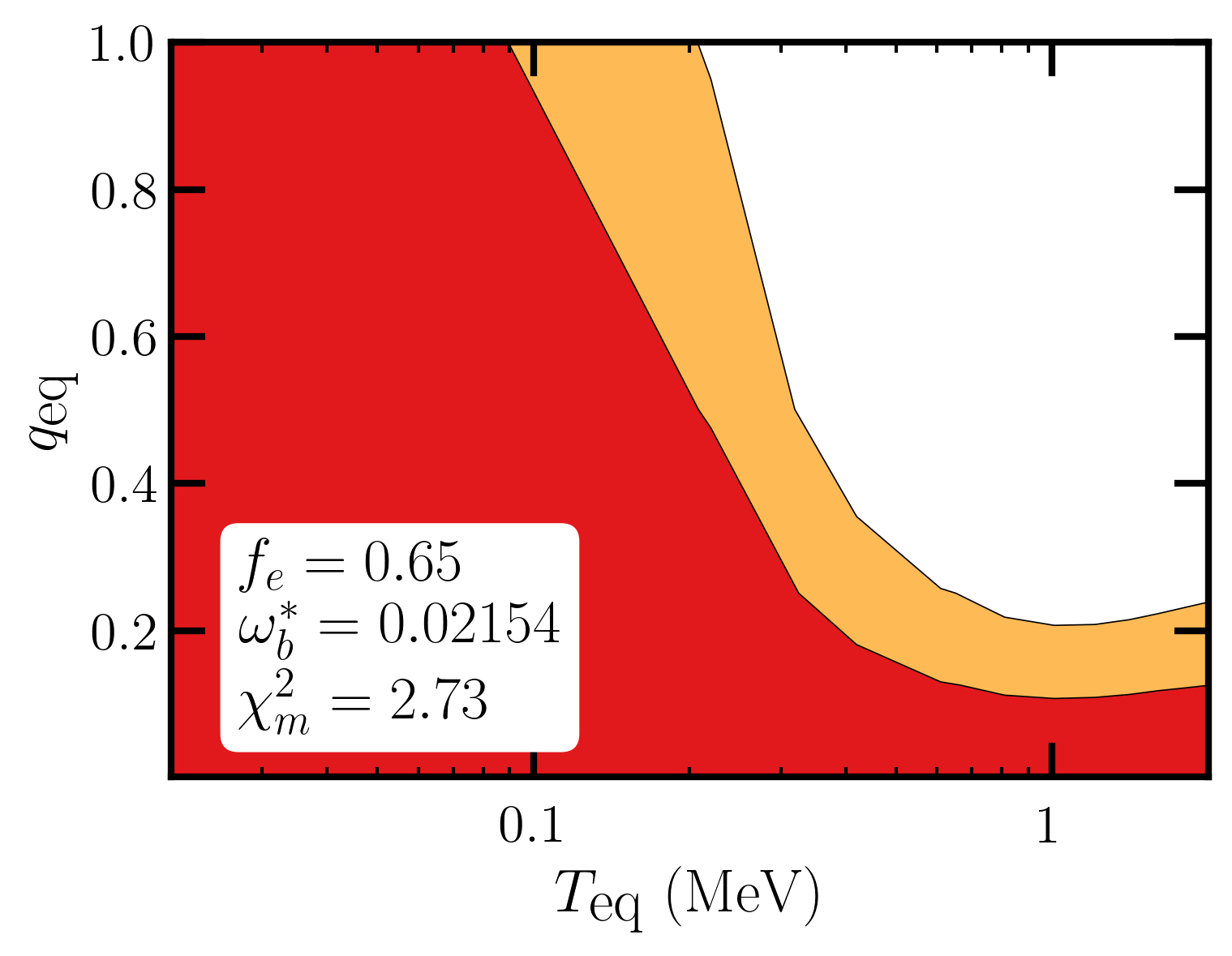}
    \includegraphics[width=0.2\textwidth]{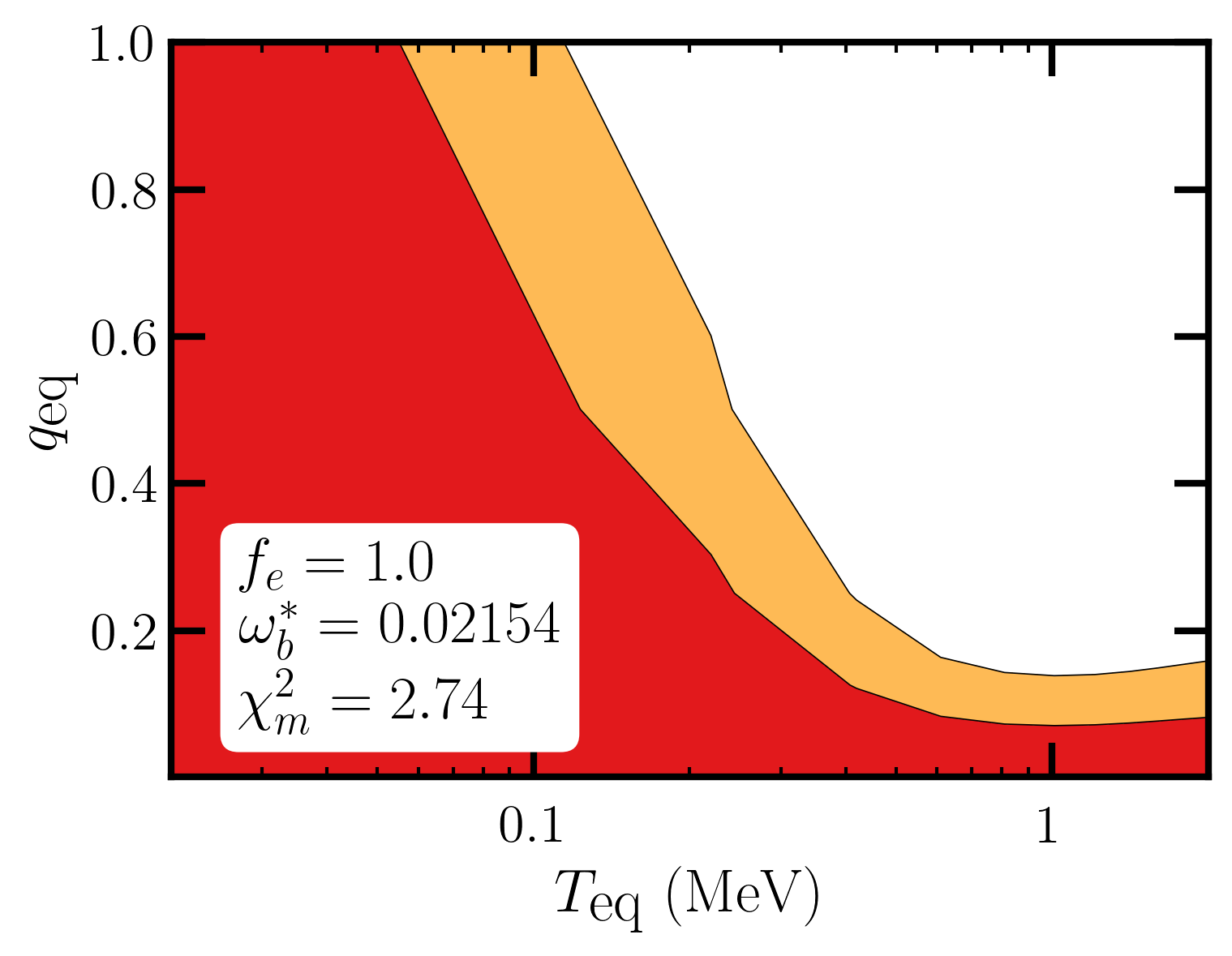}
    \caption{Limits on the cooling scenario, calculated with the ``combined" network, using $\omega_b^*=0.02154$, for different values of $f_e$.}
\end{figure*}

\begin{figure*}[h]
    \centering
    \includegraphics[width=0.2\textwidth]{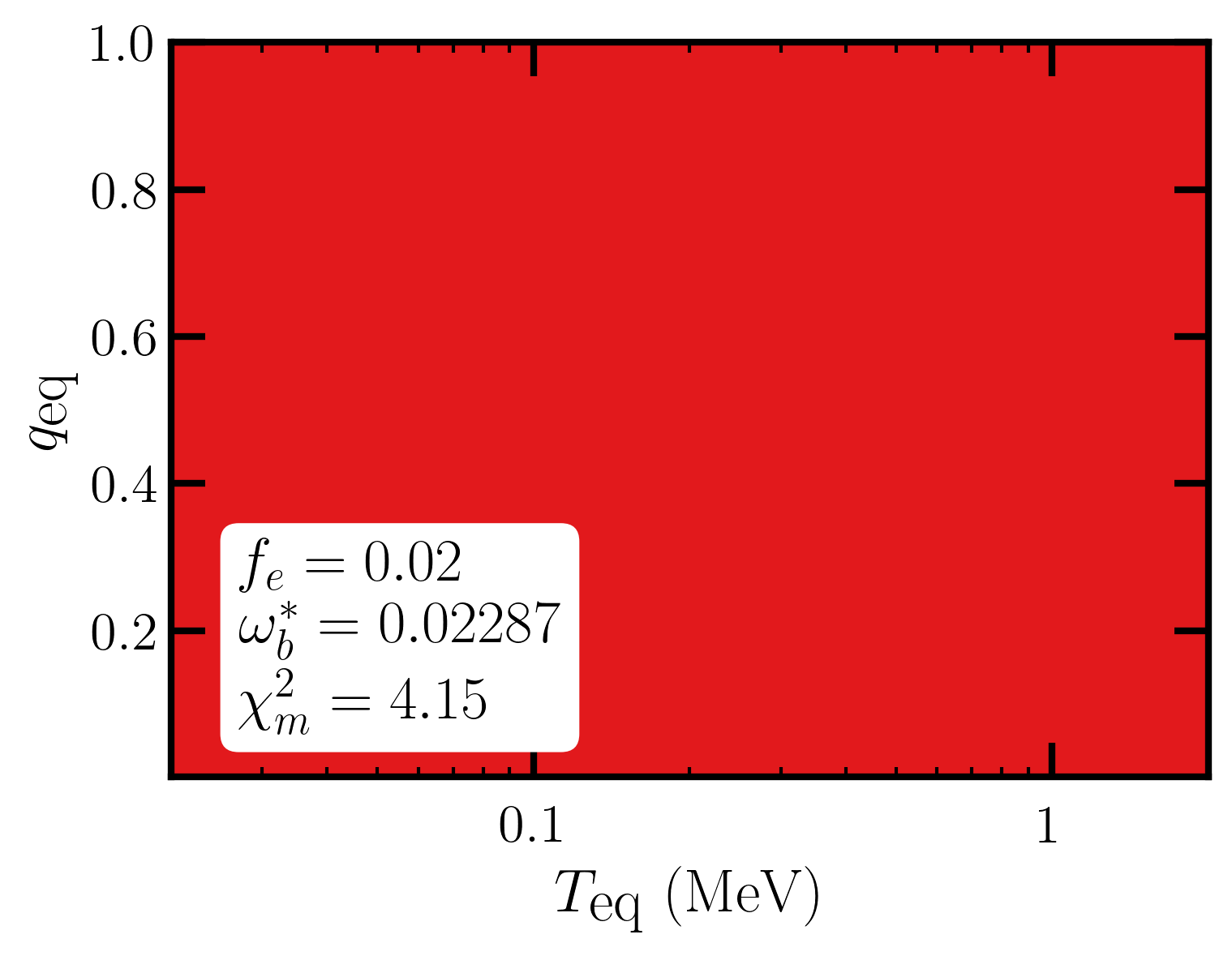}
    \includegraphics[width=0.2\textwidth]{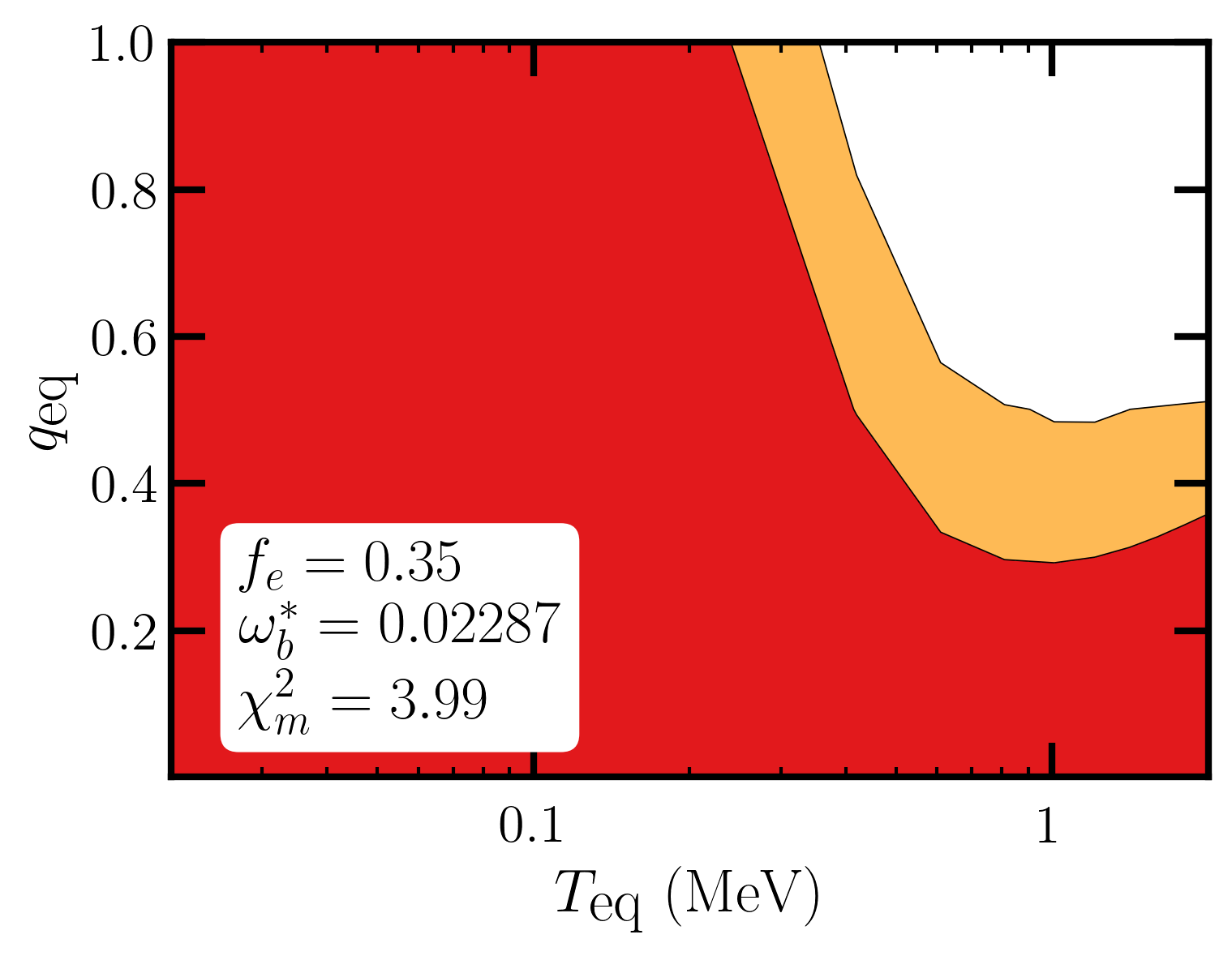}
    \includegraphics[width=0.2\textwidth]{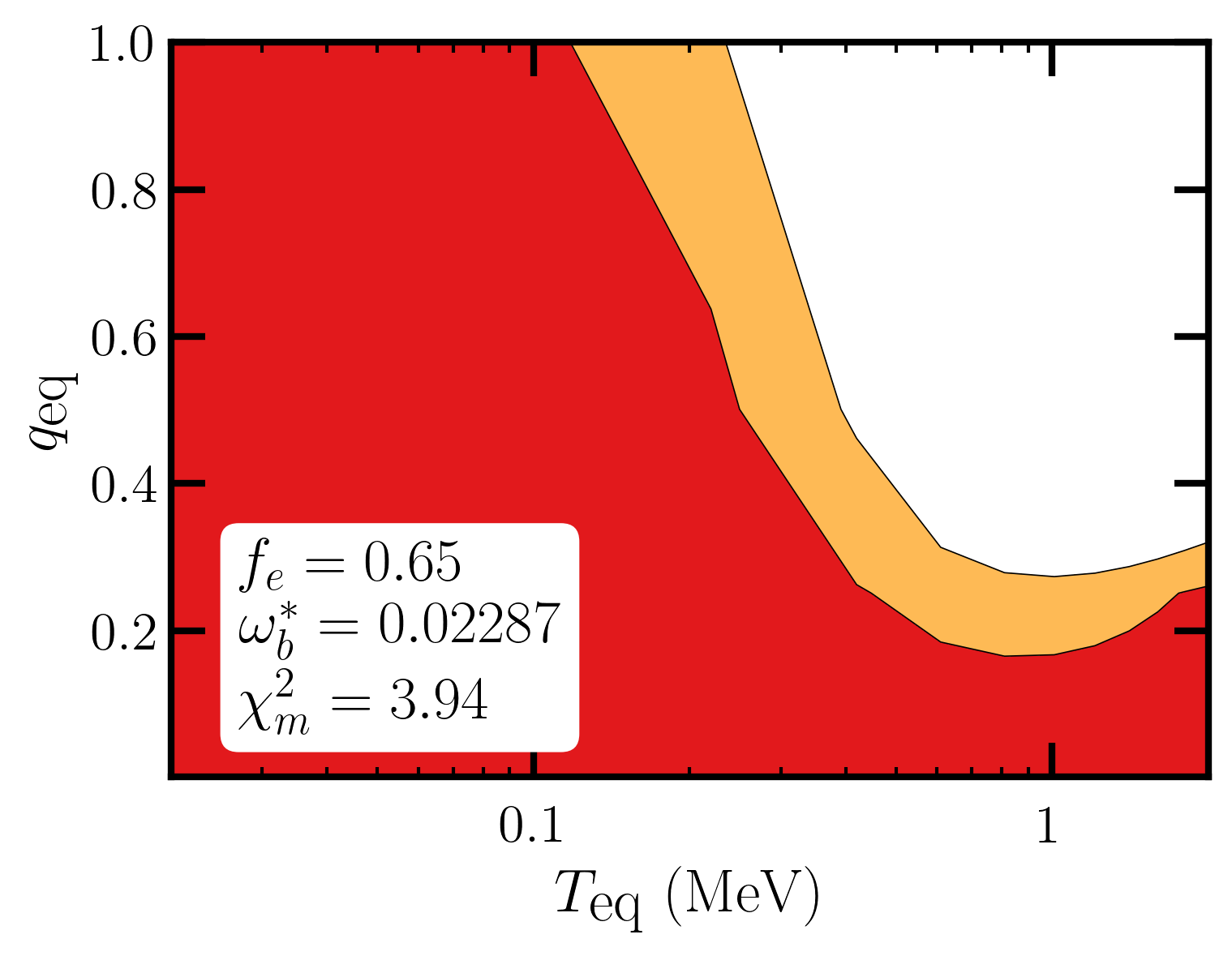}
    \includegraphics[width=0.2\textwidth]{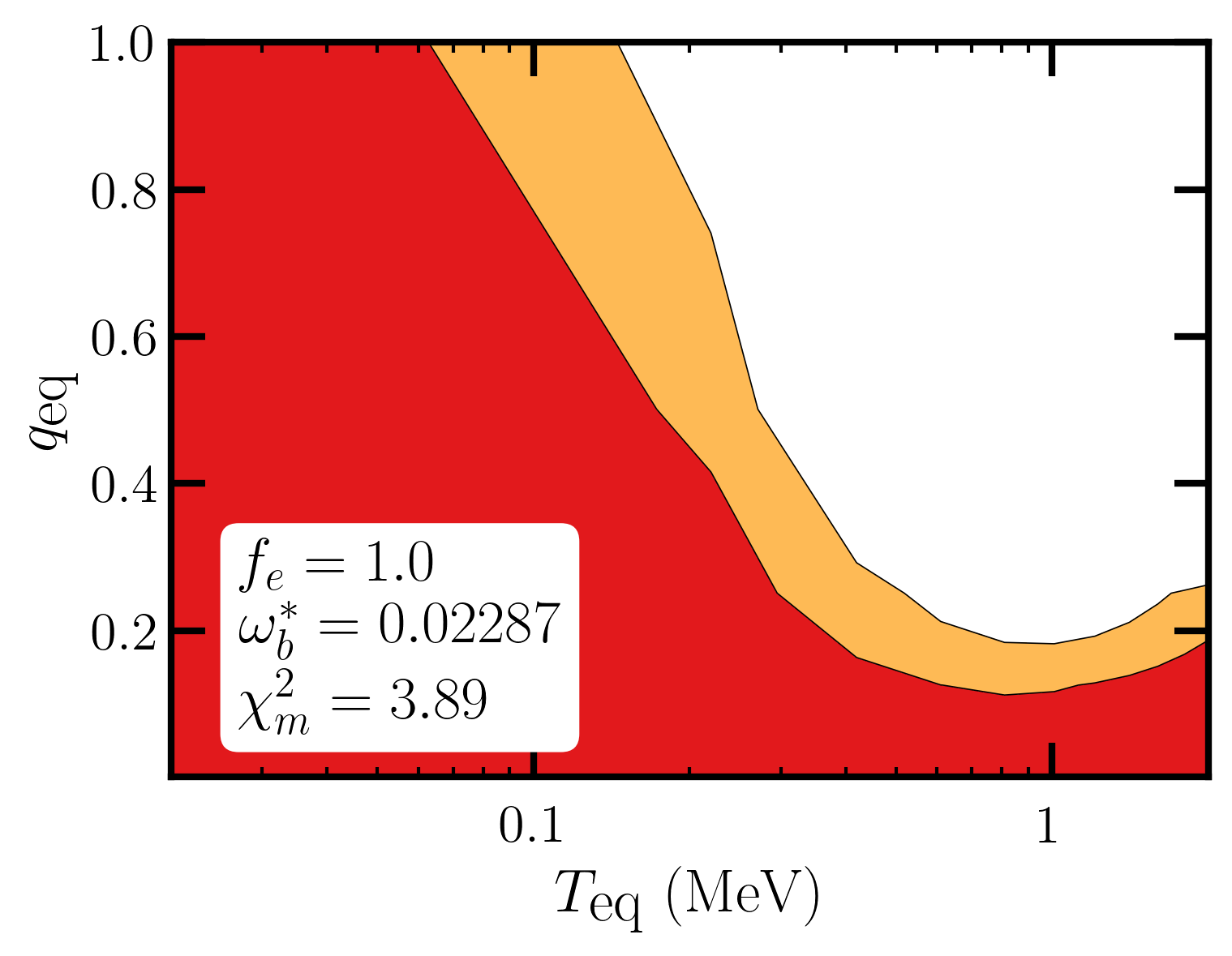}
    \caption{Limits on the cooling scenario, calculated with the ``combined" network, using $\omega_b^*=0.02287$, for different values of $f_e$.}
\end{figure*}

\begin{figure*}[h]
    \centering
    \includegraphics[width=0.2\textwidth]{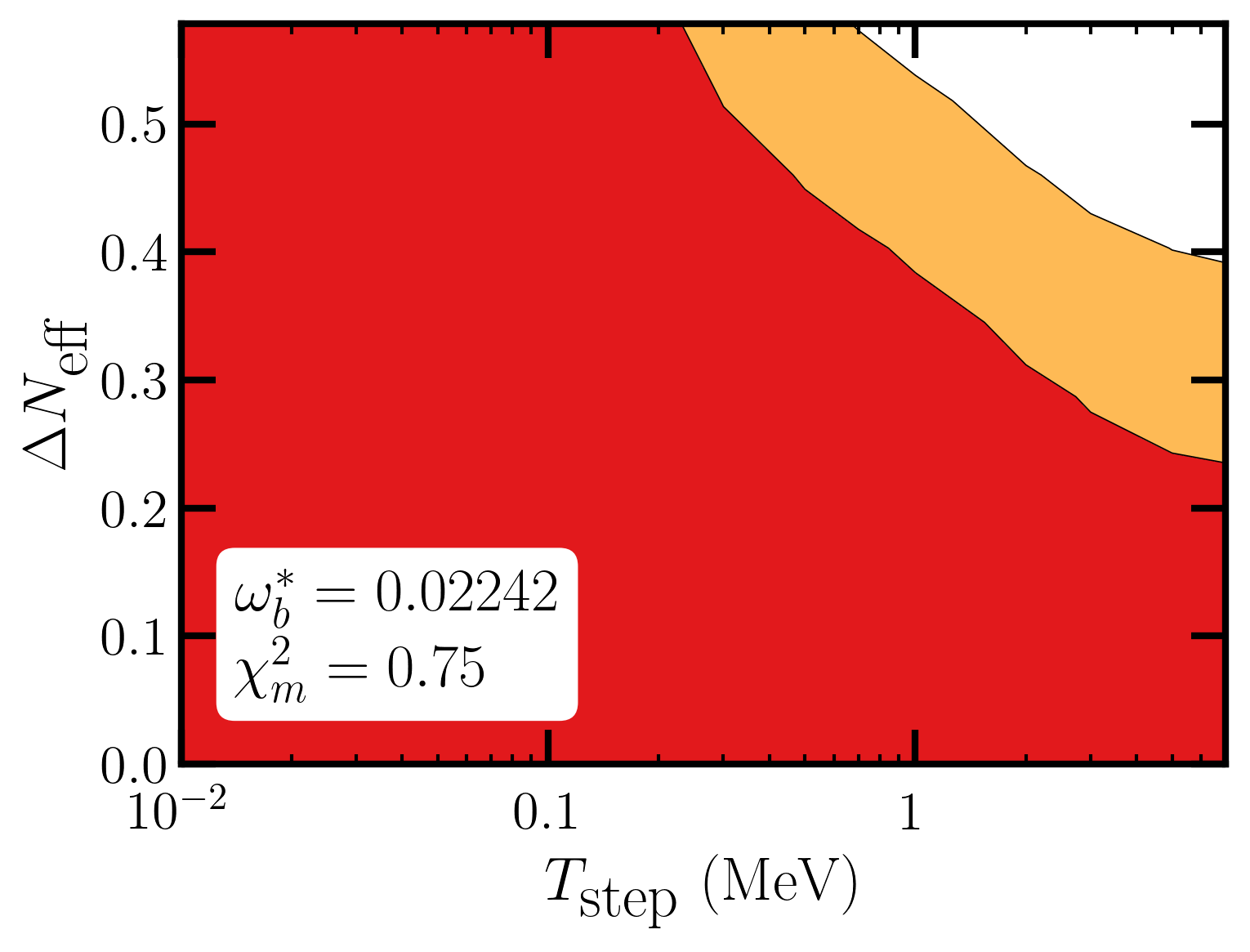}
    \includegraphics[width=0.2\textwidth]{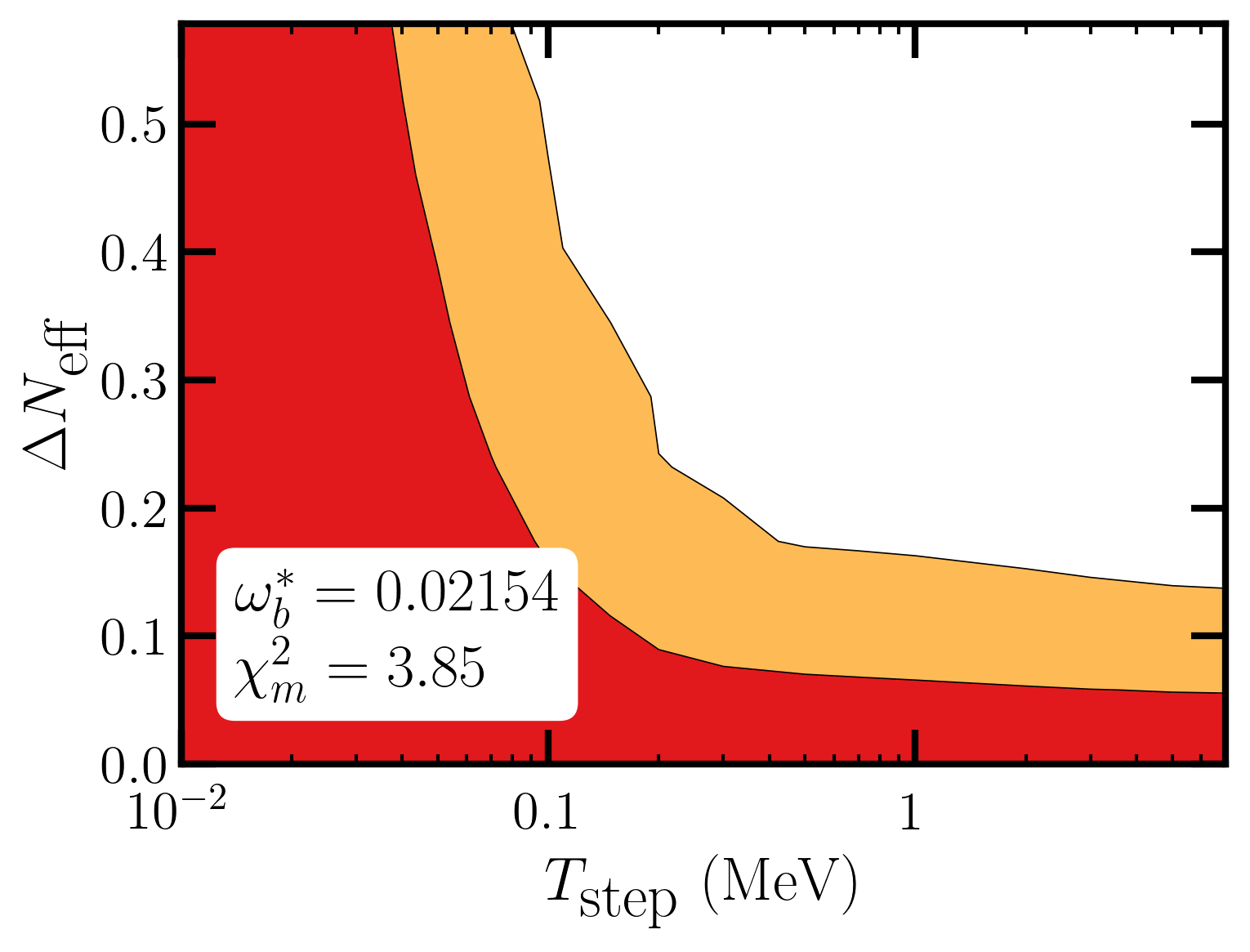}
    \includegraphics[width=0.2\textwidth]{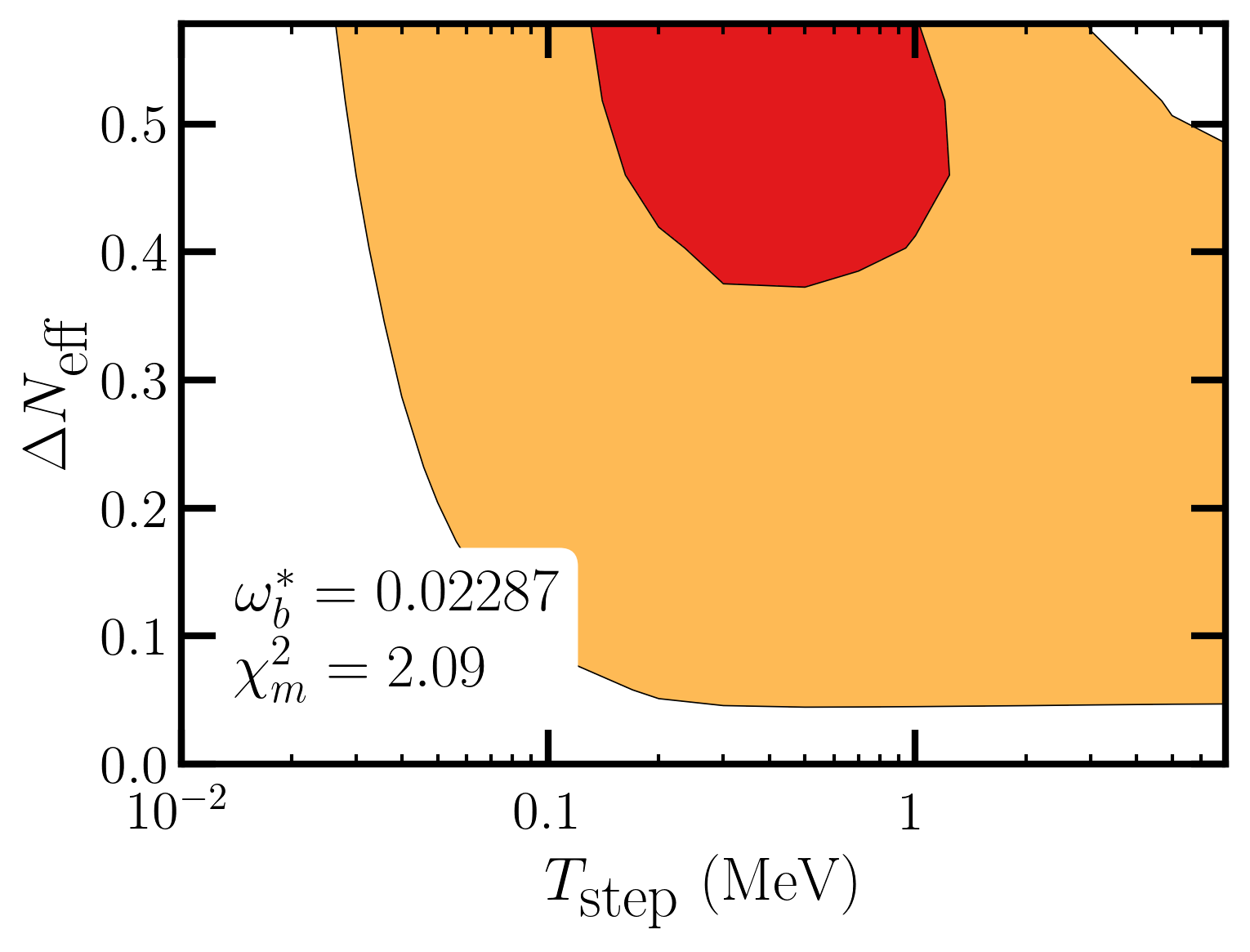}
    \caption{Limits on the step scenario, calculated with the PRIMAT network, for different values of $\omega_b^*$.}
    \label{fig:PRIMAT_step}
\end{figure*}

\begin{figure*}[h]
    \centering
    \includegraphics[width=0.2\textwidth]{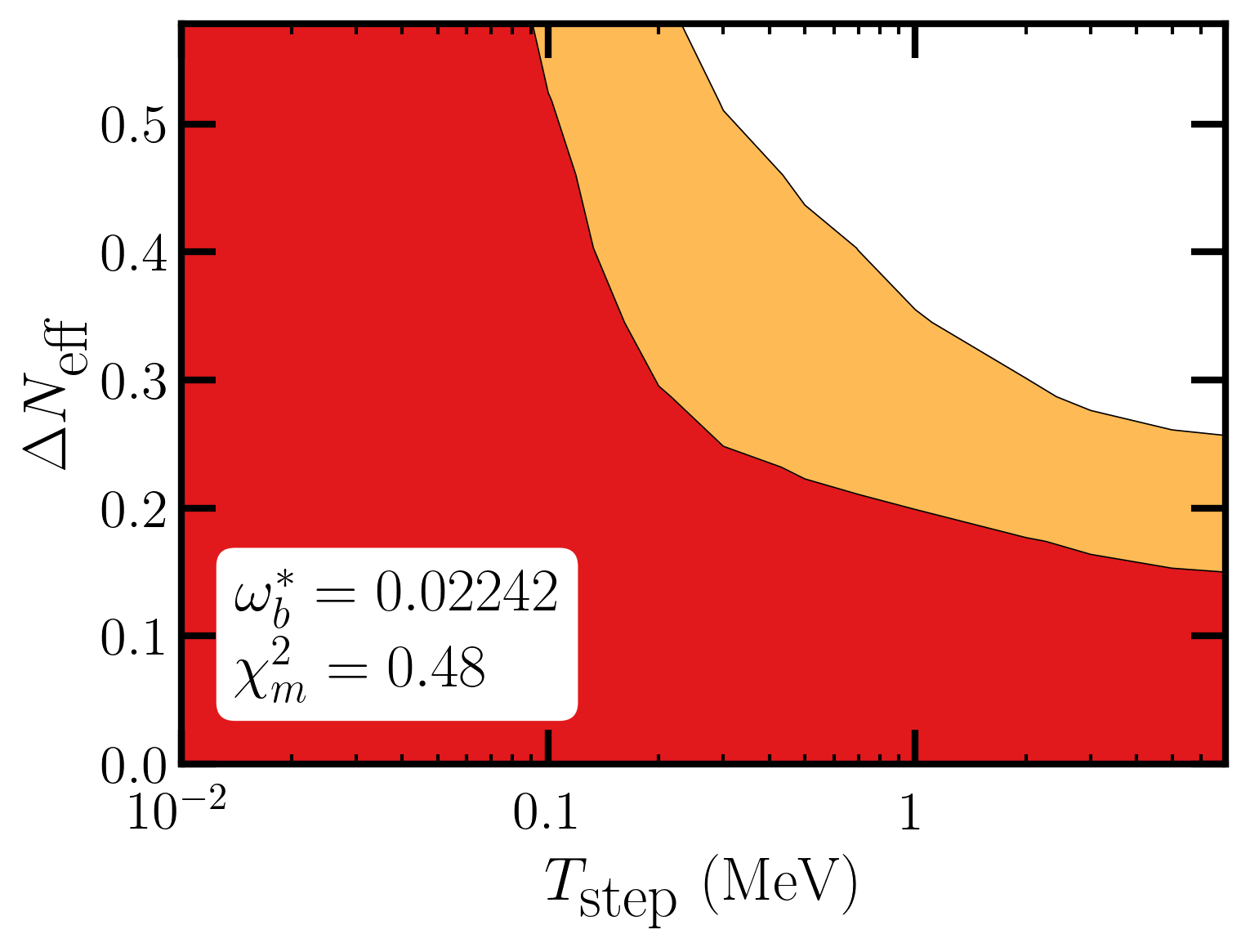}
    \includegraphics[width=0.2\textwidth]{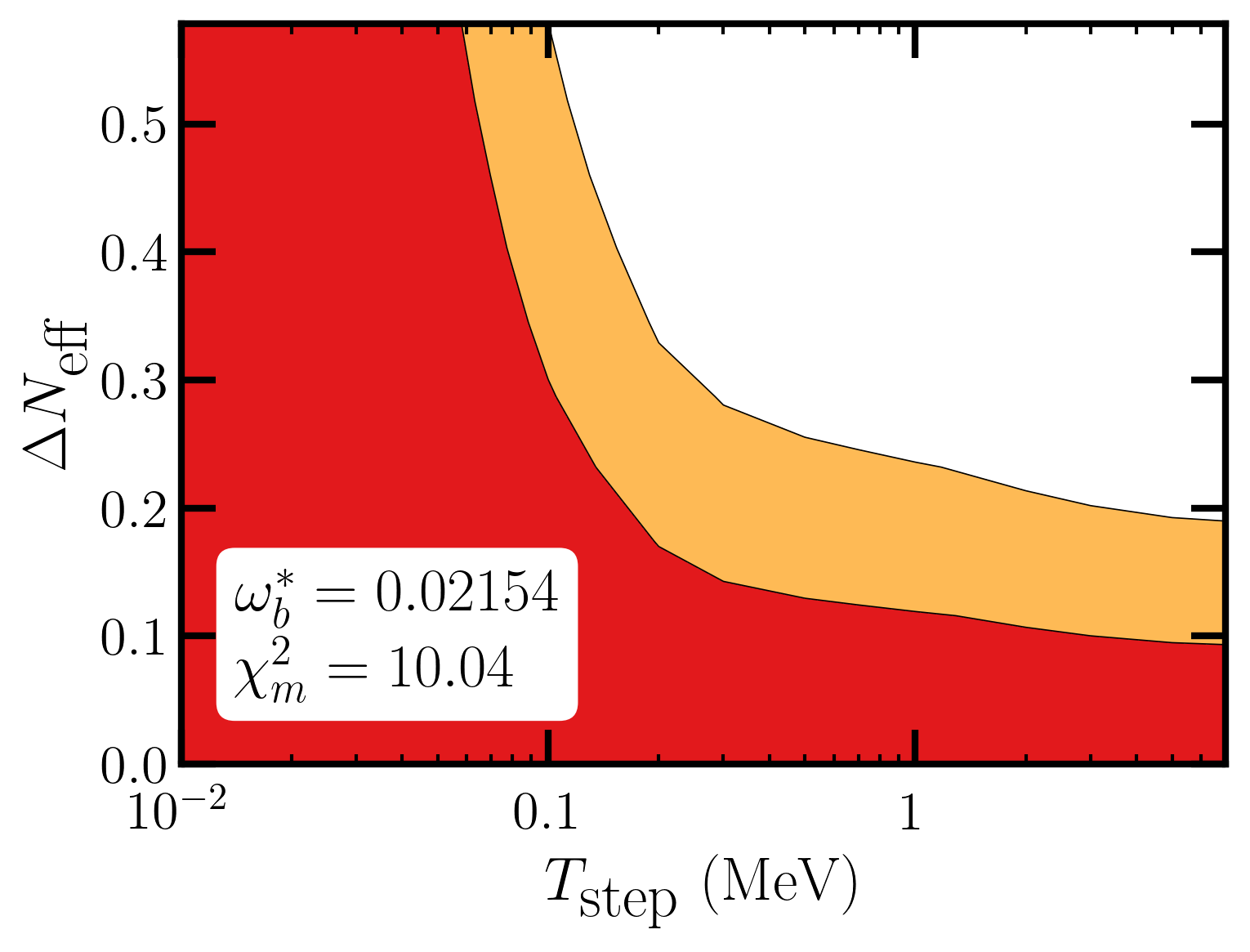}
    \includegraphics[width=0.2\textwidth]{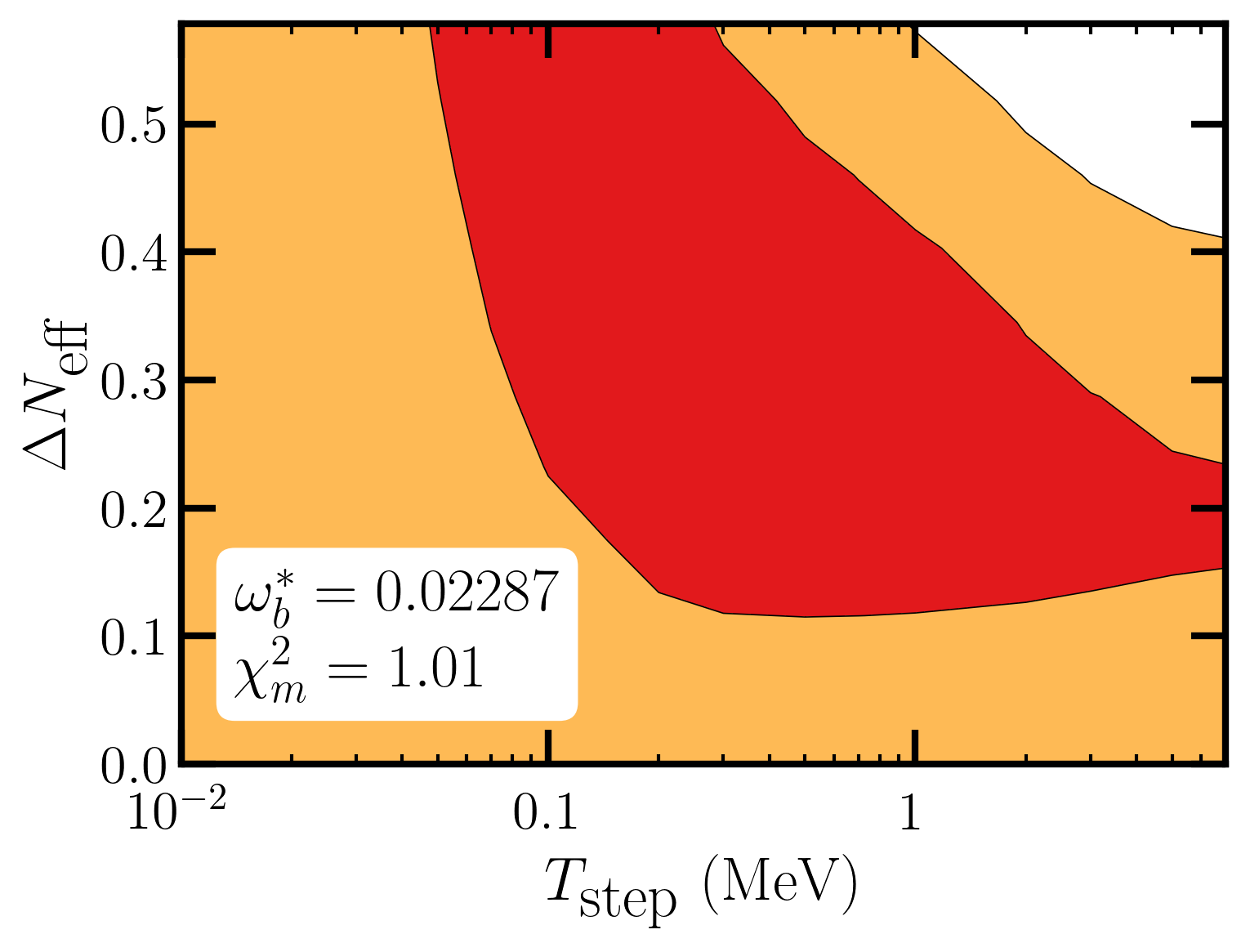}
    \caption{Limits on the step scenario, calculated with the PArthENoPE network, for different values of $\omega_b^*$.}
\end{figure*}

\begin{figure*}[h]
    \centering
    \includegraphics[width=0.2\textwidth]{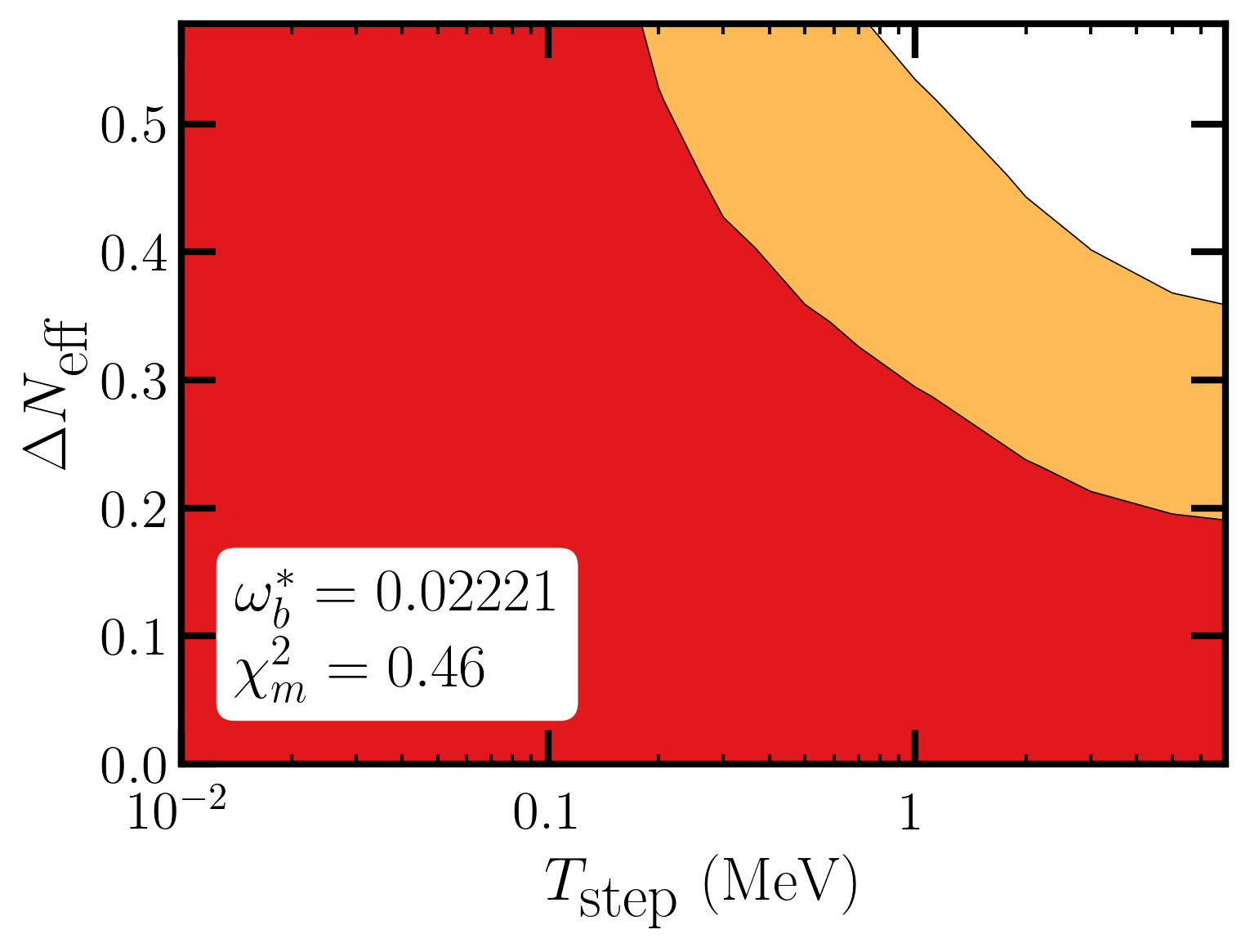}
    \includegraphics[width=0.2\textwidth]{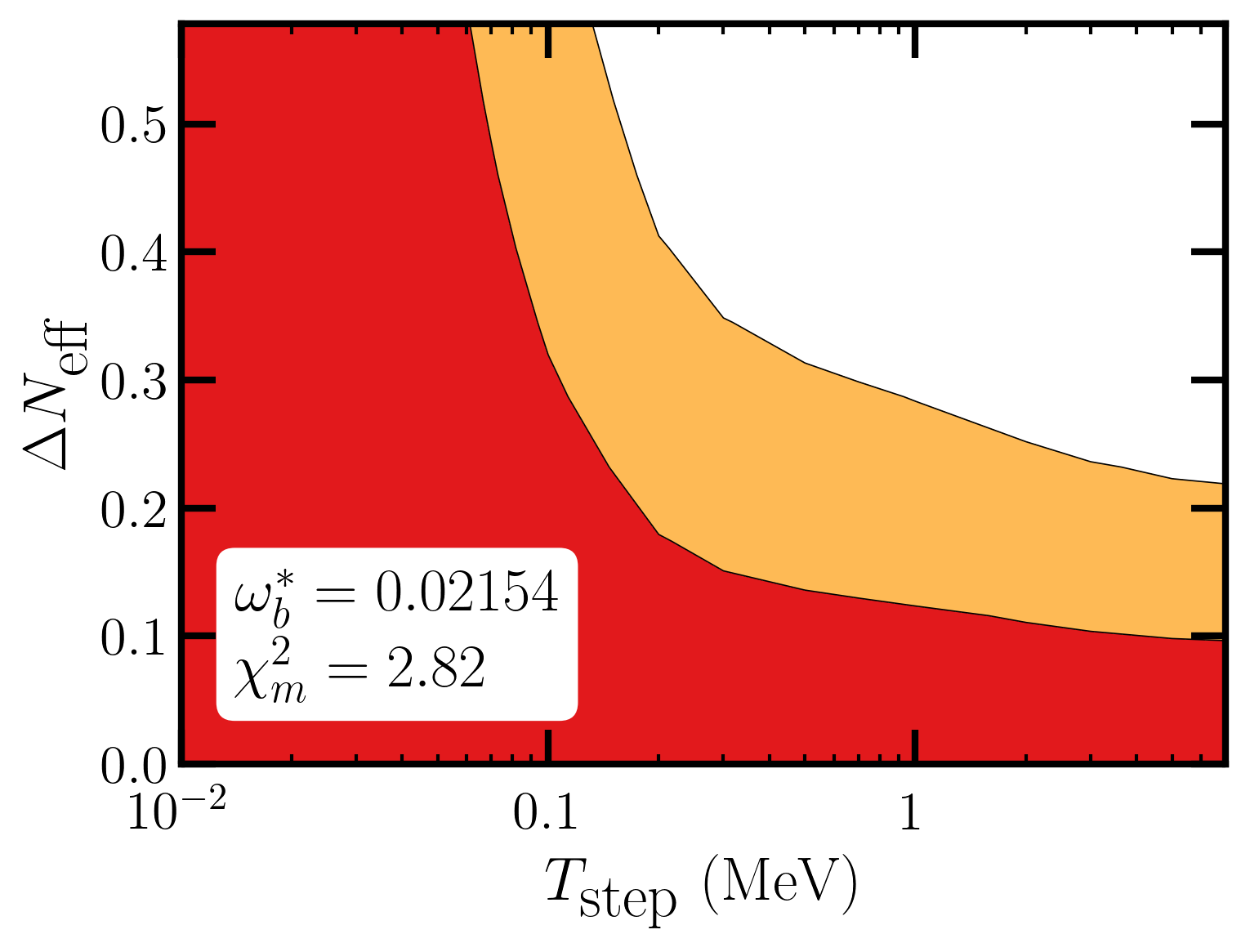}
    \includegraphics[width=0.2\textwidth]{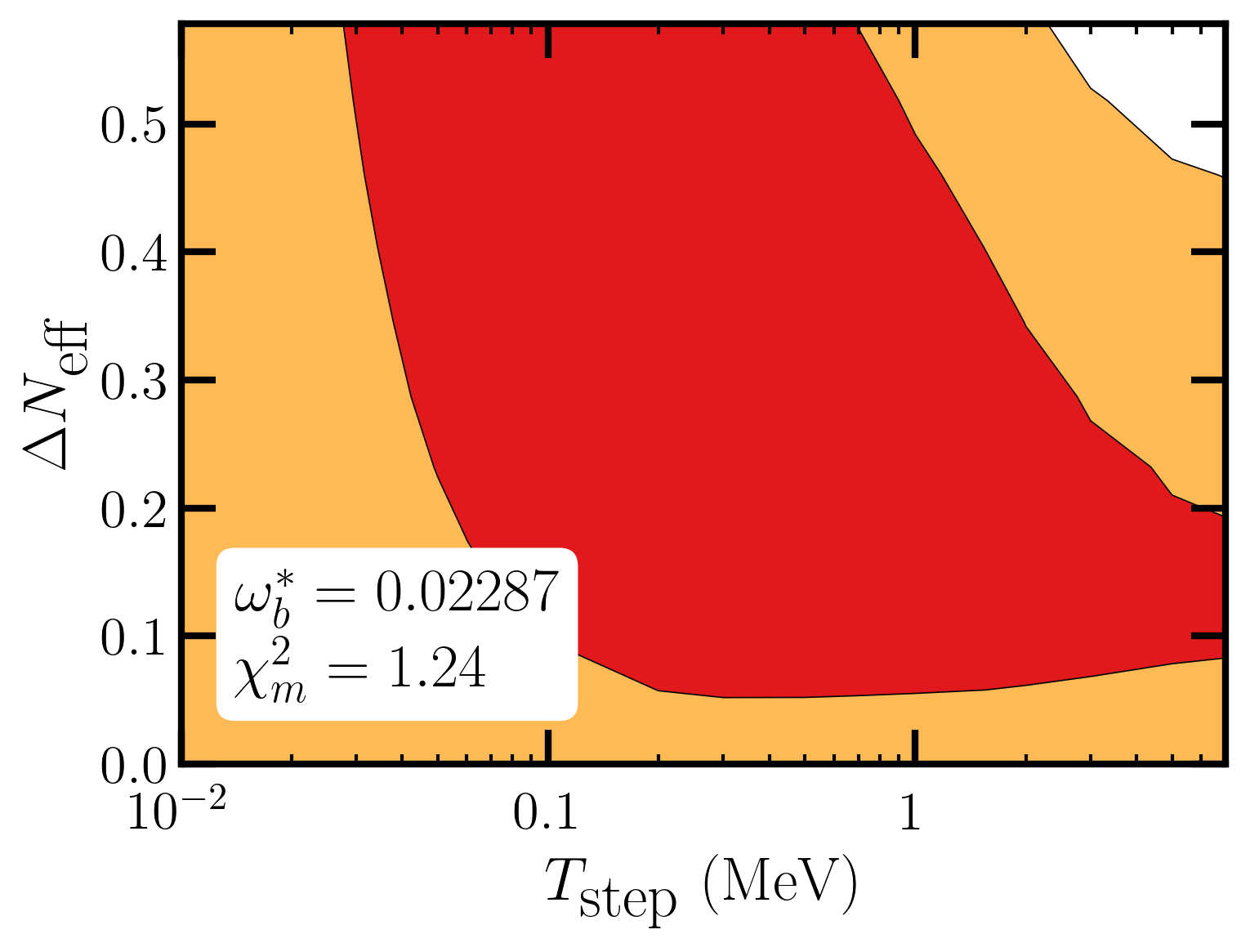}
    \caption{Limits on the step scenario, calculated with the ``combined" network, for different values of $\omega_b^*$.}
\end{figure*}

\end{document}